\documentclass[english]{article}
\usepackage{lmodern}

\usepackage[T1]{fontenc}
\usepackage[latin9]{inputenc}
\usepackage{geometry}
\usepackage{color}
\usepackage{babel}
\usepackage{verbatim}
\usepackage{rotating}
\usepackage{float}
\usepackage{units}
\usepackage{multirow}
\usepackage{amsmath}
\usepackage{amsthm}
\usepackage{amssymb}
\usepackage{graphicx}
\usepackage[unicode=true,pdfusetitle,
 bookmarks=true,bookmarksnumbered=false,bookmarksopen=false,
 breaklinks=false,pdfborder={0 0 1},backref=page,colorlinks=true]
 {hyperref}
\usepackage{breakurl}

\makeatletter

\floatstyle{ruled}
\newfloat{algorithm}{tbp}{loa}
\providecommand{\algorithmname}{Algorithm}
\floatname{algorithm}{\protect\algorithmname}

 \theoremstyle{definition}
   \newtheorem{defn}{\protect\defnname}
  \newtheorem{eg}{\protect\examplename}
  \theoremstyle{plain}
  
\theoremstyle{plain}
\newtheorem{thm}{\protect\theoremname}
\theoremstyle{plain}

\theoremstyle{plain}
\theoremstyle{plain}

\theoremstyle{plain}
\theoremstyle{plain}
\newtheorem{prop}{\protect\propositionname}
\theoremstyle{plain}

\usepackage{dsfont}
\hypersetup{pdftitle={Controlled Sequential Monte Carlo},pdfauthor={Heng et al},linkcolor=RoyalBlue,citecolor=RoyalBlue}
\usepackage[dvipsnames,svgnames,x11names,hyperref]{xcolor}

\@ifundefined{showcaptionsetup}{}{%
 \PassOptionsToPackage{caption=false}{subfig}}
\usepackage{subfig}
\makeatother

\providecommand{\defnname}{Definition}
\providecommand{\examplename}{Example}
\providecommand{\lemmaname}{Lemma}
\providecommand{\theoremname}{Theorem}
\providecommand{\assumptionname}{Assumption}
\providecommand{\propertyname}{Property}
\providecommand{\propositionname}{Proposition}

\begin{document}

\title{Controlled Sequential Monte Carlo}

\author{Jeremy Heng\thanks{ESSEC Business School; heng@essec.edu},  
Adrian N. Bishop\thanks{CSIRO and University of Technology Sydney; adrian.bishop@uts.edu.au}, 
George Deligiannidis$^\ddag$
and Arnaud Doucet\thanks{University of Oxford and the Alan Turing Institute; deligian@stats.ox.ac.uk, doucet@stats.ox.ac.uk} }
\date{}

\maketitle
\begin{abstract}
Sequential Monte Carlo methods, also known as particle methods, are a popular set of techniques for approximating high-dimensional probability distributions and their normalizing constants. These methods have found numerous applications in statistics and related fields; e.g. for inference in non-linear non-Gaussian state space models, and in complex static models. Like many Monte Carlo sampling schemes, they rely on proposal distributions which crucially impact their performance. We introduce here a class of controlled sequential Monte Carlo algorithms, where the proposal distributions are determined by approximating the solution to an associated optimal control problem using an iterative scheme. This method builds upon a number of existing algorithms in econometrics, physics, and statistics for inference in state space models, and generalizes these methods so as to accommodate complex static models. We provide a theoretical analysis concerning the fluctuation and stability of this methodology that also provides insight into the properties of related algorithms. We demonstrate significant gains over state-of-the-art methods at a fixed computational complexity on a variety of applications.
\end{abstract}
\textbf{\small{}Keywords}{\small{}: State space models, annealed importance sampling, normalizing constants, optimal control, approximate dynamic programming, 
reinforcement learning.}{\small \par}

\section{Introduction\label{sec:Introduction}}

Sequential Monte Carlo (SMC) methods have found a wide range of applications
in many areas of statistics as they can be used, among others things, to perform inference for
dynamic non-linear non-Gaussian state space models \cite{Liu_Chen_1998,Pitt_Shephard_1999,Liu_2001} but also
for complex static models \cite{Neal_2001,Chopin_2002,DelMoral_Doucet_Jasra_2006}; see
\cite{Chen_2013,Doucet_Johansen_2011,Kunsch_2013} for recent reviews of this active area.
Although these methods are supported by theoretical guarantees \cite{DelMoral_2004}, the
number of samples required to achieve a desired level of precision of the corresponding Monte Carlo estimators
can be prohibitively large in practice, especially so for high-dimensional problems.

The present work is one means to address the computational difficulties with SMC in offline inference settings.
In particular, we leverage ideas from optimal control theory and we seek novel SMC methods that achieve a desired level of precision at a fraction of the computational cost of state-of-the-art algorithms. We introduce a class of algorithms that will be referred to as controlled SMC, under which the sequence of SMC proposal distributions are related naturally with an associated optimal control problem. The cost functional is the Kullback--Leibler divergence from the sought after proposals to the target distributions and we may account for an arbitrary current proposal estimate. With this formulation, the optimal proposal distributions are specified by the optimal control policy of a related dynamic programming recursion. In general, this dynamic programming recursion is intractable. However, by making this connection, we can then exploit an array of methods and procedures for so-called approximate dynamic programming (ADP).
Broadly speaking, a single iteration of our proposed methodology involves:
1) based on a current sequence of proposal distributions, running a SMC method
to obtain a collection of samples that approximate the sequence of SMC target distributions;
2) using these samples as support points, we approximate intractable backward recursions using regression to
compute a new policy that specifies a new sequence of approximately optimal proposal distributions.
Continuing in this manner allows us to further refine the proposal distributions and improve our approximation of
the target distributions, via a novel iteration of SMC and ADP.

Prior influential work in \cite{Richard_Zhang_2007} proposed a motivating method in the context of importance sampling, for computing the marginal likelihood in state space models. In this contribution, the sequential structure which defines the marginal likelihood is exploited
and proposal distributions are defined by a sequence of parameterized Markov transition kernels.
A criterion based on the variance of importance weights is introduced to optimize these parameters and
an iterative procedure with fixed random numbers is proposed.
The work of \cite{Schartha_Kohn_2016} extends \cite{Richard_Zhang_2007} by employing these optimized proposal distributions within
a SMC methodology.
In particular, \cite{Schartha_Kohn_2016} identified the appropriate importance weights one should use for resampling,
which is crucial to ensure that the variance of the marginal likelihood estimator remains controlled.
Moreover, \cite{Schartha_Kohn_2016} also recommends relaxing the use of common random variables across iterations.
Recent work in \cite{Guarniero_Lee_Johansen_2016}, again with a focus on discrete time state space models,
may be viewed as an extension of \cite{Schartha_Kohn_2016} where resampling is performed at every iteration, instead of just the last. The resulting algorithm is numerically much more stable than \cite{Richard_Zhang_2007,Schartha_Kohn_2016}.
Although the iterative procedures in \cite{Richard_Zhang_2007,Schartha_Kohn_2016,Guarniero_Lee_Johansen_2016} are
similar in spirit to our proposed methodology, the main and important difference is that all these works employ an optimality criterion, to learn proposal distributions, that is not adjusted across iterations to account for any improvements made in prior iterations.
Finally, we highlight related ideas in \cite{Kappen_Ruiz_2016,Ruiz_Kappen_2017} where the focus is partially observed diffusion models and the algorithms proposed therein are based on other strategies to learn a parameterized additive control directly. Such ideas have also been exploited in physics to perform rare event simulation for diffusions \cite{Nemoto_2016}.

Our work extends these contributions in the following ways. Firstly, these preceding works \cite{Richard_Zhang_2007,Schartha_Kohn_2016,Guarniero_Lee_Johansen_2016} consider only state space models. In contrast, the methodology proposed here allows us to perform inference for static models; a direct extension of these prior methods \cite{Richard_Zhang_2007,Schartha_Kohn_2016,Guarniero_Lee_Johansen_2016} to static models is infeasible, as it leads to algorithms which are not implementable.

Secondly, in contrast to the methodology in \cite{Richard_Zhang_2007,Schartha_Kohn_2016,Guarniero_Lee_Johansen_2016}, the Kullback--Leibler optimality criterion at each iteration in our approach, is dependent on the approximately optimal proposal distributions computed at the preceding iteration; i.e.
we seek to minimize the residual discrepancy between any previously estimated proposals and the target distributions.
This difference allows us to elucidate the effect each iteration in our method has on refining proposal distributions
and improves algorithmic performance as illustrated in Section \ref{sec:lorenz}.
See also \cite{Theodorou_Todorov_2012,Thijssen_Kappen_2015} for related iterative procedures in continuous-time optimal control approximation.

The controlled SMC methodology is one of the main contributions of this work. Another contribution is to provide a detailed theoretical analysis of various aspects of our methodology.
In Proposition \ref{prop:moment_bound}, we provide a backward recursion that characterizes the error of policies estimated using our ADP procedure. This error is given naturally in terms of function approximation errors with finite samples and the stability properties of the dynamic programming recursion defining the optimal policy, which is addressed in Proposition \ref{prop:stability}.
These results show that we can obtain good approximations of the optimal policy and hence the optimal proposal distributions, if the function classes employed are `rich' enough and the number of samples used to learn policies is sufficiently large.
In Theorem \ref{thm:LimitShorten}, we then establish a central limit theorem for our ADP algorithm as the number of samples used in the policy learning goes to infinity. This reveals that the algorithm concentrates around an idealized ADP algorithm and provides a precise characterization of how Monte Carlo errors correlate over time. These preceding results concern a single iteration of our proposed method and may be applied to the existing algorithms discussed above, e.g. \cite{Richard_Zhang_2007,Schartha_Kohn_2016,Guarniero_Lee_Johansen_2016}. Using the notion of iterated random functions, we introduce a novel framework in Theorem \ref{thm:iADP} to understand the asymptotic behaviour of our algorithm as the number of iterations converges to infinity. This elucidates the need for iterating the ADP procedure and provides insight into the number of iterations required in practice. The discussion surrounding Theorem \ref{thm:iADP} also emphasizes a key difference between the newly proposed method and existing work in \cite{Richard_Zhang_2007,Schartha_Kohn_2016,Guarniero_Lee_Johansen_2016}. After the first version of this work appeared, a similar approach was developed for generic stochastic control problems in \cite{gupta_jain_glynn_2018}. Our results hold under strong assumptions but appear to capture our experimental results remarkably well.

The rest of this paper is organized as follows. We introduce SMC methods in the framework of 
Feynman-Kac models \cite{DelMoral_2004} in Section \ref{sec:motivating_models_smc} and 
twisted variants in Section \ref{sec:twisted_models_smc}, as this affords us the generality to 
cover both state space models and static models. We then identify the optimal policy that induces an
optimal SMC method in Section \ref{sec:optimal_policies}.
We describe general methods to approximate the optimal policy in Section
\ref{sec:ADP} and develop an iterative scheme to refine
policies in Section \ref{sec:policy_refinement}.
The proposed methodology is illustrated on a neuroscience application in Section \ref{sec:neuroscience}.
We present the results of our analysis in Section \ref{sec:analysis}
and conclude with applications in Sections \ref{sec:application_statesspacemodels}-\ref{sec:application_staticmodels}. All proofs are given in the
Supplementary Material which also includes three additional applications. MATLAB code to reproduce all numerical results is available online\footnote{Link: \url{https://github.com/jeremyhengjm/controlledSMC}}.

\section{Motivating models and sequential Monte Carlo\label{sec:motivating_models_smc}}

\subsection{Notation}
We first introduce notation used throughout the article.
Given integers $n\leq m$ and a sequence $(x_{t})_{t\in\mathbb{N}}$,
we define the set $[n:m]=\left\{ n,\ldots,m\right\}$ and write the subsequence
$x_{n:m}=(x_{n},\ldots,x_{m})$. When $n<m$, we use the convention $\prod_{t=m}^{n}x_{t}=1$.
Let $(\mathsf{E},\mathcal{E})$ be an arbitrary measurable space.
We denote the set of all finite signed measures by $\mathcal{S}(\mathsf{E})$,
the set of all probability measures by $\mathcal{P}(\mathsf{E})\subset\mathcal{S}(\mathsf{E})$,
and the set of all Markov transition kernels on $(\mathsf{E},\mathcal{E})$
by $\mathcal{M}(\mathsf{\mathsf{E}})$.
Given $\mu,\nu\in\mathcal{P}(\mathsf{E})$, we write $\mu\ll\nu$ if $\mu$ is absolutely
continuous w.r.t. $\nu$ and denote the corresponding Radon-Nikodym derivative
as $\mathrm{d}\mu/\mathrm{d}\nu$.
For any $x\in\mathsf{E}$, $\delta_{x}$ denotes the Dirac measure
at $x$.
The set of all real-valued, $\mathcal{E}$-measurable, lower bounded, bounded
or continuous functions on $\mathsf{E}$ are denoted by $\mathcal{L}(\mathsf{E})$, $\mathcal{B}(\mathsf{E})$  and $\mathcal{C}(\mathsf{E})$ respectively.
Given $\gamma\in\mathcal{S}(\mathsf{E})$ and $M\in\mathcal{M}(\mathsf{E})$,
we define $(\gamma\otimes M)(\mathrm{d}x,\mathrm{d}y)=\gamma(\mathrm{d}x)M(x,\mathrm{d}y)$
and $(M\otimes\gamma)(\mathrm{d}x,\mathrm{d}y)=M(y,\mathrm{d}x)\gamma(\mathrm{d}y)$
as  finite signed measures on the product space $\mathsf{E}\times\mathsf{E}$,
equipped with the product $\sigma$-algebra $\mathcal{E}\times\mathcal{E}$.
Given $\gamma\in\mathcal{S}(\mathsf{E})$, $M\in\mathcal{M}(\mathsf{E})$,
$\varphi\in\mathcal{B}(\mathsf{E})$, $\xi\in\mathcal{B}(\mathsf{E}\times\mathsf{E})$,
we define the integral $\gamma(\varphi)=\int_{\mathsf{E}}\varphi(x)\gamma({\rm d}x)$,
the signed measure $\gamma M(\cdot)=\int_{\mathsf{E}}\gamma({\rm d}x)M(x,\cdot)\in\mathcal{S}(\mathsf{E})$
and functions $M(\varphi)(\cdot)=\int_{\mathsf{E}}\varphi(y)M(\cdot,{\rm d}y)\in\mathcal{B}(\mathsf{E})$,
$M(\xi)(\cdot)=\int_{\mathsf{E}}\xi(\cdot,y)M(\cdot,{\rm d}y)\in\mathcal{B}(\mathsf{E})$.

\subsection{Feynman-Kac models\label{sec:feynman_kac}}
We begin by introducing Feynman-Kac models \cite{DelMoral_2004}
and defer a detailed discussion of their applications to
Sections \ref{sec:statespace_models}-\ref{sec:static_models}.
Consider a non-homogeneous Markov chain of length $T+1\in\mathbb{N}$
on a measurable space $(\mathsf{X},\mathcal{X})$, associated with
an initial distribution $\mu\in\mathcal{P}(\mathsf{X})$, and
a collection of Markov transition kernels $M_{t}\in\mathcal{M}(\mathsf{X})$ for $t\in[1:T]$.
We denote the law of the Markov chain on path space $\mathsf{X}^{T+1}$,
equipped with the product $\sigma$-algebra $\mathcal{X}^{T+1}$,
with
\begin{equation}
\mathbb{Q}(\mathrm{d}x_{0:T})=\mu(\mathrm{d}x_{0})\prod_{t=1}^{T}M_{t}(x_{t-1},\mathrm{d}x_{t})\label{eq:Q}
\end{equation}
and denote expectations w.r.t. $\mathbb{Q}$ by $\mathbb{E}_{\mathbb{Q}}$,
whereas we write $\mathbb{E}_{\mathbb{Q}}^{t,x}$ for conditional expectations
on the event $X_{t}=x\in\mathsf{X}$. Given a sequence of strictly
positive functions $G_{0}\in\mathcal{B}(\mathsf{X})$ and $G_{t}\in \mathcal{B}(\mathsf{X}\times\mathsf{X})$ for $t\in[1:T]$,
we define the Feynman-Kac path measure
\begin{equation}
\mathbb{P}(\mathrm{d}x_{0:T})=Z^{-1}G_{0}(x_{0})\prod_{t=1}^{T}G_{t}(x_{t-1},x_{t})\,\mathbb{Q}(\mathrm{d}x_{0:T})\label{eq:P}
\end{equation}
where $Z:=\mathbb{E}_{\mathbb{Q}}\left[G_{0}(X_{0})\prod_{t=1}^{T}G_{t}(X_{t-1},X_{t})\right]$
denotes the normalizing constant. Equation (\ref{eq:P}) can be understood
as the probability measure obtained by repartitioning the probability
mass of $\mathbb{Q}$ with the potential functions $\left(G_{t}\right)_{t\in[0:T]}$.

To examine the time evolution of (\ref{eq:P}), we define the following
sequence of positive signed measures $\gamma_{t}\in\mathcal{S}(\mathsf{X})$ for $t\in[0:T]$
by
\begin{equation}
\gamma_{t}(\varphi)=\mathbb{E}_{\mathbb{Q}}\bigg[\varphi(X_{t})G_{0}(X_{0})\prod_{s=1}^{t}G_{s}(X_{s-1},X_{s})\bigg]\label{eq:unnorm_FKmodel}
\end{equation}
and their normalized counterparts $\eta_{t}\in\mathcal{P}(\mathsf{X})$
by
\begin{equation}
\eta_{t}(\varphi)=\gamma_{t}(\varphi)/Z_{t}\label{eq:norm_FKmodel}
\end{equation}
for $\varphi\in\mathcal{B}(\mathsf{X})$, $t\in[0:T]$, where $Z_{t}:=\gamma_{t}(\mathsf{X})$.
Equations (\ref{eq:unnorm_FKmodel}) and (\ref{eq:norm_FKmodel})
are known as the unnormalized and normalized (updated) Feynman-Kac
models respectively  \cite[Definition 2.3.2]{DelMoral_2004}. These
models are determined by the triple $\left\{ \mu,(M_{t})_{t\in[1:T]},(G_{t})_{t\in[0:T]}\right\} $,
which depends on the specific application of interest.
The measure $\eta_{T}$ is the terminal time marginal distribution of $\mathbb{P}$ and
$Z=Z_{T}=\mu(G_{0})\prod_{t=1}^{T}\eta_{t-1}(M_{t}(G_{t}))$.

\subsection{State space models\label{sec:statespace_models}}
Consider an $\mathsf{X}$-valued hidden Markov chain $(X_{t})_{t\in[0:T]}$, whose
law on $(\mathsf{X}^{T+1},\mathcal{X}^{T+1})$ is given by
\[
\mathbb{H}(\mathrm{d}x_{0:T})=\nu(\mathrm{d}x_{0})\prod_{t=1}^{T}f_{t}(x_{t-1},\mathrm{d}x_{t})
\]
where $\nu\in\mathcal{P}(\mathsf{X})$ and $f_{t}\in\mathcal{M}(\mathsf{X})$ for $t\in[1:T]$.
The $\mathsf{Y}$-valued observations $(Y_{t})_{t\in[0:T]}$ are assumed
to be conditionally independent given $(X_{t})_{t\in[0:T]}$ and the conditional distribution of $Y_{t}$ has
a strictly positive density $g_{t}(X_{t},\cdot)$ with $g_{t}\in\mathcal{B}(\mathsf{X}\times\mathsf{Y})$ for $t\in[0:T]$.
Here $\left\{ \nu,(f_{t})_{t\in[1:T]},(g_{t})_{t\in[0:T]}\right\} $ can potentially depend on unknown static parameters $\theta\in\Theta$, but this is notationally omitted
for simplicity. Given access to a realization $y_{0:T}\in\mathsf{Y}^{T+1}$ of the observation process, statistical inference for these
models relies on the marginal likelihood of $y_{0:T}$ given
$\theta$,
\[
Z(y_{0:T})=\mathbb{E}_{\mathbb{H}}\left[\prod_{t=0}^{T}g_{t}(X_{t},y_{t})\right],
\]
and/or the smoothing distribution, i.e. the conditional distribution of
$X_{0:T}$ given $Y_{0:T}=y_{0:T}$ and $\theta$
\begin{equation}
\mathbb{P}(\mathrm{d}x_{0:T}|y_{0:T})=Z(y_{0:T})^{-1}\prod_{t=0}^{T}g_{t}(x_{t},y_{t})\,\mathbb{H}(\mathrm{d}x_{0:T}).\label{eq:smoothing_distribution}
\end{equation}
If we set  $\mathbb{Q}\in\mathcal{P}(\mathsf{X}^{T+1})$  defined in (\ref{eq:Q})  equal to $\mathbb{H}$, we recover the Feynman-Kac path measure representation (\ref{eq:P}) by defining $G_{t}(x_{t-1},x_t)=g_{t}(x_t,y_{t})$ for all $t\in[0:T]$. However, this representation is not unique. Indeed any $\mathbb{Q}$ satisfying $\mathbb{H}\ll\mathbb{Q}$ provides a Feynman-Kac path measure representation of (\ref{eq:P}) by defining the potentials
\[
G_{0}(x_{0})=\frac{\mathrm{d}(\nu\cdot g_{0})}{\mathrm{d}\mu}(x_{0}),\quad G_{t}(x_{t-1},x_{t})=\frac{\mathrm{d}(f_{t}\cdot g_{t})(x_{t-1,}\cdot)}{\mathrm{d}M_{t}(x_{t-1},\cdot)}(x_{t}),\quad t\in[1:T].
\]
As outlined in \cite{Doucet_Johansen_2011}, most SMC algorithms available at present correspond to the same basic mechanism applied to different Feynman-Kac representations of a given target probability measure. The bootstrap particle filter (BPF) presented in \cite{Gordon_Salmond_Smith_1993} corresponds to $\mathbb{Q}=\mathbb{H}$, i.e. $M_{t}(x_{t-1},\mathrm{d}x_{t})=f_{t}(x_{t-1},\mathrm{d}x_{t})$ for $t\in[1:T]$, while the popular `fully adapted' auxiliary particle filter (APF) of \cite{Pitt_Shephard_1999} uses $M_{t}(x_{t-1},\mathrm{d}x_{t})=\mathbb{P}(\mathrm{d}x_{t}|x_{t-1},y_{t})\propto f_{t}(x_{t-1},\mathrm{d}x_{t}) g_{t}(x_{t},y_{t})$.

\begin{figure}
\begin{centering}
\includegraphics[scale=0.4]{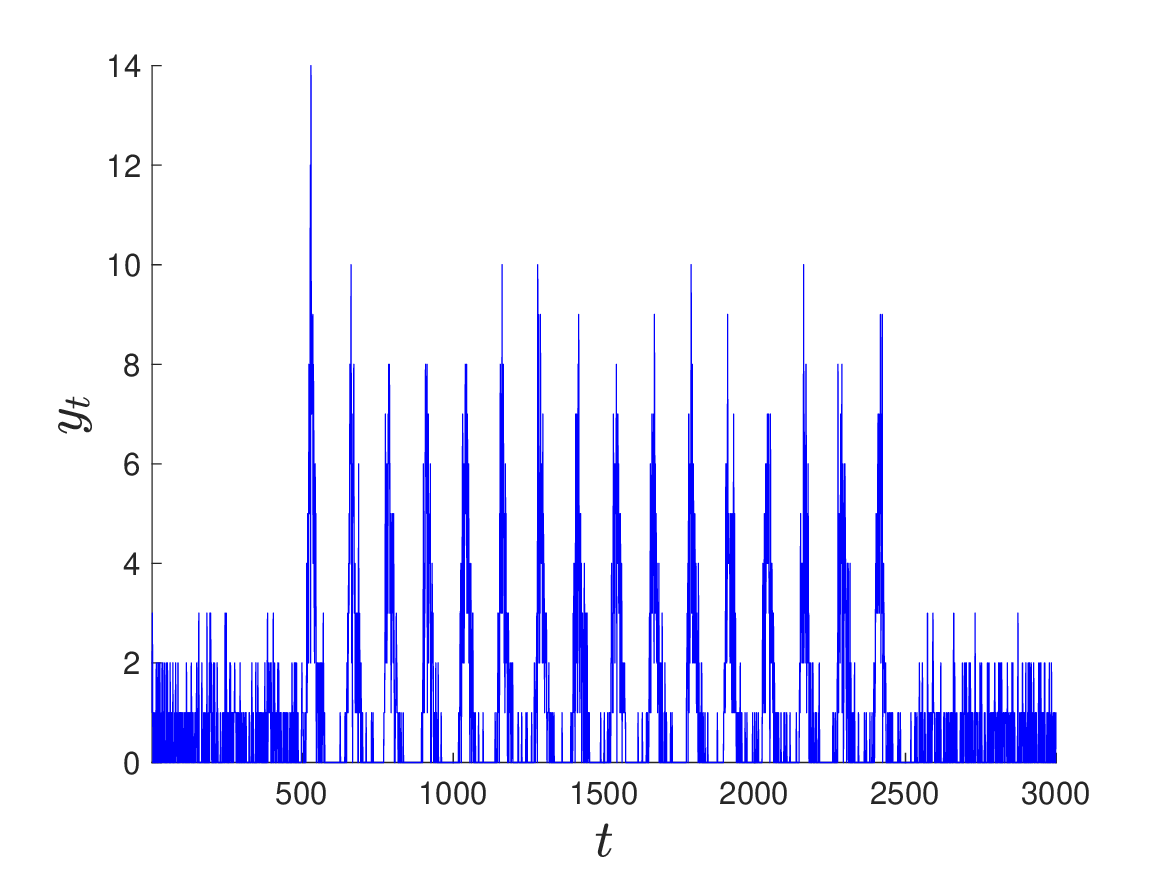}\includegraphics[scale=0.4]{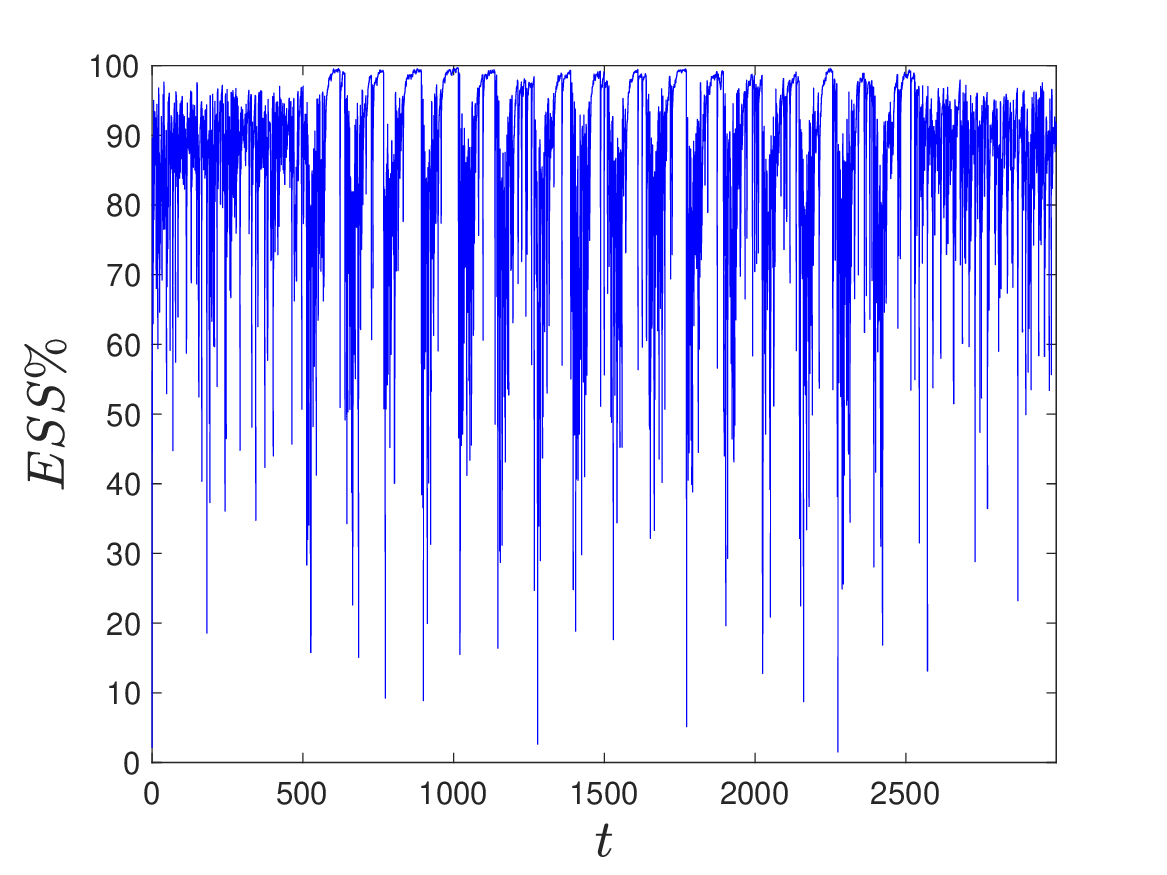}
\par\end{centering}
\caption{\label{fig:neuroscience_obs}
Number of activated neurons over $M=50$ repeated experiments with time (\emph{left}) and
effective sample size of bootstrap particle filter with $N=1024$ particles (\emph{right}) for the neuroscience model with parameters
$\alpha=0.99$ and $\sigma^2=0.11$.}
\end{figure}

As a motivating example, we consider a model for $T+1=3000$ measurements collected from a
neuroscience experiment \cite{Temereanca_Brown_Simons_2008}.
The observation $y_t\in\mathsf{Y}=[0:M]$ at each time instance $t\in[0:T]$, shown in the left panel of
Figure \ref{fig:neuroscience_obs},
represents the number of activated neurons over $M=50$ repeated experiments
and is modelled as a binomial distribution with probability of success
$p_{t}\in[0,1]$.
We will write its probability mass function as $y_{t}\mapsto\mathrm{Bin}(y_{t};M,p_{t})$.
To model the time varying behaviour of activation probabilities,
it is assumed that $p_{t}=\kappa(X_{t})$ where $\kappa(u):=(1+\exp(-u))^{-1}$
for $u\in\mathbb{R}$ is the logistic link function and $(X_{t})_{t\in[0:T]}$ is a real-valued first-order autoregressive process.
This corresponds to a time homogeneous state space model on $\mathsf{X}=\mathbb{R}$,
equipped with its Borel $\sigma$-algebra $\mathcal{X}=\mathfrak{B}(\mathbb{R})$, with $\nu=\mathcal{N}(0,1)$, $f(x_{t-1},\mathrm{d}x_{t})=\mathcal{N}(x_{t};\alpha x_{t-1},\sigma^{2})\mathrm{d}x_{t}$,
and $g(x_{t},y_{t})=\mathrm{Bin}(y_{t};M,\kappa(x_{t}))$ for  $t\in[1:T]$,
where we denote the Gaussian distribution on $\mathbb{R}^{d}$ with mean
vector $\xi\in\mathbb{R}^{d}$ and covariance matrix $\Sigma\in\mathbb{R}^{d\times d}$
by $\mathcal{N}(\xi,\Sigma)$ and its Lebesgue density by $x\mapsto\mathcal{N}(x;\xi,\Sigma)$.
The parameters of this model to be inferred from data are $\theta=(\alpha,\sigma^{2})\in[0,1]\times\mathbb{R}_{+}$.
  
\subsection{Static models\label{sec:static_models}}
Suppose we are interested in sampling from a target distribution
$\eta(\mathrm{d}x)=Z^{-1}\gamma(\mathrm{d}x)\in\mathcal{P}(\mathsf{X})$
and/or estimating its normalizing constant $Z=\gamma(\mathsf{X})$.
To facilitate inference, we introduce a sequence of probability measures
$(\eta_{t})_{t\in[0:T]}$ in $\mathcal{P}(\mathsf{X})$ that bridges
a simple distribution $\eta_{0}=\mu$ to the target distribution $\eta_{T}=\eta$ with $\eta\ll\mu$.
Our implementation in Section \ref{sec:application_staticmodels}
adopts the geometric path \cite{Gelman_Meng_1998,Neal_2001,DelMoral_Doucet_Jasra_2006}
\begin{equation}
\gamma_{t}(\mathrm{d}x):=\mu(\mathrm{d}x) \left(\frac{\mathrm{d}\gamma}{\mathrm{d}\mu}(x)\right)^{\lambda_{t}},\quad\eta_{t}(\mathrm{d}x):=\gamma_{t}(\mathrm{d}x)/Z_{t},\quad t\in[0:T],\label{eq:geometric_path}
\end{equation}
where $Z_{t}:=\gamma_{t}(\mathsf{X})$ and $(\lambda_{t})_{t\in[0:T]}\in [0,1]^{T+1}$
is an increasing sequence satisfying $\lambda_{0}=0$ and $\lambda_{T}=1$; see \cite[Section 2.3.1]{DelMoral_Doucet_Jasra_2006} for choices in other inference settings.
In order to define $\mathbb{Q}$, we introduce a sequence of `forward' Markov transition kernels $M_{t}\in\mathcal{M}(\mathsf{X})$ for $t=[1:T]$
where $\eta_{t-1}M_{t}$ approximates $\eta_{t}$. One expects
the distribution $\hat{\eta}=\eta_{0}M_{1}\cdots M_{T}$ of samples
drawn from a non-homogeneous Markov chain with initial distribution
$\eta_{0}$ and transition kernels $(M_{t})_{t\in[1:T]}$ to be close
to $\eta_{T}=\eta$. However, importance sampling cannot be employed
to correct for the discrepancy between $\hat{\eta}$ and $\eta$, as
$\hat{\eta}$ is typically analytically intractable.

SMC samplers
described in \cite{DelMoral_Doucet_Jasra_2006} circumvent this difficulty
by performing importance sampling on path space $(\mathsf{X}^{T+1},\mathcal{X}^{T+1})$
using an artificial extended target distribution of the form
\[
\mathbb{P}(\mathrm{d}x_{0:T})=\eta(\mathrm{d}x_{T})\prod_{t=1}^{T}L_{t-1}(x_{t},\mathrm{d}x_{t-1}),
\]
where $L_{t}\in\mathcal{M}(\mathsf{X})$ for $t\in[0:T-1]$ is a
sequence of auxiliary `backward' Markov transition kernels. Assuming
that we have $L_{t-1}\otimes\gamma_{t}\ll\gamma_{t-1}\otimes M_{t}$
with strictly positive and bounded Radon-Nikodym derivative for all
$t\in[1:T]$, the Feynman-Kac path measure representation (\ref{eq:P})
can be recovered by defining
\begin{equation}
G_{0}(x_{0})=1,\quad G_{t}(x_{t-1},x_{t})=\frac{\mathrm{d}(L_{t-1}\otimes\gamma_{t})}{\mathrm{d}(\gamma_{t-1}\otimes M_{t})}(x_{t-1},x_{t}),\quad t\in[1:T].\label{eq:SMCsampler_potentials}
\end{equation}
Under these potentials, the normalized Feynman-Kac models (\ref{eq:norm_FKmodel})
act as the sequence of bridging distributions $(\eta_{t})_{t\in[0:T]}$
in this setting.
In annealed importance sampling (AIS) \cite{Neal_2001} and the sequential
sampler proposed in \cite{Chopin_2002}, one selects $M_{t}\in\mathcal{M}(\mathsf{X})$
as a Markov chain Monte Carlo (MCMC) kernel that is $\eta_{t}$-invariant
and $L_{t-1}\in\mathcal{M}(\mathsf{X})$ as its time reversal, i.e.
$L_{t-1}\otimes\eta_{t}=\eta_{t}\otimes M_{t}$, so the potentials
in (\ref{eq:SMCsampler_potentials}) simplify to
\begin{equation}
G_{0}(x_{0})=1,\quad G_{t}(x_{t-1})=\frac{\gamma_{t}(x_{t-1})}{\gamma_{t-1}(x_{t-1})},\quad t\in[1:T].\label{eq:AIS_potentials}
\end{equation}

\section{Twisted models and sequential Monte Carlo\label{sec:twisted_models_smc}}

\subsection{Twisted Feynman-Kac models\label{sec:twisted}}
SMC methods can perform poorly when the discrepancy between $\mathbb{P}$ and $\mathbb{Q}$
is large. The right panel of Figure \ref{fig:neuroscience_obs} illustrates that this is the case when
we employ BPF on the neuroscience application in Section \ref{sec:statespace_models}: the effective sample size (ESS), a common criterion used to assess
 the quality of a particle approximation \cite[p. 34--35]{Liu_2001}, falls below $20\%$ when the data change abruptly.
This is because the kernel $M_{t}(x_{t-1},\mathrm{d}x_{t})=f_{t}(x_{t-1},\mathrm{d}x_{t})$ used to sample particles at time $t$ does not take the observations into account. Better performance could be obtained using observation-dependent kernels. Indeed, in the context of state space
models, the smoothing distribution (\ref{eq:smoothing_distribution}) can be written as $\mathbb{P}(\mathrm{d}x_{0:T}|y_{0:T}) = \mathbb{P}(\mathrm{d}x_{0}|y_{0:T})\prod_{t=1}^T\mathbb{P}(\mathrm{d}x_{t}|x_{t-1},y_{t:T})$ with
\begin{align}
	\mathbb{P}(\mathrm{d}x_{0}|y_{0:T}) = \frac{\nu(\mathrm{d}x_0)\psi_0^*(x_0)}{\nu(\psi_0^*)}, \quad
	\mathbb{P}(\mathrm{d}x_{t}|x_{t-1},y_{t:T}) = \frac{f_t(x_{t-1},\mathrm{d}x_t)\psi_t^*(x_t)}{f_t(\psi_t^*)(x_{t-1})}, \quad t\in[1:T],\label{eq:BackwardinfoDecompo}
\end{align}
where the kernel $f_t(x_{t-1},\cdot)$ is \textit{twisted} using the so-called backward information
filter \cite{Bresler_1986,Briers_Doucet_Maskell_2010}, given by $\psi_t^*(x_t)=\mathbb{P}(y_{t:T}|x_t)$, for $t\in[0:T]$.

The backward information filter can also be defined using the backward recursion
\begin{equation}\label{eq:BackwardinfoDecompo2}
\begin{split}
\psi_T^*(x_T)&=g_T(x_T,y_T),\\
\psi_t^*(x_t)&=g_t(x_t,y_t)f_{t+1}(\psi_{t+1}^*)(x_t),\quad t\in[0:T-1].
\end{split}
\end{equation}
We can exploit this to obtain an approximation $\hat{\psi}_t(x_t), t\in[0:T]$ using regression \cite{Richard_Zhang_2007,Schartha_Kohn_2016,Guarniero_Lee_Johansen_2016}.
We can then sample particles at time $t$ using a proposal
$M_t^{\hat{\psi}}(x_{t-1},\mathrm{d}x_t)\propto f_t(x_{t-1},\mathrm{d}x_t)\hat{\psi}_t(x_t)$ that
approximates $\mathbb{P}(\mathrm{d}x_{t}|x_{t-1},y_{t:T})$.

Abstracting the above discussion from state space models to general Feynman--Kac models, where the potential $G_t$ might depend on both $x_{t-1}$ and $x_{t}$, motivates the following definitions.

\begin{defn}\label{def:admissible}
(Admissible policies) A sequence of functions $\psi=\left(\psi_{t}\right)_{t\in[0:T]}$ is an admissible policy if
these functions are strictly positive and satisfy $\psi_{0}\in\mathcal{B}(\mathsf{X})$, $\psi_{t}\in\mathcal{B}(\mathsf{X}\times\mathsf{X})$ for all $t\in[1:T]$. The set of all admissible policies will be denoted
as $\Psi$.
\end{defn}
\begin{defn}
\label{def:twisted_pathmeasures}(Twisted path measures) Given a policy
$\psi\in\Psi$ and a path measure $\mathbb{F}\in\mathcal{P}(\mathsf{X}^{T+1})$
of the form $\mathbb{F}(\mathrm{d}x_{0:T})=\nu(\mathrm{d}x_{0})\prod_{t=1}^{T}K_{t}(x_{t-1},\mathrm{d}x_{t})$
for some $\nu\in\mathcal{P}(\mathsf{X})$ and $K_{t}\in\mathcal{M}(\mathsf{X})$ for $t\in[1:T]$,
the $\psi$-twisted path measure of $\mathbb{F}$ is defined as
$\mathbb{F}^{\psi}(\mathrm{d}x_{0:T})=\nu^{\psi}(\mathrm{d}x_{0})\prod_{t=1}^{T}K_{t}^{\psi}(x_{t-1},\mathrm{d}x_{t})$
where
\begin{align}\label{eqn:twisted_kernels}
\nu^{\psi}(\mathrm{d}x_{0}):=\frac{\nu(\mathrm{d}x_{0})\psi_{0}(x_{0})}{\nu(\psi_{0})},\quad K_{t}^{\psi}(x_{t-1},\mathrm{d}x_{t}):=\frac{K_{t}(x_{t-1},\mathrm{d}x_{t})\psi_{t}(x_{t-1},x_{t})}{K_{t}(\psi_{t})(x_{t-1})},\quad t\in[1:T].
\end{align}
\end{defn}

For any policy $\psi\in\Psi$, since $\mathbb{P}\ll\mathbb{Q}\ll\mathbb{Q}^{\psi}$
by positivity of $\psi$, we can rewrite the measure $\mathbb{P}$ defined in (\ref{eq:P}) as
\begin{equation}
\mathbb{P}(\mathrm{d}x_{0:T})=Z^{-1}G_{0}^{\psi}(x_{0})\prod_{t=1}^{T}G_{t}^{\psi}(x_{t-1},x_{t})\,\mathbb{Q}^{\psi}(\mathrm{d}x_{0:T})\label{eq:P_twistedmodel}
\end{equation}
where the twisted potentials associated with the twisted path measure $\mathbb{Q}^{\psi}$ are given by
\begin{align}
 & G_{0}^{\psi}(x_{0}):=\frac{\mu(\psi_{0})G_{0}(x_{0})M_{1}(\psi_{1})(x_{0})}{\psi_{0}(x_{0})},\label{eq:twisted_potentials}\\
 & G_{t}^{\psi}(x_{t-1},x_{t}):=\frac{G_{t}(x_{t-1},x_{t})M_{t+1}(\psi_{t+1})(x_{t})}{\psi_{t}(x_{t-1},x_{t})},\quad t\in[1:T-1],\nonumber \\
 & G_{T}^{\psi}(x_{T-1},x_{T}):=\frac{G_{T}(x_{T-1},x_{T})}{\psi_{T}(x_{T-1},x_{T})}.\nonumber
\end{align}
Note from (\ref{eq:P_twistedmodel}) that
$Z=\mathbb{E}_{\mathbb{Q}^{\psi}}\Big[G_{0}^{\psi}(X_{0})\prod_{t=1}^{T}G_{t}^{\psi}(X_{t-1},X_{t})\Big]$
by construction, whereas the triple $\big\{ \mu^{\psi},(M_{t}^{\psi})_{t\in[1:T]},$ $(G_{t}^{\psi})_{t\in[0:T]}\big\} $
induces the $\psi$-twisted Feynman-Kac models given by
\begin{equation}
\gamma_{t}^{\psi}(\varphi)=\mathbb{E}_{\mathbb{Q}^{\psi}}\left[\varphi(X_{t})G_{0}^{\psi}(X_{0})\prod_{s=1}^{t}G_{s}^{\psi}(X_{s-1},X_{s})\right],\quad\eta_{t}^{\psi}(\varphi)=\gamma_{t}^{\psi}(\varphi)/Z_{t}^{\psi},\label{eq:twisted_FK_models}
\end{equation}
for $\varphi\in\mathcal{B}(\mathsf{X})$, $t\in[0:T]$, where $Z_{t}^{\psi}:=\gamma_{t}^{\psi}(\mathsf{X})$.
For $t\in[0:T-1]$, the marginal distributions of the twisted model are given by
\begin{equation}
\eta_{t}^{\psi}(\mathrm{d}x_{t})=\eta_{t}(\mathrm{d}x_{t})M_{t+1}(\psi_{t+1})(x_{t})Z_{t}/Z_{t}^{\psi}\label{eq:relate_FK_Models}
\end{equation}
and do not generally coincide  with the ones of the original model (\ref{eq:norm_FKmodel}). However, we stress that they
coincide at time $T$ as
\begin{equation}
Z=Z_{T}^{\psi}=\mu^{\psi}(G_{0}^{\psi})\prod_{t=1}^{T}\eta_{t-1}^{\psi}(M_{t}^{\psi}(G_{t}^{\psi})).\label{eq:Z_prodform_twisted}
\end{equation}

To illustrate the effect of twisting models in the static setting of Section \ref{sec:static_models},
rewriting the twisted potentials (\ref{eq:twisted_potentials}) using (\ref{eq:relate_FK_Models}) as
\[
G_{0}^{\psi}(x_{0})=\frac{\mathrm{d}\eta_{0}^{\psi}}{\mathrm{d}\mu^{\psi}}(x_{0}),\quad G_{t}^{\psi}(x_{t-1},x_{t})=\frac{\mathrm{d}(L_{t-1}\otimes\gamma_{t}^{\psi})}{\mathrm{d}(\gamma_{t-1}^{\psi}\otimes M_{t}^{\psi})}(x_{t-1},x_{t}),\quad t\in[1:T],
\]
shows that this corresponds to employing
the same backward kernels $(L_{t})_{t\in[0:T-1]}$, but altered bridging distributions
$(\eta_{t}^{\psi})_{t\in[0:T]}$, initial distribution $\mu^{\psi}$ and forward kernels $(M_{t}^{\psi})_{t\in[1:T]}$.

\subsection{Twisted sequential Monte Carlo\label{sec:SMC}}
Consider a policy $\psi\in\Psi$ such that sampling from the
initial distribution $\mu^{\psi}\in\mathcal{P}(\mathsf{X})$ and the
transition kernels $(M_{t}^{\psi})_{t\in[1:T]}$ in $\mathcal{M}(\mathsf{X})$
is feasible and evaluation of the twisted potentials (\ref{eq:twisted_potentials})
is tractable.
We can now construct the $\psi$-twisted SMC method as simply the standard sampling-resampling SMC
algorithm applied to $\psi$-twisted Feynman-Kac models \cite{Doucet_Johansen_2011}.
The resulting algorithm provides approximations of the
probability measures $(\eta_{t}^{\psi})_{t\in[0:T]}$, normalizing
constant $Z$ and path measure $\mathbb{P}$, by simulating an
interacting particle system of size $N\in\mathbb{N}$.
An algorithmic description is detailed in Algorithm \ref{alg:twistedSMC},
where $\mathcal{R}\left(w_{1},\ldots,w_{N}\right)$
refers to a resampling operation based on a vector of unnormalized
weights $\left(w_{n}\right)_{n\in[1:N]}\in {\mathbb{R}^{N}_{+}}$. For
example, this is the categorical distribution on $[1:N]$ with probabilities $(w_{n}/\sum_{m=1}^{N}w_{m})_{n\in[1:N]}$,
when multinomial resampling is employed; other lower variance and
adaptive resampling schemes can also be considered \cite{Gerber_Chopin_Whiteley_2019}. All simulations
presented in this article employ the systematic resampling scheme.

\begin{algorithm}
\caption{$\psi$-twisted sequential Monte Carlo \label{alg:twistedSMC}}
\begin{flushleft}
\textbf{Input: }number of particles $N\in\mathbb{N}$ and policy $\psi\in\Psi$.
\end{flushleft}
\begin{enumerate}
\item At time $t=0$ and particle $n\in[1:N]$:
\begin{enumerate}
\item sample $X_{0}^{n}\sim\mu^{\psi}$;
\item sample ancestor index $A_{0}^{n}\sim\mathcal{R}\big(G_{0}^{\psi}(X_{0}^{1}),\ldots,G_{0}^{\psi}(X_{0}^{N})\big)$.
\end{enumerate}
\item For time $t\in[1:T]$ and particle $n\in[1:N]$:
\begin{enumerate}
\item sample $X_{t}^{n}\sim M_{t}^{\psi}(X_{t-1}^{A_{t-1}^{n}},\cdot)$;
\item sample ancestor index $A_{t}^{n}\sim\mathcal{R}\Big(G_{t}^{\psi}(X_{t-1}^{A_{t-1}^{1}},X_{t}^{1}),\ldots,G_{t}^{\psi}(X_{t-1}^{A_{t-1}^{N}},X_{t}^{N})\Big)$.
\end{enumerate}
\end{enumerate}
\begin{flushleft}
\textbf{Output: }trajectories $\left(X_{t}^{n}\right)_{(t,n)\in[0:T]\times[1:N]}$
and ancestries $\left(A_{t}^{n}\right)_{(t,n)\in[0:T]\times[1:N]}$.
\end{flushleft}
\end{algorithm}

Given the output of the algorithm, i.e. an array of $\mathsf{X}$-valued
position variables $\left(X_{t}^{n}\right)_{(t,n)\in[0:T]\times[1:N]}$
and an array of $[1:N]$-valued ancestor variables $\left(A_{t}^{n}\right)_{(t,n)\in[0:T]\times[1:N]}$,
we have a particle approximation of $\eta_{t}^{\psi}$ given by the weighted random
measure
\[
\eta_{t}^{\psi,N}=\sum_{n=1}^{N}W_{t}^{\psi,n}\delta_{X_{t}^{n}},\quad W_{t}^{\psi,n}:=\frac{G_{t}^{\psi}(X_{t-1}^{A_{t-1}^{n}},X_{t}^{n})}{\sum_{m=1}^{N}G_{t}^{\psi}(X_{t-1}^{A_{t-1}^{m}},X_{t}^{m})},
\]
for $t\in[1:T]$ (similar expression for $t=0$) and an unbiased estimator
of $Z$ resembling the form of (\ref{eq:Z_prodform_twisted})
\begin{equation}
Z^{\psi,N}=\left\{ \frac{1}{N}\sum_{n=1}^{N}G_{0}^{\psi}(X_{0}^{n})\right\} \prod_{t=1}^{T}\left\{ \frac{1}{N}\sum_{n=1}^{N}G_{t}^{\psi}(X_{t-1}^{A_{t-1}^{n}},X_{t}^{n})\right\} .\label{eq:twisted_Z_estimator}
\end{equation}
With stored trajectories \cite{Jacob_Murray_Rubenthaler_2015},
we can also form a particle approximation of $\mathbb{P}$ with
$\mathbb{P}^{\psi,N}=N^{-1}\sum_{n=1}^{N}\delta_{X_{0:T}^{n}}$,
where $X_{0:T}^{n}$ denotes the path obtained by tracing the ancestral
lineage of particle $X_{T}^{n}$, i.e. $X_{0:T}^{n}:=(X_{t}^{B_{t}^{n}})_{t\in[0:T]}$
with $B_{T}^{n}:=A_{T}^{n}$ and $B_{t}^{n}:=A_{t}^{B_{t+1}^{n}}$
for $t\in[0:T-1]$. Many convergence results are available for these
approximations as the size $N$ of the particle system increases \cite{DelMoral_2004}.
However, depending on the choice of $\psi\in\Psi$, the quality of these approximations
may be inadequate for practical values of $N$;
for example, the large variance of
(\ref{eq:twisted_Z_estimator}) often hinders its use within particle MCMC schemes \cite{Andrieu_Doucet_Holenstein_2010} and
the approximation $\mathbb{P}^{\psi,N}$ could degenerate quickly
with $T$.
The choice of an optimal policy is addressed in the following section.

\section{Controlled sequential Monte Carlo}
\subsection{Optimal policies\label{sec:optimal_policies}}
Suppose we have an arbitrary current policy $\psi\in\Psi$, initially given by a sequence of constant functions. We
would like to twist the path measure $\mathbb{Q}^{\psi}\in\mathcal{P}(\mathsf{X}^{T+1})$ further
with a policy $\phi\in\Psi$, so that the resulting twisted path measure
$(\mathbb{Q}^{\psi})^{\phi}\in\mathcal{P}(\mathsf{X}^{T+1})$ is in some sense `closer' to the target Feynman-Kac
measure $\mathbb{P}$.
Note from Definition \ref{def:twisted_pathmeasures} that $(\mathbb{Q}^{\psi})^{\phi}=\mathbb{Q}^{\psi\cdot\phi}$,
where $\psi\cdot\phi=\left(\psi_{t}\cdot\phi_{t}\right)_{t\in[0:T]}$
denotes element-wise multiplication, is simply the $(\psi\cdot\phi)$-twisted
path measure of $\mathbb{Q}$. From (\ref{eq:twisted_potentials}),
the corresponding twisted potentials are given by
\begin{align}
 & G_{0}^{\psi\cdot\phi}(x_{0})=\frac{\mu^{\psi}(\phi_{0})G_{0}^{\psi}(x_{0})M_{1}^{\psi}(\phi_{1})(x_{0})}{\phi_{0}(x_{0})},\label{eq:further_twisted_potentials}\\
 & G_{t}^{\psi\cdot\phi}(x_{t-1},x_{t})=\frac{G_{t}^{\psi}(x_{t-1},x_{t})M_{t+1}^{\psi}(\phi_{t+1})(x_{t})}{\phi_{t}(x_{t-1},x_{t})},\quad t\in[1:T-1],\nonumber \\
 & G_{T}^{\psi\cdot\phi}(x_{T-1},x_{T})=\frac{G_{T}^{\psi}(x_{T-1},x_{T})}{\phi_{T}(x_{T-1},x_{T})}.\nonumber
\end{align}

The choice of $\phi$ that optimally refines an arbitrary policy $\psi$ is given by the following optimality result.

\begin{prop}
\label{prop:Optimality}For any $\psi\in\Psi$, under the policy $\phi^{*}=\left(\phi_{t}^{*}\right)_{t\in[0:T]}$
defined recursively as
\begin{align}
 & \phi_{T}^{*}(x_{T-1},x_{T})=G_{T}^{\psi}(x_{T-1},x_{T}),\label{eq:optimal_phi}\\
 & \phi_{t}^{*}(x_{t-1},x_{t})=G_{t}^{\psi}(x_{t-1},x_{t})M_{t+1}^{\psi}(\phi_{t+1}^{*})(x_{t}),\quad t\in[1:T-1],\nonumber \\
 & \phi_{0}^{*}(x_{0})=G_{0}^{\psi}(x_{0})M_{1}^{\psi}(\phi_{1}^{*})(x_{0}),\nonumber
\end{align}
the refined policy $\psi^{*}:=\psi\cdot\phi^{*}$ satisfies the following
properties:
\begin{enumerate}
\item the
twisted path measure $\mathbb{Q}^{\psi^{*}}$ coincides with the Feynman-Kac path measure $\mathbb{P}$;
\item the normalized Feynman-Kac model $\eta_{t}^{\psi^{*}}$ is the
time $t$-marginal distribution of $\mathbb{P}$ and its normalizing
constant $Z_{t}^{\psi^{*}}=Z$ for all $t\in[0:T]$;
\item the normalizing constant estimator $Z^{\psi^{*},N}=Z$ almost
surely for any $N\in\mathbb{N}$.
\end{enumerate}
Moreover, if $G_{0}^{\psi}\in\mathcal{B}(\mathsf{X})~and~G_{t}^{\psi}\in\mathcal{B}(\mathsf{X}\times\mathsf{X})$ for $t\in[1:T]$
then $\phi^{*}\in\Psi$.
\end{prop}

This proposition implies that SMC sampling with the optimal $\psi^{*}$-twisted version of Algorithm \ref{alg:twistedSMC} ensures that the normalizing constant estimator is constant over the entire time horizon, and is equal to the desired normalizing constant. This follows because the SMC weights themselves are almost surely constant; one can see this by substituting the optimal choice (\ref{eq:optimal_phi}) into (\ref{eq:further_twisted_potentials}). The variance of the SMC weights and the constancy of the normalizing constant estimator can both be used (as described later) as measures of performance evaluation or adaptive tuning.

In a state space context, (\ref{eq:optimal_phi}) corresponds to the recursion satisfied by the backward information filter introduced in (\ref{eq:BackwardinfoDecompo2}) when  $\psi\in\Psi$ are constant functions, i.e. $\mu^{\psi}=\mu=\nu$ and
$M_{t}^{\psi}=M_{t}=f_{t},t\in[1:T]$;
see, e.g., \cite{Bresler_1986,Briers_Doucet_Maskell_2010}.

As it can be shown that $\phi^{*}$ is the optimal policy of an associated Kullback--Leibler
optimal control problem (Supplementary Material, Section \ref{sec:control_connection}), we shall
refer to it as the optimal policy w.r.t. $\mathbb{Q}^{\psi}$, although the optimality properties
in Proposition \ref{prop:Optimality} only identify a policy up to normalization factors.
An application of this result gives us the optimal policy $\psi^{*}=\psi\cdot\phi^{*}$ w.r.t.
$\mathbb{Q}$,
which is admissible if the original potentials $(G_{t})_{t\in[0:T]}$ are bounded\footnote{For ease of presentation,
the notion of admissibility adopted in Definition \ref{def:admissible} is more stringent than necessary as
non-admissible optimal policies can still lead to valid optimal SMC methods.}.

\subsection{Approximate dynamic programming\label{sec:ADP}}

Equation (\ref{eq:optimal_phi}) may be viewed as a dynamic programming backward recursion. The optimal policy $\phi^{*}$ w.r.t. $\mathbb{Q}^{\psi}$ will give rise to an optimally controlled SMC algorithm via a $\psi^{*}=(\psi\cdot\phi^{*})$-twisted version of Algorithm \ref{alg:twistedSMC}.
In all but simple cases, the recursion (\ref{eq:optimal_phi})
defining $\phi^{*}$ is intractable.
We now exploit the connection to optimal control
by adapting numerical methods (i.e. approximate dynamic programming) for finite horizon control problems
\cite[p. 329--331]{Bertsekas_Tsitsiklis_1996} to our setup. The resulting
methodology approximates $\phi^{*}$ by combining function approximation
and iterating the backward recursion (\ref{eq:optimal_phi}).

In the following, we will approximate $V_t^*:=-\log\phi_t^*,t\in[0:T]$ as this corresponds
to learning the optimal value functions of the associated control problem. Compared to learning optimal policies directly,
as considered in \cite{Guarniero_Lee_Johansen_2016},
the latter choice is often more desirable as computing in logarithmic
scale offers more numerical stability and the minimization is additionally analytically tractable in important scenarios.
Moreover, this allows us to relate regression errors to performance properties of the resulting
twisted SMC method in the next section.

Let $\left(X_{t}^{n}\right)_{(t,n)\in[0:T]\times[1:N]}$ and $\left(A_{t}^{n}\right)_{(t,n)\in[0:T]\times[1:N]}$ denote
the trajectories and ancestries, obtained by running a $\psi$-twisted SMC.
At time $T$, to approximate $V_T^*:=-\log\phi_T^*=-\log G_T^{\psi}$, we consider the least squares problem
\begin{align}\label{eqn:adp_step1}
\hat{V}_{T}=\arg\min_{\varphi\in\mathsf{F}_{T}}\sum_{n=1}^{N}\left(\varphi(X_{T-1}^{A_{T-1}^{n}},X_{T}^{n})
+ \log G_{T}^{\psi}(X_{T-1}^{A_{T-1}^{n}},X_{T}^{n})\right)^{2},
\end{align}
where $\mathsf{F}_T$ is a pre-specified function class. An approximation of $\phi_T^*$ can then be obtained by
taking $\hat{\phi}_T := \exp(-\hat{V}_{T})$. To iterate the backward recursion
$\phi_{T-1}^*=G_{T-1}^{\psi}M_{T}^{\psi}(\phi^*_T)$,
we set $\xi_{T-1}:=G_{T-1}^{\psi}M_{T}^{\psi}(\hat{\phi}_T)$
by plugging in the approximation $\hat{\phi}_T\approx\phi_T^*$ and
consider the least squares problem 
\begin{align}\label{eqn:adp_step2}
\hat{V}_{T-1}=\arg\min_{\varphi\in\mathsf{F}_{T-1}}\sum_{n=1}^{N}\left(\varphi(X_{T-2}^{A_{T-2}^{n}},X_{T-1}^{n})
+\log \xi_{T-1}(X_{T-2}^{A_{T-2}^{n}},X_{T-1}^{n})\right)^{2},
\end{align}
where $\mathsf{F}_{T-1}$ is another function class to be specified. As before, we form
the approximation $\hat{\phi}_{T-1}:=\exp(-\hat{V}_{T-1})$. Continuing in this manner until time $0$ gives
us an approximation $\hat{\phi}=(\hat{\phi}_{t})_{t\in[0:T]}$ of $\phi^*$.
We shall refer to this procedure as
the approximate dynamic programming algorithm and provide a
detailed description in Algorithm \ref{alg:ADP}.

Restricting the
function classes $(\mathsf{F}_{t})_{t\in[0:T]}$ to contain only lower bounded functions ensures that
the estimated policy $\hat{\phi}$ lies in $\Psi$, hence the refined policy $\psi\cdot\hat{\phi}$ also lies
in $\Psi$.
We defer a detailed discussion on the choice of function classes and
shall assume for now this is such that under the refined policy $\psi\cdot\hat{\phi}\in\Psi$,
sampling from initial distribution $\mu^{\psi\cdot\hat{\phi}}\in\mathcal{P}(\mathsf{X})$,
transition kernels $(M_{t}^{\psi\cdot\hat{\phi}})_{t\in[1:T]}$ in $\mathcal{M}(\mathsf{X})$
is feasible and evaluation of twisted potentials $(G_{t}^{\psi\cdot\hat{\phi}})_{t\in[0:T]}$
is tractable.

As the size of the particle system $N$ increases, it is natural to
expect $\hat{\phi}$ to converge (in a suitable sense) to a policy
defined by an idealized algorithm that performs the least squares approximations
in (\ref{eqn:adp_step1})-(\ref{eqn:adp_step2}) using $L^2$-projections.
This will be established in Section \ref{sec:LimitTheorems}
for a common choice of function class. It follows that the quality
of $\hat{\phi}$, as an approximation of the optimal policy $\phi^{*}$,
will depend on the number of particles $N$ and the `richness' of
chosen function classes $(\mathsf{F}_{t})_{t\in[0:T]}$. A more
precise characterization of the ADP error in terms of approximate
projection errors will be given in Section \ref{sec:PolicyLearning}.

\begin{algorithm}
\caption{Approximate dynamic programming \label{alg:ADP}}
\begin{flushleft}
\textbf{Input:} policy $\psi\in\Psi$ and output of $\psi$-twisted
SMC method (Algorithm \ref{alg:twistedSMC}).
\end{flushleft}
\begin{enumerate}
\item Initialization: set $M_{T+1}^{\psi}(\hat{\phi}_{T+1})(X_{T}^{n})=1$
for $n\in[1:N]$.
\item For time $t\in[1:T]$:
\begin{enumerate}
\item set $\xi_{t}(X_{t-1}^{A_{t-1}^{n}},X_{t}^{n})=G_{t}^{\psi}(X_{t-1}^{A_{t-1}^{n}},X_{t}^{n})M_{t+1}^{\psi}(\hat{\phi}_{t+1})(X_{t}^{n})$
for $n\in[1:N]$;
\item fit $\hat{V}_{t}=\arg\min_{\varphi\in\mathsf{F}_{t}}\sum_{n=1}^{N}\left(\varphi(X_{t-1}^{A_{t-1}^{n}},X_{t}^{n})+\log\xi_{t}(X_{t-1}^{A_{t-1}^{n}},X_{t}^{n})\right)^{2}$;
\item set $\hat{\phi}_{t}=\exp(-\hat{V}_{t})$.
\end{enumerate}
\item At time $t=0$:
\begin{enumerate}
\item set $\xi_{0}(X_{0}^{n})=G_{0}^{\psi}(X_{0}^{n})M_{1}^{\psi}(\hat{\phi}_{1})(X_{0}^{n})$
for $n\in[1:N]$;
\item fit $\hat{V}_{0}=\arg\min_{\varphi\in\mathsf{F}_{0}}\sum_{n=1}^{N}\left(\varphi(X_{0}^{n})+\log\xi_{0}(X_{0}^{n})\right)^{2}$;
\item set $\hat{\phi}_{0}=\exp(-\hat{V}_{0})$.
\end{enumerate}
\end{enumerate}
\begin{flushleft}
\textbf{Output: }policy $\hat{\phi}=(\hat{\phi}_{t})_{t\in[0:T]}\in\Psi$.
\end{flushleft}
\end{algorithm}

\subsection{Policy refinement\label{sec:policy_refinement}}

If the recursion (\ref{eq:optimal_phi}) could be performed exactly,
 no policy refinement would be necessary as we would initialize $\psi$
as a policy of constant functions and obtain the optimal policy
$\psi^{*}=\phi^{*}$ w.r.t. $\mathbb{Q}$. This will not be possible in practical scenarios. Given a current policy
$\psi\in\Psi$, we employ ADP and obtain an approximation $\hat{\phi}$
of the optimal policy $\phi^{*}$ w.r.t. $\mathbb{Q}^{\psi}$.
The residuals from the corresponding least squares approximations (\ref{eqn:adp_step1})-(\ref{eqn:adp_step2})
are given by
\begin{align*}
\varepsilon_{T}^{\psi}:=\log\hat{\phi}_{T}-\log G_{T}^{\psi},\quad \varepsilon_{t}^{\psi}:=\log\hat{\phi}_{t}-\log G_{t}^{\psi}-\log M_{t+1}^{\psi}(\hat{\phi}_{t+1}),\quad t\in[0:T-1].
\end{align*}
From (\ref{eq:further_twisted_potentials}), these residuals are related to twisted potentials of the refined policy $\psi\cdot\hat{\phi}$
via
\begin{align}\label{eq:residual_potential_relation}
\log G_{0}^{\psi\cdot\hat{\phi}} = \log\mu^{\psi}(\hat{\phi}_{0})-\varepsilon_{0}^{\psi},\quad
\log G_{t}^{\psi\cdot\hat{\phi}} = -\varepsilon_{t}^{\psi},\quad t\in[1:T].
\end{align}
Using this relation, we can monitor the efficiency of ADP via the variance of SMC weights in the $(\psi\cdot\hat{\phi})$-twisted version of Algorithm \ref{alg:twistedSMC}. It follows from (\ref{eq:residual_potential_relation}) that the Kullback--Leibler divergence from
$(\mathbb{Q}^{\psi})^{\hat{\phi}}$ to $\mathbb{P}$ is at most
\begin{equation}
|\log\mu^{\psi}(\hat{\phi}_{0})-\log Z|+\|\varepsilon_{0}^{\psi}\|_{L^{1}(\mathbb{P}_{0})}+\sum_{t=1}^{T}\|\varepsilon_{t}^{\psi}\|_{L^{1}(\mathbb{P}_{t-1,t})}\label{eq:KL_ADPerror}
\end{equation}
where $\|\cdot\|_{L^{1}}$ denotes the $L^1$-norm w.r.t. the one time
$(\mathbb{P}_{t})_{t\in[0:T]}$ and two time $(\mathbb{P}_{t,s})_{(t,s)\in[0:T-1]\times[t+1:T]}$
marginal distributions of $\mathbb{P}$.
This shows how performance of $(\psi\cdot\hat{\phi})$-twisted
SMC depends on the quality of the ADP approximation of the
optimal policy w.r.t. $\mathbb{Q}^{\psi}$.

If we further twist the path measure $\mathbb{Q}^{\psi\cdot\hat{\phi}}$
by a policy $\hat{\zeta}\in\Psi$, the subsequent ADP procedure defining $\hat{\zeta}$ would consider
the least squares problem
\begin{align}\label{eqn:further_step1}
-\log\hat{\zeta}_{T} & :=\arg\min_{\varphi\in\mathsf{F}_{T}}\sum_{n=1}^{N}\left(\varphi(X_{T-1}^{A_{T-1}^{n}},X_{T}^{n})
- \varepsilon_{T}^{\psi}(X_{T-1}^{A_{T-1}^{n}},X_{T}^{n})\right)^{2},
\end{align}
at time $T$, and for $t\in[1:T-1]$
\begin{align}\label{eqn:further_step2}
-\log\hat{\zeta}_{t} & :=\arg\min_{\varphi\in\mathsf{F}_{t}}\sum_{n=1}^{N}\left(\varphi(X_{t-1}^{A_{t-1}^{n}},X_{t}^{n})
- (\varepsilon_{t}^{\psi}-\log M_{t+1}^{\psi\cdot\hat{\phi}}(\hat{\zeta}_{t+1}))(X_{t-1}^{A_{t-1}^{n}},X_{t}^{n})\right)^{2},
\end{align}
where $\left(X_{t}^{n}\right)_{(t,n)\in[0:T]\times[1:N]}$ and $\left(A_{t}^{n}\right)_{(t,n)\in[0:T]\times[1:N]}$ denote
the output of $(\psi\cdot\hat{\phi})$-twisted SMC.
Equations (\ref{eqn:further_step1})-(\ref{eqn:further_step2})
reveal that it might be beneficial to have an iterative scheme to
refine policies as this allows repeated least squares
fitting of residuals, in the spirit of $L^{2}$-boosting methods \cite{Buhlmann_Yu_2003}.
Moreover, it follows from (\ref{eq:residual_potential_relation})-(\ref{eq:KL_ADPerror})
that errors would not accumulate over iterations. The resulting iterative
algorithm, summarized in Algorithm \ref{alg:cSMC}, will be referred
to as the controlled SMC method (cSMC).
The overall computational complexity is of order $I\times T \times (NC_{\textrm{sample}}C_{\textrm{evaluate}} + C_{\mathrm{approx}})$,
where $C_{\textrm{sample}}(d)$ is the cost of sampling from each initial distribution or transition kernel in (\ref{eqn:twisted_kernels}),
$C_{\textrm{evaluate}}(d)$ is the cost of evaluating each twisted potential in (\ref{eq:twisted_potentials}),
and $C_{\mathrm{approx}}(N,d)$ is the cost of each least squares approximation\footnote{The dependence of these costs
on their arguments will depend on the specific problem of interest and the choice of function classes. As an example, $C_{\mathrm{approx}}$ will depend linearly on $N$ in the case of linear least squares regression.}.
The first iteration of the algorithm would coincide with that of \cite{Guarniero_Lee_Johansen_2016}
for state space models, if regressions were computed on the natural scale; subsequent iterations differ
in policy refinement strategy.
To maintain a coherent terminology,
we will refer to the standard SMC method
and $\psi^{*}$-twisted SMC method as the uncontrolled and optimally
controlled SMC methods respectively. From the output of the algorithm, we can estimate $\mathbb{P}$ with $\mathbb{P}^{\psi^{(I)},N}$ and
its normalizing constant $Z$ with $Z^{\psi^{(I)},N}$ as explained in Section \ref{sec:SMC}.

It is possible to consider performance monitoring and adaptive tuning for the SMC sampling and the iterative policy refinement.
Recalling the relationship between residuals
and twisted potentials (\ref{eq:residual_potential_relation}), we
note that monitoring the variance of the SMC weights, using for example
the ESS, allows us to evaluate the effectiveness
of the ADP algorithm and to identify time instances when the approximation
is inadequate. We can also deduce if the estimated policy is far from
optimal by comparing the behaviour of the normalizing constant
estimates across time with those when the optimal policy is applied,
as detailed in Proposition \ref{prop:Optimality}. When implementing Algorithm \ref{alg:cSMC}, the number of iterations
$I\in\mathbb{N}$ can be pre-determined using preliminary runs or chosen adaptively until
successive policy refinement yields no improvement in performance.
For example, one can iterate policy refinement until the ESS across time achieves a desired minimum threshold and/or there is no improvement in ESS across iterations; see Section \ref{sec:lorenz} for a numerical illustration. In Section \ref{sec:iADP}, under appropriate regularity assumptions,
we show that this iterative scheme generates a geometrically ergodic
Markov chain on $\Psi$ and characterize its unique invariant distribution.
For all numerical examples considered in this article, we observe
that convergence happens very rapidly, so only a small number of iterations
is necessary.

\begin{algorithm}
\caption{Controlled sequential Monte Carlo \label{alg:cSMC}}
\begin{flushleft}
\textbf{Input:} number of particles $N\in\mathbb{N}$ and iterations
$I\in\mathbb{N}$.
\end{flushleft}
\begin{enumerate}
\item Initialization: set $\psi^{(0)}$ as constant one functions.
\item For iterations $i\in[0:I-1]$:
\begin{enumerate}
\item run $\psi^{(i)}$-twisted SMC method (Algorithm \ref{alg:twistedSMC});
\item perform ADP (Algorithm \ref{alg:ADP}) with SMC output to
obtain policy $\hat{\phi}^{(i+1)}$;
\item construct refined policy $\psi^{(i+1)}=\psi^{(i)}\cdot\hat{\phi}^{(i+1)}$.
\end{enumerate}
\item At iteration $i=I$:
\begin{enumerate}
\item run $\psi^{(I)}$-twisted SMC method (Algorithm \ref{alg:twistedSMC}).
\end{enumerate}
\end{enumerate}
\begin{flushleft}
\textbf{Output: }trajectories $\left(X_{t}^{n}\right)_{(t,n)\in[0:T]\times[1:N]}$
and ancestries $\left(A_{t}^{n}\right)_{(t,n)\in[0:T]\times[1:N]}$
from $\psi^{(I)}$-twisted SMC method.
\end{flushleft}
\end{algorithm}

\subsection{Illustration on neuroscience model\label{sec:neuroscience}}
We now apply our proposed methodology on the
neuroscience model introduced in Section \ref{sec:statespace_models}.
We take BPF as the uncontrolled SMC method, i.e. we set $\mu=\nu$ and $M_{t}=f$ for
$t\in[1:T]$.
Under the following choice of function classes
\begin{equation}
\mathsf{F}_{t}=\left\{ \varphi(x_{t})=a_{t}x_{t}^{2}+b_{t}x_{t}+c_{t}:(a_{t},b_{t},c_{t})\in\mathbb{R}^{3}\right\} ,\quad t\in[0:T],\label{eq:neuroscience_functionclass}
\end{equation}
the policy $\psi^{(i)}=(\psi_{t}^{(i)})_{t\in[0:T]}$ at iteration
$i\in[1:I]$ of Algorithm \ref{alg:cSMC} has the form
\[
\psi_{t}^{(i)}(x_{t})=\exp\left(-a_{t}^{(i)}x_{t}^{2}-b_{t}^{(i)}x_{t}-c_{t}^{(i)}\right),\quad t\in[0:T],
\]
where $a_{t}^{(i)}:=\sum_{j=1}^{i}a_{t}^{j},b_{t}^{(i)}:=\sum_{j=1}^{i}b_{t}^{j},c_{t}^{(i)}:=\sum_{j=1}^{i}c_{t}^{j}$
for $t\in[0:T]$ and $(a_{t}^{j+1},b_{t}^{j+1},c_{t}^{j+1})_{t\in[0:T]}$
denotes the coefficients estimated using linear least squares at iteration
$j\in[0:I-1]$.
Exact expressions of the twisted initial distribution, transition kernels and potentials,
required to implement cSMC are given in Section \ref{appendix:neuroscience} of Supplementary Material.

Figure \ref{fig:neuroscience_compare} illustrates that the
parameterization (\ref{eq:neuroscience_functionclass})
provides a good approximation of the optimal policy. We note (left panel) the significant improvement of ESS across iterations, and see how this may be used as a measure of performance evaluation. In the right panel, we can also deduce how far the estimated policy is from optimality by observing the behaviour of normalizing constant estimates as discussed previously.
Indeed, while the uncontrolled
SMC approximates $Z_{t}=Z_{t}^{\psi^{(0)}}=p(y_{0:t})$, the controlled SMC scheme approximates $Z_{t}^{\psi^*}=p(y_{0:T})$ for all $t\in[0:T]$. 

\begin{figure}
\begin{centering}
\includegraphics[scale=0.4]{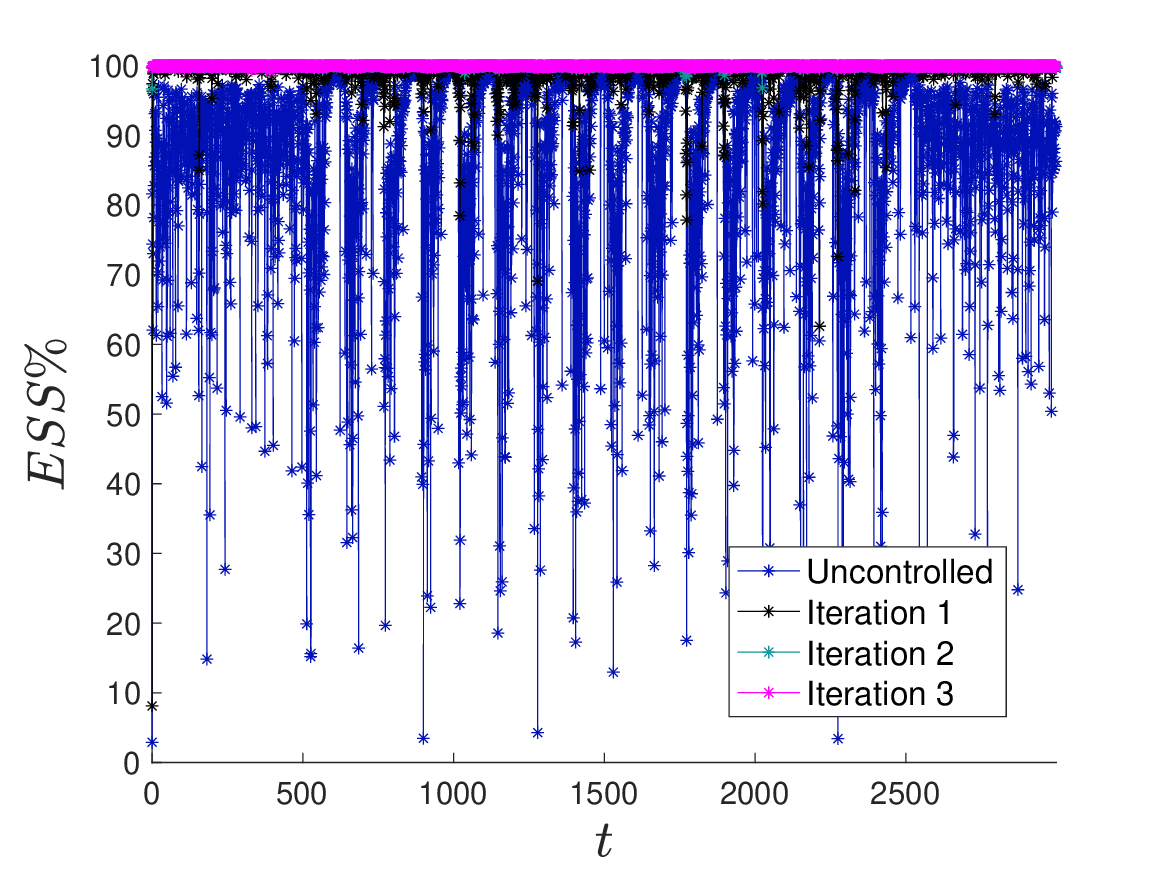}\includegraphics[scale=0.4]{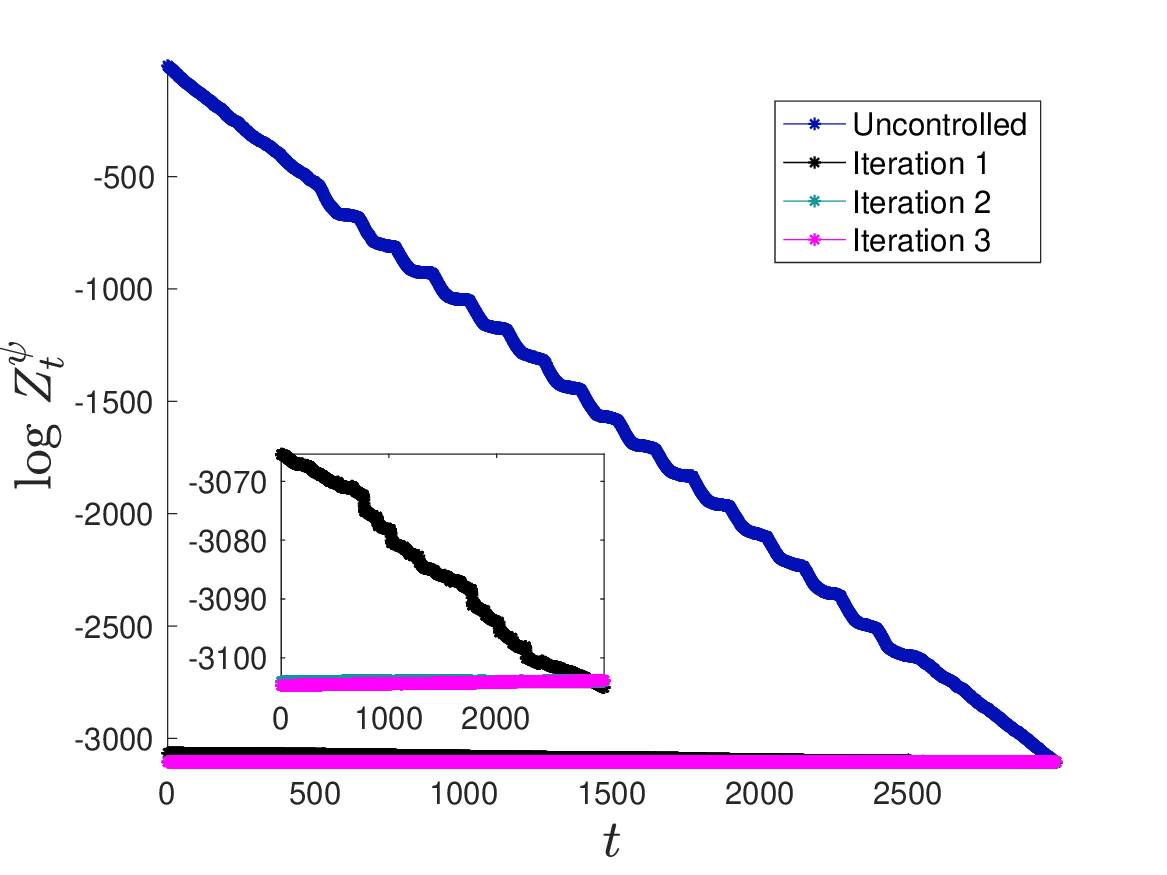}
\par\end{centering}
\caption{\label{fig:neuroscience_compare}Comparison of uncontrolled and controlled
SMC methods in terms of effective sample size (\emph{left}) and normalizing
constant estimation (\emph{right}) on the
neuroscience model introduced in Section \ref{sec:statespace_models}. The parameters
are $\alpha=0.99,\sigma^{2}=0.11$ and the algorithmic settings
of cSMC are $I=3,N=128$. }
\end{figure}

\begin{figure}
\begin{centering}
\includegraphics[scale=0.4]{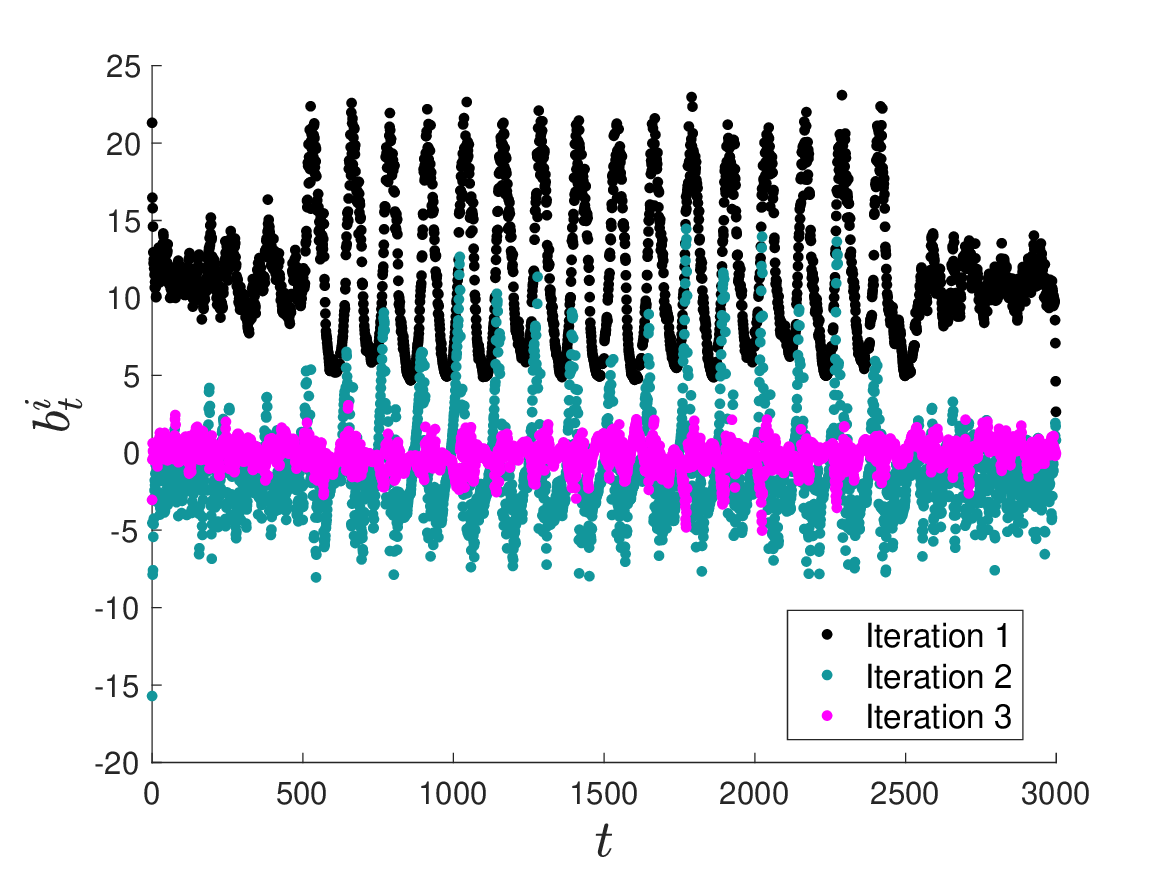}\includegraphics[scale=0.4]{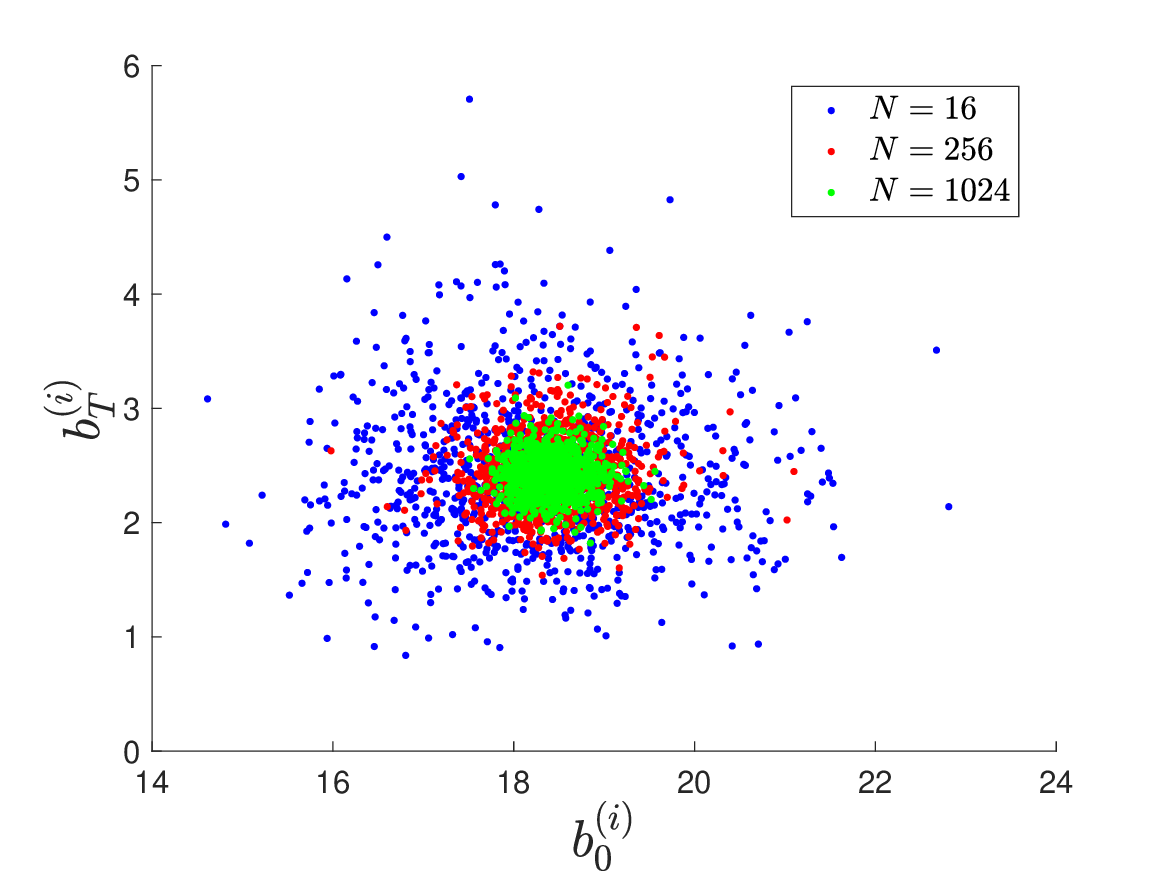}
\par\end{centering}
\caption{\label{fig:neuroscience_coefficients}Applying controlled SMC method on
the neuroscience model introduced in Section \ref{sec:statespace_models}:
coefficients estimated at each iteration with $N=128$ particles (\emph{left}) and
invariant distribution of coefficients with various number of particles (\emph{right}).}
\end{figure}

Moreover, we see from the left panel of Figure \ref{fig:neuroscience_coefficients} that the improvement
in performance is reflected in the estimated policy's ability to capture
abrupt changes in the data.
This plot also demonstrates the effect of policy
refinement: by refitting residuals from previous iterations (\ref{eqn:further_step1})-(\ref{eqn:further_step2}),
the magnitude of estimated coefficients decreases with iterations
as the residuals can be adequately approximated by simpler functions.
Lastly, in the right panel of Figure \ref{fig:neuroscience_coefficients}, we illustrate the
invariant distribution of coefficients estimated by cSMC using a long
run of $I=1000$ iterations, with the first 10 iterations discarded
as burn-in. These plots show that the distribution concentrates as
the size of the particle system $N$ increases, which is consistent
with our findings presented in Section \ref{sec:iADP}.

\section{Analysis\label{sec:analysis}}
This section considers several theoretical aspects of the proposed methodology, and may be skipped without affecting the methodological developments thus far and the experimental results that follow.
\subsection{Policy learning\label{sec:PolicyLearning}}
The goal of this section is to characterize the error of ADP (Algorithm \ref{alg:ADP})
for learning the optimal policy (\ref{eq:optimal_phi})
in terms of regression errors.
We first define, for any
$\mu\in\mathcal{P}(\mathsf{E})$, 
 the set
$\mathfrak{L}^{2}(\mu)$ of $\mathcal{E}$-measurable
functions $\varphi:\mathsf{E}\rightarrow\mathbb{R}^{d}$ such that
$\|\varphi\|_{L^{2}(\mu)}:=(\int_{\mathsf{E}}|\varphi(x)|^{2}\mu(\mathrm{d}x))^{1/2}<\infty$,
and $L^{2}(\mu)$ as the set of equivalence classes of functions in $\mathfrak{L}^{2}(\mu)$ that agree $\mu$-almost everywhere.
To simplify notation, we introduce some operators.

\begin{defn}
\label{def:bellman_operators}(Bellman operators) Given $\psi\in\Psi$
such that $G_{0}^{\psi}\in\mathcal{B}(\mathsf{X})$ and $G_{t}^{\psi}\in\mathcal{B}(\mathsf{X}\times\mathsf{X})$ for $t\in[1:T]$,
we define the operators $Q_{t}^{\psi}:L^{2}(\nu_{t+1}^{\psi})\rightarrow L^{2}(\nu_{t}^{\psi})$
for $t\in[0:T-1]$ as
\begin{align*}
Q_{0}^{\psi}(\varphi)(x) & =G_{0}^{\psi}(x)M_{1}^{\psi}(\varphi)(x),\quad\varphi\in L^{2}(\nu_{1}^{\psi}),\\
Q_{t}^{\psi}(\varphi)(x,y) & =G_{t}^{\psi}(x,y)M_{t+1}^{\psi}(\varphi)(y),\quad\varphi\in L^{2}(\nu_{t+1}^{\psi}),
\end{align*}
where $\nu_{0}^{\psi}:=\mu^{\psi}\in\mathcal{P}(\mathsf{X})$ and
$\nu_{t}^{\psi}:=\eta_{t-1}^{\psi}\otimes M_{t}^{\psi}\in\mathcal{P}(\mathsf{X}\times\mathsf{X})$
for $t\in[1:T]$. For notational convenience
define $Q_{T}^{\psi}(\varphi)(x,y)=G_{T}^{\psi}(x,y)$ for any $\varphi$ (take
$\nu_{T+1}^{\psi}$ as an arbitrary element in $\mathcal{P}(\mathsf{X}\times\mathsf{X})$).
\end{defn}
Although these operators are typically used to define unnormalized
predictive Feynman-Kac models \cite[Proposition 2.5.1]{DelMoral_2004},
we shall adopt terminology from control literature and refer to them
as Bellman operators.
It can be shown that these Bellman operators are well-defined and are in fact bounded
linear operators -- see Proposition \ref{prop:stability}. In this notation,
we can rewrite (\ref{eq:optimal_phi}) more succinctly as
\begin{align}\label{eq:backward_phi}
\phi_{T}^{*} = G_{T}^{\psi},\quad
\phi_{t}^{*} = Q_{t}^{\psi}\phi_{t+1}^{*},\quad t\in[0:T-1].
\end{align}
To understand how regression errors propagate in time,
for $-1\leq s\leq t\leq T$, we define the Feynman-Kac semigroup $Q_{s,t}^{\psi}:L^{2}(\nu_{t+1}^{\psi})\rightarrow L^{2}(\nu_{s+1}^{\psi})$
associated to a policy $\psi\in\Psi$ as
\begin{equation}
Q_{s,t}^{\psi}(\varphi)=\begin{cases}
\varphi, & s=t,\\
Q_{s+1}^{\psi}\circ\cdots\circ Q_{t}^{\psi}(\varphi), & s<t,
\end{cases}\label{eq:feynmanKac_semigroup}
\end{equation}
for $\varphi\in L^{2}(\nu_{t+1}^{\psi})$.
To describe regression steps taken to approximate the intractable recursion (\ref{eq:backward_phi}),
we introduce the following operations.

\begin{defn}
\label{def:log_projection}(Logarithmic projection)
On a measurable space $(\mathsf{E},\mathcal{E})$, let $\nu\in\mathcal{P}(\mathsf{E})$,
$\xi:\mathsf{E}\rightarrow\mathbb{R}_{+}$ be a $\mathcal{E}$-measurable
function such that $-\log\xi\in L^{2}(\nu)\cap\mathcal{L}(\mathsf{E})$,
and $\mathsf{F}\subset\mathcal{L}(\mathsf{E})$ be a closed linear
subspace of $L^{2}(\nu)$. We define the $(\mathsf{F},\nu)$-projection
operator $P^{\nu}:\mathcal{B}(\mathsf{E})\rightarrow\mathcal{B}(\mathsf{E})$
as
\begin{equation}
P^{\nu}\xi=\exp\Big(-\arg\min_{\varphi\in\mathsf{F}}\|\varphi+\log\xi\|_{L^{2}(\nu)}^{2}\Big).\label{eq:log_projection}
\end{equation}
\end{defn}
The projection theorem gives existence of a unique $P^{\nu}\xi$.
We have chosen to define $-\log P^{\nu}\xi$ as the orthogonal
projection of $-\log\xi$ onto $\mathsf{F}$, as this corresponds
to learning the optimal value functions of the associated control problem.
Since projections are typically
intractable, a practical implementation will involve a Monte Carlo
approximation of (\ref{eq:log_projection}).
\begin{defn}(Approximate projection) Following notation in Definition \ref{def:log_projection},
given a consistent approximation $\nu^{N}$ of $\nu$, i.e. $\nu^{N}(\varphi)\rightarrow\nu(\varphi)$
almost surely for any $\varphi\in L^{1}(\nu)$, we define the approximate
$(\mathsf{F},\nu)$-projection operator $P^{\nu,N}:\mathcal{B}(\mathsf{E})\rightarrow\mathcal{B}(\mathsf{E})$
as the $(\mathsf{F},\nu^{N})$-projection operator. We additionally
assume that the function class $\mathsf{F}$ is such that $P^{\nu,N}\xi$
is a random function for all $\xi\in\mathcal{B}(\mathsf{E})$.
\end{defn}

If $\psi\in\Psi$ is the current policy, we use the output of
$\psi$-twisted SMC (Algorithm \ref{alg:twistedSMC})
to learn the optimal policy $\phi^{*}$, through the empirical
measures
\begin{equation}
\nu_{0}^{\psi,N}=\frac{1}{N}\sum_{n=1}^{N}\delta_{X_{0}^{n}},\quad\nu_{t}^{\psi,N}=\frac{1}{N}\sum_{n=1}^{N}\delta_{\big(X_{t-1}^{A_{t-1}^{n}}, \, X_{t}^{n}\big)},\quad t\in[1:T],\label{eq:empirical_measures}
\end{equation}
which are consistent approximations of $(\nu_{t}^{\psi})_{t\in[0:T]}$
\cite{DelMoral_2004}, defined in Definition \ref{def:bellman_operators}.
Given pre-specified closed and linear function
classes $\mathsf{F}_{0}\subset L^{2}(\nu_{0}^{\psi})\cap\mathcal{L}(\mathsf{X})$,
$\mathsf{F}_{t}\subset L^{2}(\nu_{t}^{\psi})\cap\mathcal{L}(\mathsf{X}^{2})$,
$t\in[1:T]$, we denote the approximate $(\mathsf{F}_{t},\nu_{t}^{\psi})$-projection
operator by $P_{t}^{\psi,N}$ for $t\in[0:T]$.
We can now write our ADP algorithm detailed in Algorithm \ref{alg:ADP} succinctly as
\begin{align}\label{eq:ADP}
\hat{\phi}_{T} = P_{T}^{\psi,N}G_{T}^{\psi}, \quad
\hat{\phi}_{t} = P_{t}^{\psi,N}Q_{t}^{\psi}\hat{\phi}_{t+1},\quad t\in[0:T-1].
\end{align}
The following result characterizes how well (\ref{eq:ADP}) can approximate (\ref{eq:backward_phi}).

\begin{prop}\label{prop:moment_bound}
Suppose that we have a policy $\psi\in\Psi$, number of particles
$N$ and closed, linear function classes $\mathsf{F}_{0}\subset L^{2}(\nu_{0}^{\psi})\cap\mathcal{L}(\mathsf{X})$,
$\mathsf{F}_{t}\subset L^{2}(\nu_{t}^{\psi})\cap\mathcal{L}(\mathsf{X}^{2})$,
$t\in[1:T]$ such that:

\textbf{[A1]} the Feynman-Kac semigroup defined in (\ref{eq:feynmanKac_semigroup})
satisfies
\begin{equation}
\|Q_{s,t}^{\psi}(\varphi)\|_{L^{2}(\nu_{s+1}^{\psi})}\leq C_{s,t}^{\psi}\|\varphi\|_{L^{2}(\nu_{t+1}^{\psi})},\quad-1\leq s < t\leq T-1,\label{eq:stability_semigroup}
\end{equation}
for some $C_{s,t}^{\psi}\in[0,\infty)$ and all $\varphi\in L^{2}(\nu_{t+1}^{\psi})$;

\textbf{[A2]} the approximate $(\mathsf{F}_{t},\nu_{t}^{\psi})$-projection
operator satisfies
\[
\sup_{\xi\in\mathsf{S}_{t}^{\psi}}\mathbb{E}^{\psi,N}\|P_{t}^{\psi,N}\xi-\xi\|_{L^{2}(\nu_{t}^{\psi})}\leq e_{t}^{\psi,N}<\infty
\]
where $\mathsf{S}_{t}^{\psi}:=\{Q_{t}^{\psi}\exp(-\varphi):\varphi\in\mathsf{F_{t+1}}\}$
for $t\in[0:T-1]$ and $\mathsf{S}_{T}^{\psi}:=\{G_{T}^{\psi}\}$.
Then the policy $\hat{\phi}\in\Psi$ generated by ADP algorithm (\ref{eq:ADP})
satisfies
\begin{equation}
\mathbb{E}^{\psi,N}\|\hat{\phi}_{t}-\phi_{t}^{*}\|_{L^{2}(\nu_{t}^{\psi})}\leq\sum_{u=t}^{T}C_{t-1,u-1}^{\psi}e_{u}^{\psi,N},\quad t\in[0:T],\label{eq:ADP_bound}
\end{equation}
where $C_{t-1,t-1}^{\psi}=1$ and $\mathbb{E}^{\psi,N}$ denotes expectation w.r.t. the law of
the $\psi$-twisted SMC method (Algorithm \ref{alg:twistedSMC}).
\end{prop}

Equation (\ref{eq:ADP_bound}) reveals how function approximation
errors propagate backwards in time. If the choice of function class
is `rich' enough and the number of particles is sufficiently large,
then these errors can be kept small and ADP
provides a good approximation of the optimal policy. If the number of particles is taken to infinity, the projection errors are driven solely by the choice of function class (as the latter dictates $e_{t}^{\psi,\infty}$). Moreover, observe
that these errors are also modulated by stability constants of the
Feynman-Kac semigroup in (\ref{eq:stability_semigroup}). We now establish
the inequality (\ref{eq:stability_semigroup}). For $\varphi\in\mathcal{B}(\mathsf{E})$,
we write its supremum norm as $\|\varphi\|_{\infty}=\sup_{x\in\mathsf{E}}|\varphi(x)|$.

\begin{prop}
\label{prop:stability}
Suppose $\psi\in\Psi$ is such that $G_{0}^{\psi}\in\mathcal{B}(\mathsf{X})$, $G_{t}^{\psi}\in\mathcal{B}(\mathsf{X}\times\mathsf{X})$ for $t\in[1:T]$ and
let $\delta:=\max_{t\in[0:T]}\|G_{t}^{\psi}\|_{\infty}$ (and $Z_{-1}^{\psi}:=1$).
Then (\ref{eq:stability_semigroup}) holds with
\vspace{-6pt}
\begin{align}
\quad C_{s,t}^{\psi}
&=\bigg(Z_{t}^{\psi}/Z_{s}^{\psi}\prod_{u=s+1}^{t}\|G_{u}^{\psi}\|_{\infty}\bigg)^{1/2} \notag\\
&\leq
\left(Z_{t}^{\psi}/Z_{s}^{\psi}\right)^{1/2}\delta^{(t-s)/2},\quad-1\leq s < t\leq T-1.\label{eq:stability_naive}
\end{align}
\noindent
For the case $G_{t}^{\psi}(x,y)=G_{t}^{\psi}(y)$
for all $x,y\in\mathsf{X}$ and $t\in[1:T]$, if we assume additionally
for each $t\in[1:T]$ that:

\noindent\textbf{{[}A3{]}} there exist $\sigma_{t}^\psi\in \mathcal{P}(\mathsf{X})$ and $\kappa_{t}^\psi \in (0,\infty)$ such that for all $x \in \mathsf{X}$ we have
\begin{equation}
M_t^\psi(x, \mathrm{d} y) \leq \kappa_t^\psi \sigma_t^\psi ( \mathrm{d} y).\label{eq:mixing_assumption}
\end{equation}
Then inequality (\ref{eq:stability_semigroup}) holds with
\begin{equation}
C_{s,t}^{\psi}= \bigg[\kappa_{s+2}^\psi \, \|G_{s+1}^\psi\|_{\infty}\, \sigma_{s+2}^\psi \big(Q_{s+1,t}^\psi (1)\big) \,\frac{Z_t^\psi}{Z_s^\psi}\bigg]^{1/2},\quad-1\leq s < t\leq T-1.\label{eq:stability_mixing}
\end{equation}
\end{prop}

The assumption of bounded potentials is typical in similar analyses
of ADP errors \cite[Section 8.3.3]{Gobet_2016} and stability
of SMC methods \cite{DelMoral_2004}.
The second part of Proposition \ref{prop:stability}
shows that it is possible to exploit regularity properties
of the transition kernels to obtain better constants $C_{s,t}^{\psi}$.
Conditions such as (\ref{eq:mixing_assumption}) are common in the filtering literature, see for example
\cite[Eq.~(9)]{delMoralGuionnet2001} and \cite[ch. 4]{DelMoral_2004}.

\subsection{Limit theorems}\label{sec:LimitTheorems}
We now study the asymptotic behaviour of the ADP algorithm (\ref{eq:ADP}),
with a current policy $\psi\in\Psi$, as the size of the particle
system $N$ grows to infinity. For a common choice of function class,
we will establish convergence
to a policy $\tilde{\phi}=(\tilde{\phi}_{t})_{t\in[0:T]}$,
defined by the idealized algorithm
\begin{align}\label{eq:ideal_ADP}
\tilde{\phi}_{T} = P_{T}^{\psi}G_{T}^{\psi},\quad
\tilde{\phi}_{t} = P_{t}^{\psi}Q_{t}^{\psi}\tilde{\phi}_{t+1},\quad t\in[0:T-1],
\end{align}
where $P_{t}^{\psi}$ denotes the $(\mathsf{F}_{t},\nu_{t}^{\psi})$-projection
operator for $t\in[0:T]$.
In particular, we consider logarithmic
projections that are defined by linear least squares approximations;
this corresponds to function classes of the form
\begin{equation}
\mathsf{F}_{t}:=\left\{ \Phi_{t}^{T}\beta:\beta\in\mathbb{R}^{M}\right\} ,\quad t\in[0:T],\label{eq:linear_functionclass}
\end{equation}
where $\Phi_{0}\subset L^{2}(\nu_{0}^{\psi})\cap\mathcal{L}(\mathsf{X})$,
$\Phi_{t}\subset L^{2}(\nu_{t}^{\psi})\cap\mathcal{L}(\mathsf{X}^{2})$,
$t\in[1:T]$ are vectors of $M\in\mathbb{N}$ pre-specified basis
functions. We will treat $M$ as fixed in our analysis and refer to
\cite[Theorem 8.2.4]{Gobet_2016} for results on how $M$ should increase
with $N$ to balance the tradeoff between enriching (\ref{eq:linear_functionclass}) and the need for more samples
to achieve the same estimation precision. We denote by $\tilde{\phi}:=(\tilde{\phi}_{t})_{t\in[0:T]}$ the policy generated by the idealized algorithm (\ref{eq:ideal_ADP})
where $\tilde{\phi}_{t}:=\exp(-\Phi_{t}^{T}\beta_{t}^{\psi})$, $\beta_{t}^{\psi}$ being the corresponding least squares estimate.
This result builds upon the central limit theorem for particle methods established in \cite{Chopin_2004,DelMoral_2004,Kunsch_2005}.
\begin{thm}\label{thm:LimitShorten}
Consider the ADP algorithm (\ref{eq:ADP}) with current policy $\psi\in\Psi$,
under linear least squares approximations (\ref{eq:linear_functionclass}).
Under appropriate regularity conditions, for all $x\in\mathsf{X}^{2T+1}$,
the estimated policy $\hat{\phi}(x)$ converges in probability to the policy
$\tilde{\phi}(x)$ as $N\rightarrow\infty$.
Moreover, for all $x\in\mathsf{X}^{2T+1}$,
\begin{equation}
\sqrt{N}\left(\hat{\phi}(x)-\tilde{\phi}(x)\right)\stackrel{\mathtt{d}}{\longrightarrow}\mathcal{N}\left(0_{(T+1)},\Omega^{\psi}(x)\right)
\label{eq:CLT_policy_shorten}
\end{equation}
for some $\Omega^{\psi}:\mathsf{X}^{2T+1}\rightarrow\mathbb{R}^{(T+1)\times(T+1)}$, where $\stackrel{\mathtt{d}}{\longrightarrow}$ denotes convergence in distribution and $0_{p}=(0,\ldots,0)^{T}\in\mathbb{R}^{p}$
is the zero vector.
\end{thm}
A precise mathematical statement of this result and its proof are given in Section \ref{sec:supp_limittheorems} of Supplementary Material. Note that the proof relies on a technical central limit theorem on path space that 
can be deduced in the case of multinomial resampling from \cite[Theorem 9.7.1]{DelMoral_2004}. 
The exact form of $\Omega^{\psi}$ reveals how errors correlate over time and suggests that
we may expect the variance of the estimated policy to be larger at earlier
times, due to the inherent backward nature of the ADP approximation.

\subsection{Iterated approximate dynamic programming\label{sec:iADP}}
We provide here a theoretical framework to understand
the qualitative behaviour of policy $\psi^{(I)}$, estimated by Algorithm
\ref{alg:cSMC}, as the number of iterations $I$ grows to infinity.
This offers a novel perspective of iterative algorithms for finite
horizon optimal control problems and may be of general interest.

To do so, we require the set of all admissible policies to be a complete
separable metric space. This follows if we impose that $\mathsf{X}$
is a compact metric space and work with $\Psi:=\mathcal{C}(\mathsf{X})\prod_{t=1}^{T}\mathcal{\mathcal{C}}(\mathsf{X}\times\mathsf{X})$,
equipped with the metric $\rho(\varphi,\xi):=\sum_{t=0}^{T}\|\varphi_{t}-\xi_{t}\|_{\infty}$
for $\varphi=(\varphi_{t})_{t\in[0:T]},\xi=(\xi_{t})_{t\in[0:T]}\in\Psi$;
non-compact state spaces can also be accommodated with a judicious choice
of metric (see e.g. \cite[p. 380]{Bichteler_2002}).

We begin by writing the iterative algorithm with $N\in\mathbb{N}$
particles as an iterated random function $F^{N}:\mathsf{U}\times\Psi\rightarrow\Psi$,
defined by $F_{U}^{N}(\psi)=\psi\cdot\hat{\phi}$, where $\hat{\phi}$
is the output of ADP algorithm (\ref{eq:ADP}) and $U\in\mathsf{U}$
encodes all uniform random variables needed to simulate a $\psi$-twisted
SMC method (Algorithm \ref{alg:twistedSMC}). As the uniform variables
$(U^{(I)})_{I\in\mathbb{N}}$ used at every iteration are independent
and identically distributed, iterating $F^{N}$ defines a Markov chain
$(\psi^{(I)})_{I\in\mathbb{N}}$ on $\Psi$. We
will write $\mathbb{E}$ to denote expectation w.r.t.\ the law of $(U^{(I)})_{I\in\mathbb{N}}$
and $\pi^{(I)}\in\mathcal{P}(\Psi)$ to denote the law of $\psi^{(I)}$.
Similarly, we denote the iterative scheme with exact projections by
$F:\Psi\rightarrow\Psi$, defined as $F(\psi)=\psi\cdot\tilde{\phi}$,
where $\tilde{\phi}$ is the output of the idealized ADP algorithm
(\ref{eq:ideal_ADP}).
We denote by $\varphi^*\in\Psi$ a fixed point (if it exists) of $F$, i.e.
$F(\varphi^*)=\varphi^*$.
The following is based on results developed
in \cite{Diaconis_Freedman_1999}.
\begin{thm}
\label{thm:iADP}Assume that the iterated random function $F^{N}$
satisfies:

\textbf{{[}A4{]}} $\mathbb{E}\left[\rho(F_{U}^{N}(\varphi_{0}),\varphi_{0})\right]<\infty$
for some $\varphi_{0}\in\Psi$,

\textbf{{[}A5{]}} there exists a measurable function $L^{N}:\mathsf{U}\rightarrow\mathbb{R}_{+}$
with $\mathbb{E}\left[L_{U}^{N}\right]<\alpha$ for some $\alpha\in[0,1)$
such that $\rho(F_{U}^{N}(\varphi),F_{U}^{N}(\xi))\leq L_{U}^{N}\rho(\varphi,\xi)$ for all $\varphi,\xi\in\Psi$.

Then the $\Psi$-valued Markov chain $(\psi^{(I)})_{I\in\mathbb{N}}$
generated by Algorithm \ref{alg:cSMC} admits a unique invariant distribution
$\pi\in\mathcal{P}(\Psi)$ and 
\begin{equation}
\varrho(\pi^{(I)},\pi)\leq C(\psi^{(0)})r^{I},\quad I\in\mathbb{N},\label{eq:geometric_convergence}
\end{equation}
\noindent for some $C:\Psi\rightarrow\mathbb{R}_{+}$ and $r\in(0,1)$, where
$\varrho$ denotes the Prohorov metric on $\mathcal{P}(\Psi)$ induced
by the metric $\rho$. If we suppose in addition that:

\textbf{{[}A6{]}} for each $\psi\in\Psi$, $\rho(F_{U}^{N}(\psi),F(\psi))\leq N^{-1/2}E_{U}^{\psi,N}$
where $(E_{U}^{\psi,N})_{N\in\mathbb{N}}$ is a uniformly integrable
sequence of non-negative random variables with finite mean that converges
in distribution to a limiting distribution with support on $\mathbb{R}_{+}$,
then we also have that
\begin{equation}
\mathbb{E}_{\pi}\left[\rho(\psi,\varphi^{*})\right]\leq N^{-1/2}\mathbb{E}\left[E_{U}^{\varphi^{*},N}\right](1-\alpha)^{-1}\label{eq:characterize_invariant}
\end{equation}
where $\varphi^{*}$ is a fixed point of $F$ and $\mathbb{E}_{\pi}$
denotes expectation w.r.t. $\psi\sim\pi$.
\end{thm}
Assumption A5 requires the ADP procedure to be sufficiently regular:
i.e. for two policies $\varphi,\xi\in\Psi$ that are close, given
the same uniform random variables $U$ to simulate a $\varphi$-twisted
and $\xi$-twisted SMC method, the policies $\hat{\varphi}$ (w.r.t.
$\mathbb{Q}^{\varphi}$) and $\hat{\xi}$ (w.r.t. $\mathbb{Q}^{\xi}$)
estimated by (\ref{eq:ADP}) should also be close enough to keep the
Lipschitz constant $L_{U}^{N}$ small.
Assumption A6 is necessary
to quantify the Monte Carlo error involved when employing approximate
projections and can be deduced for example using the central limit
theorem in (\ref{eq:CLT_policy_shorten}).
See Section \ref{sec:iterativethm} of the
Supplementary Material for a discussion on when and why contraction occurs, 
and a simple example where Assumptions A4-A6 are verified.

The first part of Theorem \ref{thm:iADP}, which establishes existence
of a unique invariant distribution and geometric convergence to the
latter, follows from standard theory on iterated random functions; see, e.g., \cite{Diaconis_Freedman_1999}.
The second conclusion of Theorem \ref{thm:iADP}, which provides a
characterization of the limiting distribution, is to the best of our
knowledge novel. The fixed point $\varphi^*$ can be interpreted as a policy
for which subsequent refinement using exact (i.e. with $N\rightarrow\infty$) projections onto the same function classes yields no change.

\section{Application to state space models\label{sec:application_statesspacemodels}}
\subsection{Neuroscience model}
\begin{figure}
\begin{centering}
\includegraphics[scale=0.4]{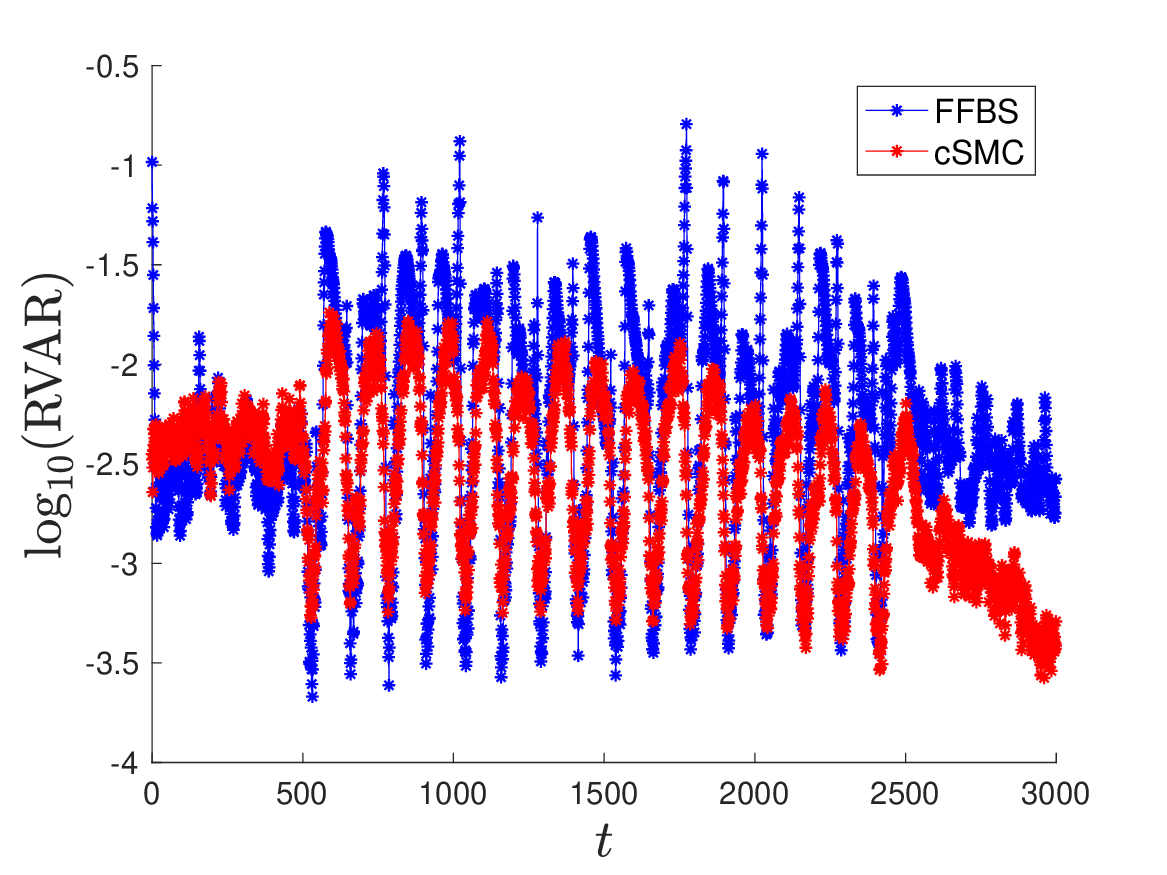}\includegraphics[scale=0.4]{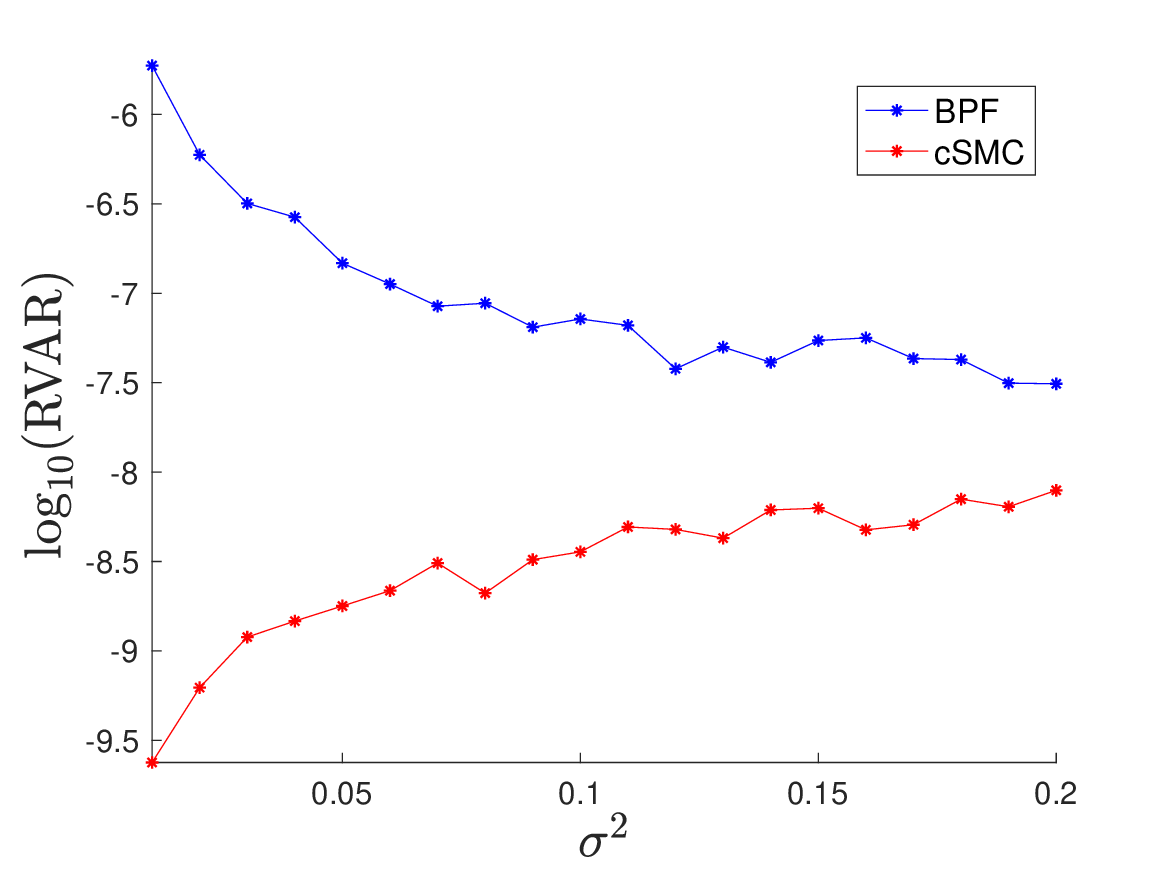}
\par\end{centering}
\caption{\label{fig:neuroscience_compareBPF} Assessing performance on the neuroscience model introduced in Section \ref{sec:statespace_models} based on $100$ independent repetitions of each algorithm:
sample relative variance of smoothing expectation (\emph{left}) and log-marginal likelihood estimates (\emph{right}).}
\end{figure}

We return to the neuroscience model introduced in Section \ref{sec:statespace_models}
and explore cSMC's utility as a smoother, with algorithmic
settings described in Section \ref{sec:neuroscience},
in comparison to the forward filtering backward smoothing (FFBS) procedure of \cite{Doucet_Godsill_Andrieu_2000,Kunsch_2013}.
We consider an approximation of the maximum likelihood estimate (MLE)
$(\alpha,\sigma^{2})=(0.99,0.11)$ as parameter value and the smoothing
functional $x_{0:T}\mapsto M(\kappa(x_{0}),...,\kappa(x_{T}))$ whose
expectation represents the expected number of activated neurons at
each time.
Although BPF's particle approximation of the smoothing distribution
degenerates quickly in time, cSMC with $I=3$ iterations offers a marked
improvement: for example, the number of distinct ancestors at the
initial time is on average $63$ times that of BPF.
We use $N=1024$ particles for cSMC and select the number
of particles in FFBS to match compute time.
The results, displayed in the left panel of Figure \ref{fig:neuroscience_compareBPF},
show some gains over FFBS and especially so at later times.

We then investigate the relative variance of the log-marginal likelihood estimates
obtained using cSMC and BPF in a neighbourhood of the approximate MLE.
As the marginal likelihood surface is rather flat in $\alpha$, we fix $\alpha=0.99$ and
vary $\sigma^{2}\in\{0.01,0.02,\ldots,0.2\}$.
We use $I=3$ iterations, $N=128$ particles for cSMC and $N=5529$
particles for BPF to match computational cost.
The results, reported in the right panel of Figure \ref{fig:neuroscience_compareBPF},
demonstrate that while the relative variance of estimates produced by BPF increases
exponentially as $\sigma^{2}$ decreases, that of cSMC is stable across the values of
$\sigma^{2}$ considered.

\begin{figure}
\begin{centering}
\includegraphics[scale=0.4]{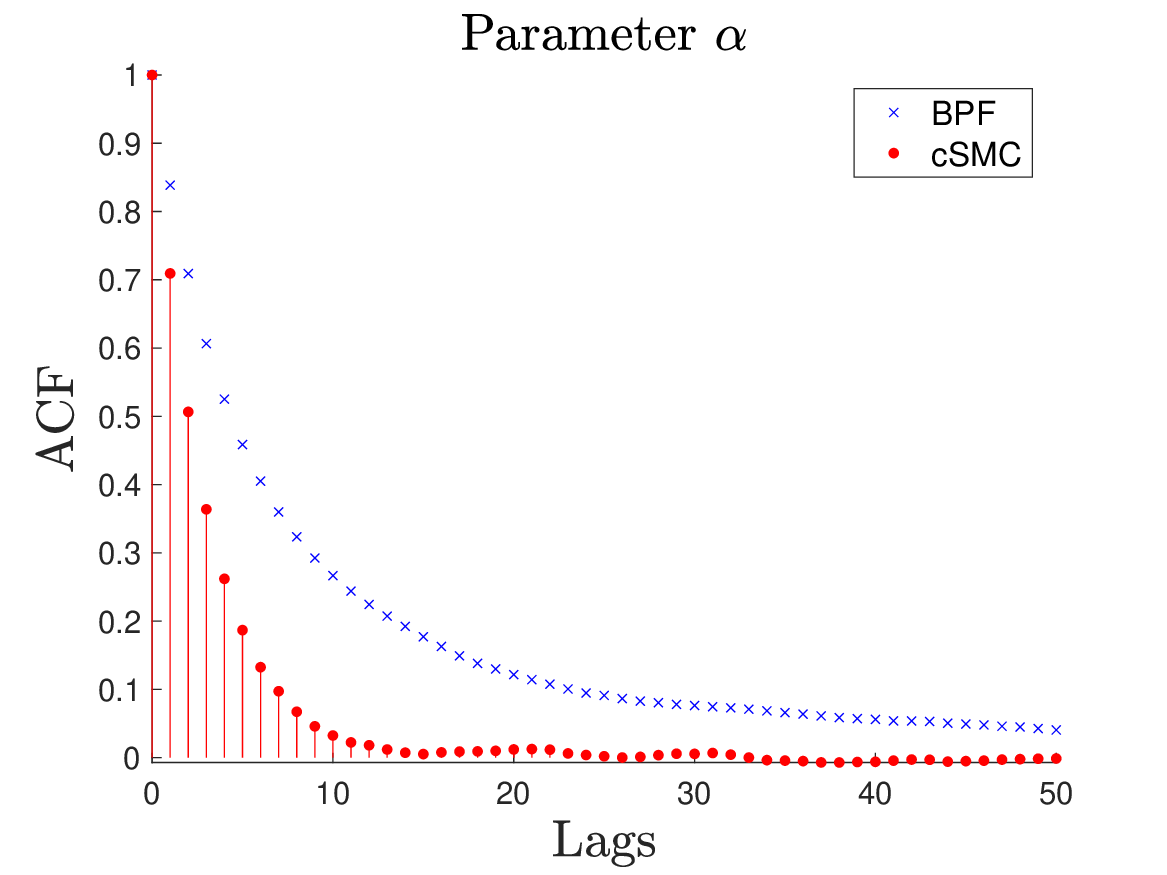}\includegraphics[scale=0.4]{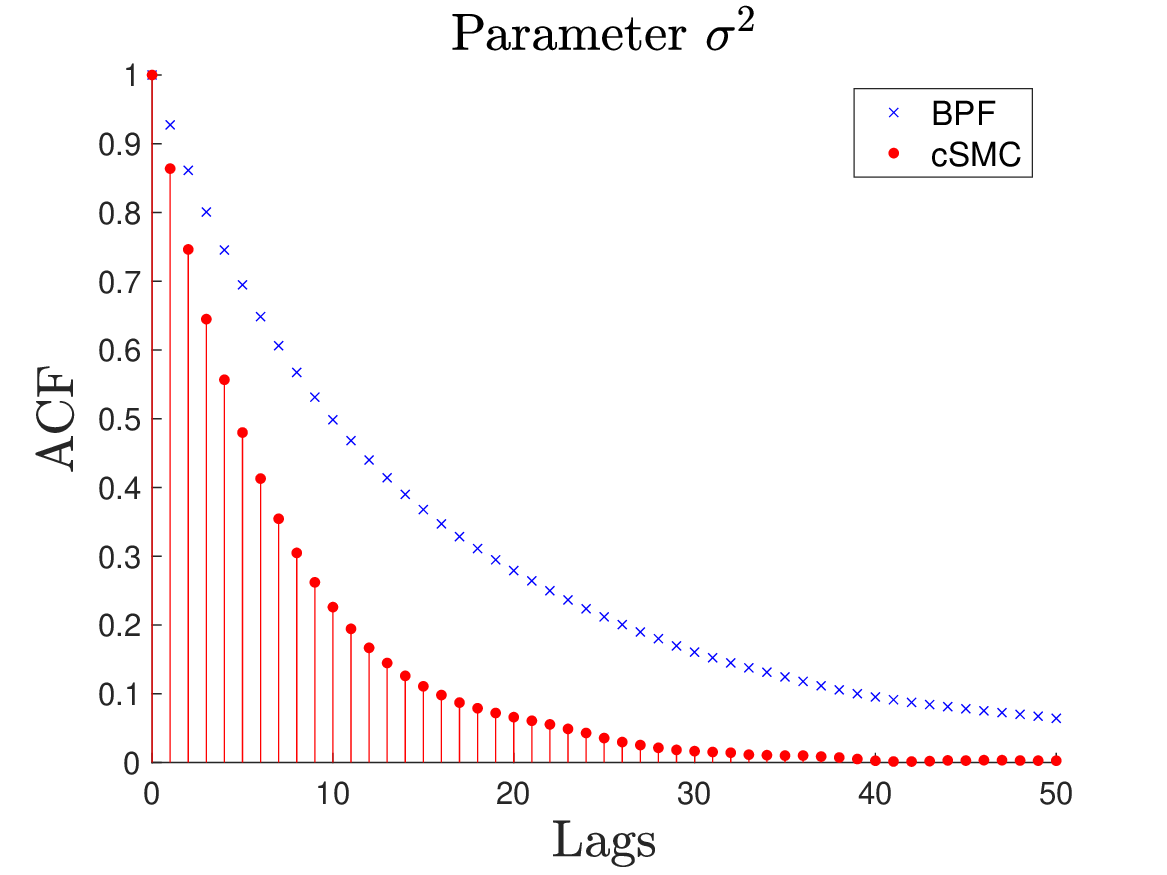}
\par\end{centering}
\caption{\label{fig:PMMH_output}Autocorrelation functions of PMMH chains, with marginal likelihood estimates produced by cSMC or BPF, for parameters of the neuroscience model introduced in Section \ref{sec:statespace_models}.}
\end{figure}

Lastly, we perform Bayesian inference on the unknown parameters $\theta=(\alpha,\sigma^{2})$
and compare the efficiency of cSMC and BPF within a particle marginal
Metropolis--Hastings (PMMH) algorithm \cite{Andrieu_Doucet_Holenstein_2010}.
We specify a uniform prior on $[0,1]$ for $\alpha$ and an independent
inverse-Gamma prior distribution $\mathcal{IG}(1,0.1)$ for $\sigma^{2}$.
Initializing at $\theta=(0.99,0.11)$, we run two PMMH chains $(\theta_{k}^{\mathrm{cSMC}})_{k\in[0:K]}$,
$(\theta_{k}^{\mathrm{BPF}})_{k\in[0:K]}$ of length $K=100,000$.
Both chains are updated using an independent Gaussian random walk
proposal with standard deviation $(0.002,0.01)$, but rely on cSMC
or BPF to produce unbiased estimates of the marginal likelihood
when computing acceptance probabilities. To ensure a fair comparison,
we use $I=3$ iterations and $N=128$ particles for cSMC which matches
the compute time taken by BPF with $N=5529$ particles, so that
both PMMH chains require very similar computational cost.
The autocorrelation functions of each PMMH chain, shown in Figure \ref{fig:PMMH_output},
reveal that the $(\theta_{k}^{\mathrm{BPF}})_{k\in[0:K]}$ chain has
poorer mixing properties.
These differences can be summarized by the effective sample size,
computed as the length of the chain $K$ divided by the estimated integrated autocorrelation
time for each parameter of interest, which was found to be $(4356,2442)$
for $(\theta_{k}^{\mathrm{BPF}})_{k\in[0:K]}$ and $(20973,13235)$ for
$(\theta_{k}^{\mathrm{cSMC}})_{k\in[0:K]}$.

\subsection{The Lorenz-96 model\label{sec:lorenz}}
Following \cite{Murray_Singh_Jacob_Lee_2016}, we consider the Lorenz-96
model \cite{Lorenz_1996} in a low noise regime, i.e. the It{\^o} process
$\xi(s)=(\xi_{i}(s))_{i\in[1:d]},s\geq0$ defined as the weak solution
of the stochastic differential equation:
\begin{equation}
\mathrm{d}\xi_{i}=\left(-\xi_{i-1}\xi_{i-2}+\xi_{i-1}\xi_{i+1}-\xi_{i}+\text{\ensuremath{\alpha}}\right)\mathrm{d}t+\sigma_{f}dB_{i},\quad i\in[1:d],\label{eq:lorenz96}
\end{equation}
where indices should be understood modulo $d$, $\alpha\in\mathbb{R}$
is a forcing parameter, $\sigma_{f}^{2}\in\mathbb{R}_{+}$ is a noise
parameter and $B(s)=(B_{i}(s))_{i\in[1:d]},s\geq0$ is a $d$-dimensional
standard Brownian motion. The initial condition is taken as $\xi(0)\sim\mathcal{N}(0_{d},\sigma_{f}^{2}I_{d})$.
We assume that the process is observed at a regular time grid of size
$h>0$ according to $Y_{t}\sim\mathcal{N}(H\xi(s_{t}),R), s_{t}=th,t\in[0:T]$,
and consider the partially observed case where $H_{ii}=1$ for $i=1,\ldots,p$ and $0$ otherwise
with $p=d-2$.

As discussed in \cite{Murray_Singh_Jacob_Lee_2016}, an efficient
discretization scheme in this low noise regime \cite[ch. 3]{Milstein_Tretyakov_2004}
is given by adding Brownian increments to the output of a high-order
numerical integration scheme on the drift of (\ref{eq:lorenz96}).
Incorporating time discretization gives a time homogenous state space
model on $(\mathsf{X},\mathcal{X})=(\mathbb{R}^{d},\mathfrak{B}(\mathbb{R}^{d}))$
with
$\nu=\mathcal{N}(0_d,\sigma_{f}^{2}I_{d})$,
$f(x_{t-1},\mathrm{d}x_{t})=\mathcal{N}(x_{t};q(x_{t-1}),\sigma_{f}^{2}hI_{d})\mathrm{d}x_{t}$
and $g(x_{t},y_{t})=\mathcal{N}(y_{t};Hx_{t},R)$
for $t\in[1:T]$, where $y_{0:T}\in\mathsf{Y}^{T+1}=(\mathbb{R}^{p})^{T+1}$
is a realization of the observation process and $q:\mathsf{X}\rightarrow\mathsf{X}$
denotes the mapping induced by a fourth order Runge--Kutta (RK4) method
on $[0,h]$.
We will take noise parameters as $\sigma_{f}^{2}=10^{-2},R=\sigma_{g}^{2}I_{p}$,
observe the process for $10$ time units, i.e. set $h=0.1$, $T=100$
and implement RK4 with a step size of $10^{-2}$.
For this application, we can employ the fully adapted APF as uncontrolled
SMC method \cite{Pitt_Shephard_1999},
i.e. set $\mu=\nu^{\psi}$ and $M_{t}=f^{\psi}$ for $t\in[1:T]$ with policy
$\psi_{t}=g,t\in[0:T]$.

Our ADP approximation will utilize the function classes
\begin{equation}\label{eq:lorenz_functionclass}
\mathsf{F}_{t}=\left\{ \varphi(x_{t})=x_{t}^{T}A_{t}x_{t}+x_{t}^{T}b_{t}+c_{t}:(A_{t},b_{t},c_{t})\in\mathbb{S}_{d}\times\mathbb{R}^{d}\times\mathbb{R}\right\}, t\in[0:T],
\end{equation}
where $\mathbb{S}_{d}=\{A\in\mathbb{R}^{d\times d}: A=A^T\}$. Under this parameterization, the policy
$\psi^{(i)}=(\psi_{t}^{(i)})_{t\in[0:T]}$ at iteration $i\in[1:I]$ of Algorithm \ref{alg:cSMC} is given by
\begin{equation}
-\log\psi_{t}^{(i)}(x_{t})=x_{t}^{T}A_{t}^{(i)}x_{t}+x_{t}^{T}b_{t}^{(i)}+c_{t}^{(i)},\quad t\in[0:T],\label{eq:lorenz_currentpolicy}
\end{equation}
where $A_{t}^{(i)}:=\sum_{j=1}^{i}A_{t}^{j},b_{t}^{(i)}:=\sum_{j=1}^{i}b_{t}^{j},c_{t}^{(i)}:=\sum_{j=1}^{i}c_{t}^{j}$
for $t\in[0:T]$ and $(A_{t}^{j+1},b_{t}^{j+1},c_{t}^{j+1})_{t\in[0:T]}$
denotes coefficients estimated using linear least squares at iteration
$j\in[0:I-1]$. Having APF as uncontrolled SMC
is also equivalent to taking BPF as uncontrolled with an initial policy
$\psi^{(0)}=(\psi_{t}^{(0)})_{t\in[0:T]}$ of the form (\ref{eq:lorenz_currentpolicy})
with $A_{t}^{(0)} := \frac{1}{2}\sigma_{g}^{-2}H^{T}H$, $b_{t}^{(0)} := -\sigma_{g}^{-2}Hy_{t}$ and
$c_{t}^{(0)} := \frac{1}{2}\sigma_{g}^{-2}y_{t}^{T}y_{t}+\frac{1}{2}p\log(2\pi)+\frac{1}{2}d\log(\sigma_{g}^{2})$
for $t\in[1:T]$.
For $A\in\mathbb{S}_{d}$, the notation $A\succ0$ refers to $A$ being positive
definite. If the constraints $(\sigma_{f}^{-2}I_{d}+2A_{0}^{(i)})^{-1}\succ0$, $(\sigma_{f}^{-2}h^{-1}I_{d}+2A_{t}^{(i)})^{-1}\succ0,t\in[1:T]$ are satisfied or imposed\footnote{In our numerical implementation,
we find that these constraints are already satisfied when the step size $h$ is sufficiently small. Otherwise, they
can be imposed by projecting onto the set of real symmetric positive definite matrices using the algorithm in \cite{Higham_1988}.},
then sampling from the
twisted initial distribution and transition kernels is feasible and evaluation of
the corresponding potentials is also tractable; see Section \ref{appendix:lorenz96} of
Supplementary Material for exact expressions.
The diagnostics discussed in Section \ref{sec:neuroscience} indicate that
(\ref{eq:lorenz_currentpolicy}) provides an adequate approximation of the
optimal policy by adapting to the chaotic behaviour of the Lorenz system.

\begin{figure}
\begin{centering}
\includegraphics[scale=0.4]{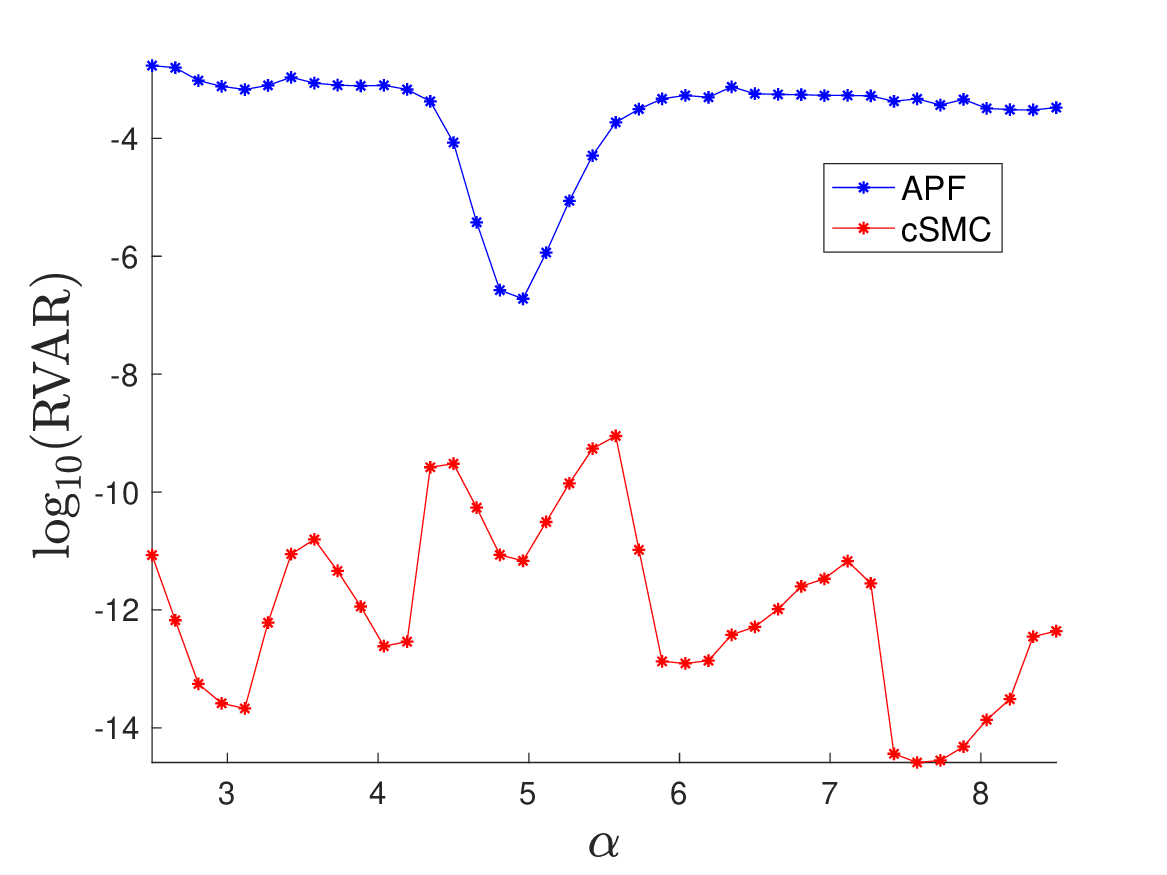}\includegraphics[scale=0.4]{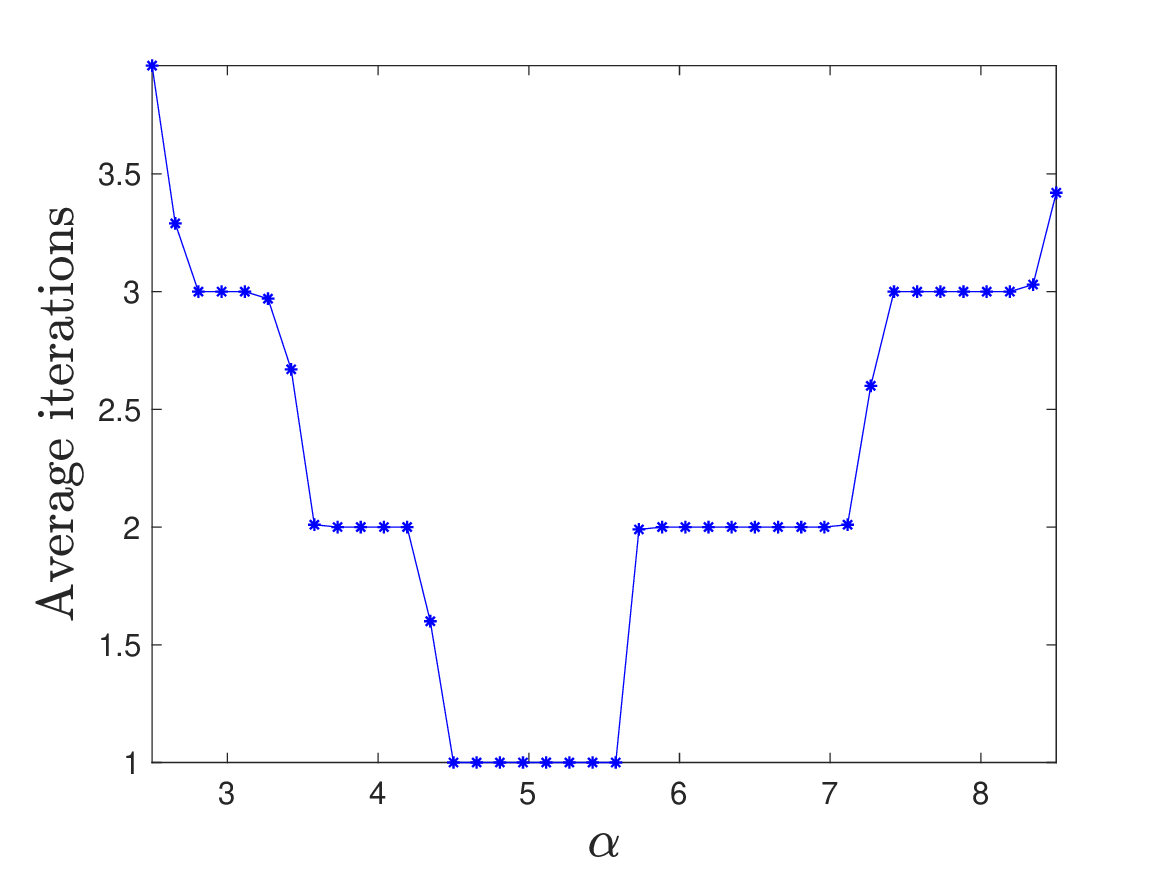}
\par\end{centering}
\caption{\label{fig:lorenz_compare}Lorenz-96 model of Section \ref{sec:lorenz} with data generating parameter
$\alpha=4.8801$: sample relative variance of log-marginal likelihood estimates
based on $100$ independent repetitions of each algorithm (\emph{left}),
average number of iterations taken by cSMC with adaptation (\emph{right}).}
\end{figure}

We begin by comparing the relative variance of the log-marginal likelihood estimates
obtained by cSMC and APF, as $\alpha$ takes values in a regular grid
between $2.5$ to $8.5$. We consider $d=8$ and simulate observations
under the model with $\alpha=4.8801,\sigma_{g}^{2}=10^{-4}$.
We employ $N=512$ particles and the following adaptive strategy within cSMC: perform policy refinement
until the minimum ESS over time is at least $90\%$, terminating at
a maximum of $4$ iterations. To ensure a fair comparison, the number of particles used in APF
is chosen to match computation time.
The results, plotted in the left panel of Figure \ref{fig:lorenz_compare},
show that cSMC offers significant variance reduction across all values
of $\alpha$ considered. Moreover, we see from the right panel of Figure \ref{fig:lorenz_compare}
that the adaptive criterion allows us to adaptively increase the number
of iterations as we move away from the data generating parameter.
We then compare cSMC against the iterated APF \cite[Algorithm 4]{Guarniero_Lee_Johansen_2016}
when function approximations are performed in the logarithmic scale (\ref{eq:lorenz_functionclass}).
Using $N=512$ particles and $I=3$ iterations with the fully adapted APF as initialization for both algorithms,
the sample variance of cSMC log-marginal likelihood estimates at $\alpha=4.8801$, based on $1,000$
independent repetitions, was smaller than iterated APF at each iteration $i\in[1:3]$,
with a relative ratio of $\{0.99, 0.94, 0.92\}$, respectively.

Next we consider configurations $(d,\sigma_{g}^{2})\in\{8,16,32,64\}\times\{10^{-4},10^{-3},10^{-2}\}$
with $\alpha=4.8801$ and generate observations under the model.
We use $I=1$ iteration for cSMC in all configurations
and increase the number of particles $N$ with $d$ for both algorithms.
As before, $N$ is chosen so that both methods require the same
compute time to ensure a fair comparison.
The relative variance of both methods are reported in
Table \ref{tab:lorenz}. These results indicate several order of magnitude
gains over APF in all configurations considered.

\begin{table}
\begin{centering}
\begin{tabular}{|c|c|c|c|c|c|c|}
\cline{5-7}
\multicolumn{1}{c}{} & \multicolumn{1}{c}{} & \multicolumn{1}{c}{} &  & \multicolumn{3}{c|}{\textbf{Observation noise}}\tabularnewline
\cline{5-7}
\multicolumn{1}{c}{} & \multicolumn{1}{c}{} & \multicolumn{1}{c}{} &  & $\sigma_{g}^{2}=10^{-4}$ & $\sigma_{g}^{2}=10^{-3}$ & $\sigma_{g}^{2}=10^{-2}$\tabularnewline
\cline{4-7}
\multicolumn{1}{c}{} & \multicolumn{1}{c}{} &  & \textit{$N$} & \textit{$\log_{10}(\mathrm{RVAR})$} & \textit{$\log_{10}(\mathrm{RVAR})$} & \textit{$\log_{10}(\mathrm{RVAR})$}\tabularnewline
\hline
\multirow{8}{*}{\begin{turn}{90}
\textbf{Algorithm}
\end{turn}} & \multirow{4}{*}{{APF}} & \textit{$d=8$} & \multicolumn{1}{c|}{\textit{$1382$}} & \textit{$-6.7263$} & \textit{$-5.6823$} & \textit{$-4.4061$}\tabularnewline
 &  & \textit{$d=16$} & \textit{$2027$} & \textit{$-7.4056$} & \textit{$-5.9009$} & \textit{$-4.4719$}\tabularnewline
 &  & \textit{$d=32$} & \textit{$4034$} & \textit{$-7.5943$} & \textit{$-5.4901$} & \textit{$-4.1039$}\tabularnewline
 &  & \textit{$d=64$} & \textit{$11,468$} & \textit{$-7.5173$} & \textit{$-5.3765$} & \textit{$-3.1057$}\tabularnewline
\cline{2-7}
 & \multirow{4}{*}{{cSMC}} & \textit{$d=8$} & \textit{$512$} & \textit{$-11.1252$} & \textit{$-10.4173$} & \textit{$-8.66563$}\tabularnewline
 &  & \textit{$d=16$} & \textit{$512$} & \textit{$-11.8899$} & \textit{$-11.1011$} & \textit{$-9.29596$}\tabularnewline
 &  & \textit{$d=32$} & \textit{$1024$} & \textit{$-12.5804$} & \textit{$-11.8622$} & \textit{$-9.6577$}\tabularnewline
 &  & \textit{$d=64$} & \textit{$4096$} & \textit{$-13.5959$} & \textit{$-12.7691$} & \textit{$-9.74631$}\tabularnewline
\hline
\end{tabular}
\par\end{centering}
\caption{\label{tab:lorenz}Algorithmic settings and performance of APF and cSMC
for each dimension $d$ and observation noise $\sigma_{g}^{2}$ considered.
Notationally, $N$ refers to the number of particles and $\mathrm{RVAR}$ is the
sample relative variance of log-marginal likelihood estimates over $100$ independent
repetitions of each method. }
\end{table}

\section{Application to static models\label{sec:application_staticmodels}}
We now detail how the proposed methodology can be applied to
static models described in Section \ref{sec:static_models}.
The framework introduced in \cite{DelMoral_Doucet_Jasra_2006}
generalizes the AIS method of \cite{Neal_2001} and the sequential sampler of \cite{Chopin_2002} by allowing arbitrary forward and backward
kernels instead of being restricted to MCMC kernels. This degree of freedom is useful here
as sampling from twisted MCMC kernels and
computing integrals w.r.t. these kernels is typically impossible.
\subsection{Setup}\label{sec:setupstaticmodels}
We consider the Bayesian framework where the target distribution of
interest is a posterior distribution $\eta(\mathrm{d}x)=Z^{-1}\thinspace\mu(\mathrm{d}x)\ell(x,y)$
defined on $(\mathsf{X},\mathcal{X})=(\mathbb{R}^{d},\mathfrak{B}(\mathbb{R}^d))$,
given by a Bayes update with a prior distribution $\mu\in\mathcal{P}(\mathsf{X})$ and a likelihood
function $\ell:\mathsf{X}\times\mathsf{Y}\rightarrow\mathbb{R}_{+}$.
In applications, the marginal likelihood $Z(y):=\int_{\mathsf{X}}\mu(\mathrm{d}x)\ell(x,y)$
of observations $y\in\mathsf{Y}$ is often also a quantity of interest.
Assuming $\eta$ has a strictly positive and continuously differentiable density
$x\mapsto\eta(x)$ w.r.t. Lebesgue measure on $\mathbb{R}^{d}$, we
select the forward kernel $M_{t}$ related to the transition kernel of an unadjusted Langevin algorithm
(ULA) \cite{Roberts_Tweedie_1996,Roberts_Stramer_2002}
targeting $\eta_{t}$ defined in (\ref{eq:geometric_path}). For e.g., we will define $M_{t}(x_{t-1},\mathrm{d}x_{t})=\mathcal{N}(x_{t};x_{t-1}+\frac{1}{2}h\Gamma\nabla\log\eta_{t}(x_{t-1}),h \Gamma)\mathrm{d}x_{t}$ where $h>0$ denotes the step size, and $\Gamma$ is a positive definite pre-conditioning matrix (which in the simplest case may be the identity $\Gamma=I_d$).

Under appropriate regularity conditions, for sufficiently small $h$,
$M_{t}$ admits an invariant distribution that is close to $\eta_{t}$
\cite{Mattingly_Stuart_Higham_2002}. Moreover, as the corresponding
Langevin diffusion is $\eta_{t}$-reversible, this suggests
that $M_{t}$ will also be approximately $\eta_{t}$-reversible for
small $h$. This prompts the choice of backward
kernel $L_{t-1}(x_{t},\mathrm{d}x_{t-1})=M_{t}(x_{t},\mathrm{d}x_{t-1})$,
in which case, we expect the potentials (\ref{eq:SMCsampler_potentials})
to be close to (\ref{eq:AIS_potentials}) when the step size is small.
We have limited the scope of this article to overdamped Langevin dynamics;
future work could consider the use of generalized Langevin dynamics and other non-reversible dynamics.

\subsection{Log-Gaussian Cox point process}\label{sec:coxprocess}
\begin{figure}
\centering{}\includegraphics[scale=0.4]{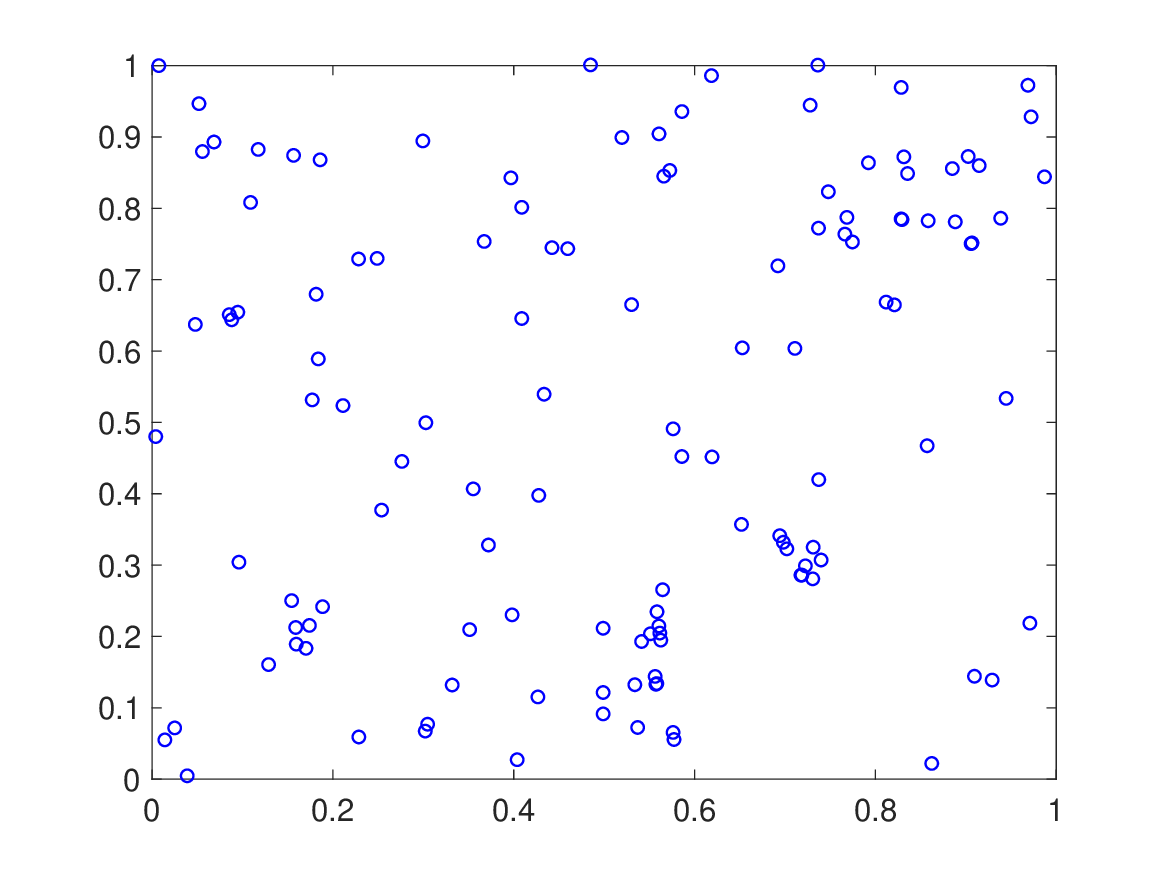}\includegraphics[scale=0.4]{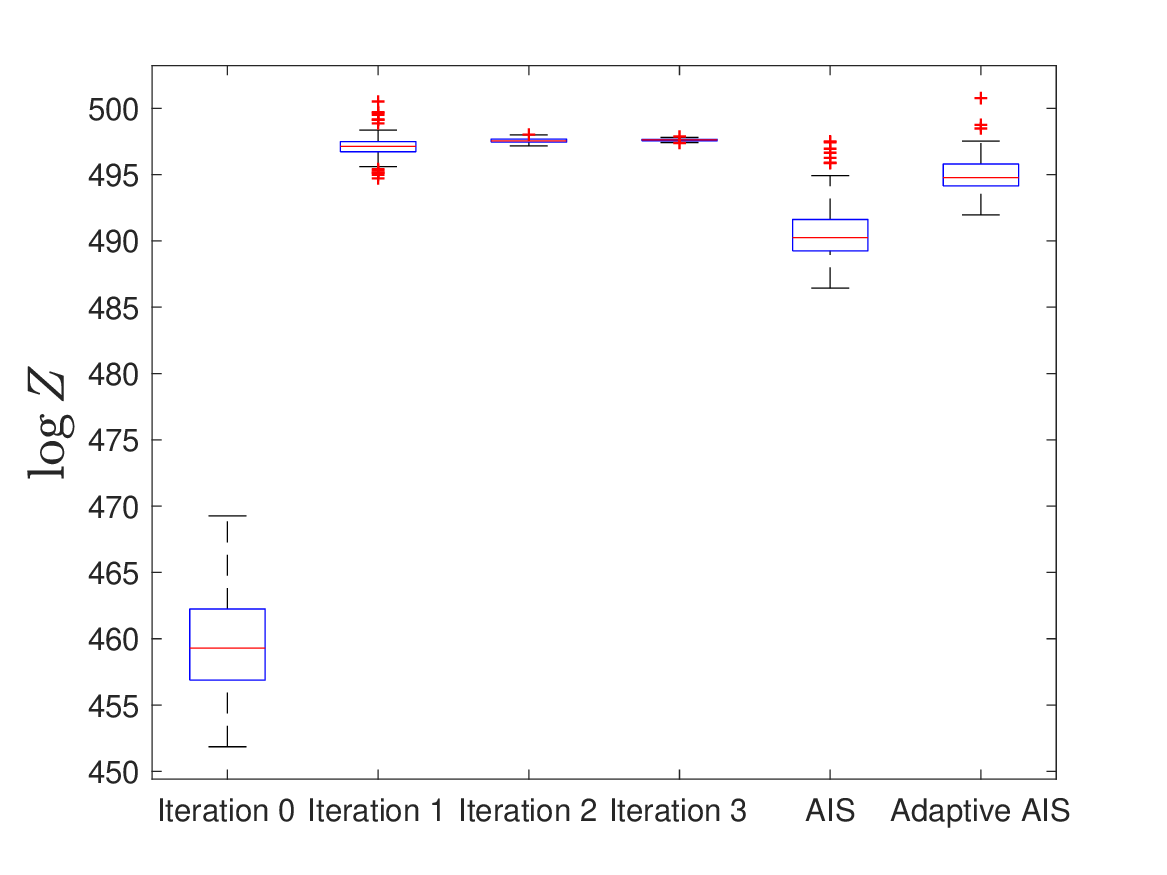}\caption{\label{fig:scotspine}Locations of $126$ Scots pine saplings in a
natural forest in Finland (\emph{left}) and log-marginal likelihood estimates
obtained with $100$ independent repetitions of cSMC, standard AIS, and adaptive AIS (\emph{right}). }
\end{figure}

We end with a challenging high dimensional application of Bayesian inference
for log-Gaussian Cox point processes on a
dataset\footnote{The dataset can be found in the $\texttt{R}$ package $\texttt{spatstat}$
as $\texttt{finpines}$.} concerning the locations of $126$ Scots pine saplings in a natural forest
in Finland \cite{Moller_Syversveen_Waagepetersen_1998,Christensen_Roberts_Rosenthal_2005,Girolami_Calderhead_2011}.
The actual square plot of $10\times10$ square metres is standardized
to the unit square and locations are plotted in the left panel
of Figure \ref{fig:scotspine}. We then discretize $[0,1]^{2}$ into
a $30\times30$ regular grid. Given a latent intensity process $\text{\ensuremath{\Lambda}}=(\Lambda_{m})_{m\in[1:30]^{2}}$,
the number of points in each grid cell $Y=(Y_{m})_{m\in[1:30]^{2}}\in\mathbb{N}^{30^{2}}$
are modelled as conditionally independent and Poisson distributed
with means $a\Lambda_{m}$, where $a=1/30^{2}$ is the area of each
grid cell. The prior distribution for $\Lambda$ is specified
by $\Lambda_{m}=\exp(X_{m})$, $m\in[1:30]^{2}$, where $X=(X_{m})_{m\in[1:30]^{2}}$
is a Gaussian process with constant mean $\mu_{0}\in\mathbb{R}$ and
exponential covariance function $\Sigma_{0}(m,n)=\sigma^{2}\exp(-|m-n|/(30\beta))$ for $m,n\in[1:30]^{2}.$
We will adopt the parameter values $\sigma^{2}=1.91$, $\beta=1/33$
and $\mu_{0}=\log(126)-\sigma^{2}/2$ estimated by \cite{Moller_Syversveen_Waagepetersen_1998}.
This application corresponds to dimension $d=900$, a prior distribution
$\mu=\mathcal{N}(\mu_{0}1_{d},\Sigma_{0})$
with $1_{d}=(1,\ldots,1)^T\in\mathbb{R}^d$ and likelihood function
$\ell(x,y)=\prod_{m\in[1:30]^{2}}\exp\left(x_{m}y_{m}-a\exp(x_{m})\right)$,
where $y=(y_{m})_{m\in[1:30]^{2}}\in\mathsf{Y}=\mathbb{N}^{d}$ is
the given dataset.

For this application, cSMC relies on pre-conditioned ULA moves with the choice of $\Gamma^{-1} = \Sigma_{0}^{-1} + a\exp(\mu_0+\sigma^2/2)I_d$ considered in \cite{Girolami_Calderhead_2011}. 
As the above choice of pre-conditioning captures the curvature of the posterior distribution, we adopt the following function classes
\begin{align}\label{eqn:functionclass_staticmodels}
\mathsf{F}_{0} & =\left\{ \varphi(x_{0})=x_{0}^{T}A_{0}x_{0}+x_{0}^{T}b_{0}+c_{0}:(A_{0},b_{0},c_{0})\in\mathbb{S}_{d}\times\mathbb{R}^{d}\times\mathbb{R}\right\},\\
\mathsf{F}_{t} & =\left\{ \varphi(x_{t-1},x_{t})=x_{t}^{T}A_{t}x_{t}+x_{t}^{T}b_{t}+c_{t}-(\lambda_{t}-\lambda_{t-1})\log\ell(x_{t-1},y)\right.\nonumber \\
 & \left.\quad\quad:(A_{t},b_{t},c_{t})\in\mathbb{S}_{d}\times\mathbb{R}^{d}\times\mathbb{R}\right\} ,\quad t\in[1:T],\nonumber
\end{align}
where $(A_{t})_{t\in[0:T]}$ are restricted to diagonal matrices to reduce the computational
overhead involved in estimating large number of coefficients for a problem
of this scale.
The rationale for approximating the $x_{t-1}$ dependency in $\psi_{t}^{*}(x_{t-1},x_{t}),t\in[1:T]$
is based on the argument that the potentials (\ref{eq:SMCsampler_potentials})
would be close to that of AIS (\ref{eq:AIS_potentials}) for sufficiently
small step size $h$.
We refer to Section \ref{sec:logistic_ADP} of Supplementary Material
for exact expressions required to implement cSMC.
As before, the diagnostics discussed in Section \ref{sec:neuroscience}
reveal that such a parameterization offers an adequate approximation of the
optimal policy.

We select as competing algorithms: 1) standard AIS
with pre-conditioned Metropolis-adjusted Langevin algorithm (MALA) moves; and, 2) an adaptive (pre-conditioned) AIS.
For both cSMC and standard AIS, we adopt the geometric path (\ref{eq:geometric_path}) with $\lambda_{t}=t/T$ and fix the number of time steps as $T=20$. We use $N=4096$ particles, $I=3$ iterations for cSMC and $5$ times more particles for standard AIS to ensure that our comparison is performed at a fixed computational complexity. Using pilot runs, we chose a step size of $0.4$ for MALA to achieve suitable acceptance probabilities, and a smaller step size of $0.05$ for ULA as this improves the approximation in (\ref{eqn:functionclass_staticmodels}). For the adaptive AIS algorithm, we also adopt (\ref{eq:geometric_path}) with $\lambda_{t}$ adapted so that the ESS$\%$ is maintained above $80\%$ \cite{Jasra_etal_2011,Schafer_Chopin_2013,Zhou_Johansen_Aston_2016} and with an adaptive step size chosen to ensure an acceptance probability within the range of $30\%$ to $50\%$ at each time step \cite{Jasra_etal_2011,Beskos_etal_2016}. Since the runtime of adaptive AIS is random, we choose the number of particles to ensure the averaged computational cost matches that of cSMC and standard AIS; this is typically on the order of $2$ times as many particles as cSMC.

The results obtained show that standard AIS performs poorly in this scenario, providing high variance estimates of the log-marginal likelihood
compared to each iteration of cSMC, as displayed in the right panel of Figure \ref{fig:scotspine}. Adaptive AIS performs better than standard AIS but it is still outperformed by cSMC. The sample variance of log-marginal likelihood estimates is $573$ times smaller for the last iteration of cSMC compared to standard AIS, and it is $200$ times smaller compared to adaptive AIS. The mean squared
error\footnote{Computed by taking reference to an estimate obtained using many repetitions
of a SMC sampler with a large number of particles.} of adaptive AIS algorithm is $920$ times larger than that of cSMC. 

\section*{Supplementary material}
The supplement contains proofs of all results, a detailed description of
the connection to Kullback--Leibler control, three more applications employing other flexible function classes,
and some model specific expressions.

\appendix

\section{Proofs of Section \ref{sec:optimal_policies}}
\begin{proof}[Proof of Proposition \ref{prop:Optimality}]
By Fubini's theorem, $\phi^{*}$ is well-defined as the integrals
in (\ref{eq:optimal_phi}) exist since $Z=\mathbb{E}_{\mathbb{Q}^{\psi}}\left[G_{0}^{\psi}(X_{0})\prod_{t=1}^{T}G_{t}^{\psi}(X_{t-1},X_{t})\right]$
is finite, and is admissible if the potentials $(G_{t}^{\psi})_{t\in[0:T]}$
are bounded. From (\ref{eq:P_twistedmodel}), the first $t^{th}$-marginal
distribution and time $t^{th}$-marginal distribution of $\mathbb{P}$
are given by
\begin{align}
\mathbb{P}(\mathrm{d}x_{0:t}) & =Z^{-1}\mu^{\psi}(\mathrm{d}x_{0})G_{0}^{\psi}(x_{0})\left\{\prod_{s=1}^{t-1}M_{s}^{\psi}(x_{s-1},\mathrm{d}x_{s})G_{s}^{\psi}(x_{s-1},x_{s})\right\}M_{t}^{\psi}(x_{t-1},\mathrm{d}x_{t})\phi_{t}^{*}(x_{t-1},x_{t})\label{eq:P_firstmarginals}
\end{align}
and
\begin{equation}
\mathbb{P}(\mathrm{d}x_{t})=Z^{-1}Z_{t}^{\psi}\eta_{t}^{\psi}(\mathrm{d}x_{t})M_{t+1}^{\psi}(\phi_{t+1}^{*})(x_{t})\label{eq:P_marginal}
\end{equation}
respectively, for $t\in[0:T]$. The representation (Property 1)
\[
\mathbb{P}(\mathrm{d}x_{0:T})=\left(\mu^{\psi}\right)^{\phi^{*}}(\mathrm{d}x_{0})\prod_{t=1}^{T}(M_{t}^{\psi})^{\phi^{*}}(x_{t-1},\mathrm{d}x_{t})=\mathbb{Q}^{\psi^{*}}(\mathrm{d}x_{0:T})
\]
follows from (\ref{eq:P_firstmarginals})-(\ref{eq:P_marginal}) by
noting that $\mu^{\psi}(\phi_{0}^{*})=Z$ and
\[
\mathbb{P}(\mathrm{d}x_{t}|x_{0:t-1})=\frac{M_{t}^{\psi}(x_{t-1},\mathrm{d}x_{t})\phi_{t}^{*}(x_{t-1},x_{t})}{M_{t}^{\psi}(\phi_{t}^{*})(x_{t-1})}
\]
for $t\in[1:T]$. Under the refined policy $\psi^{*}:=\psi\cdot\phi^{*}$,
it follows from (\ref{eq:further_twisted_potentials}) that
\begin{align*}
 & G_{0}^{\psi^{*}}(x_{0})=Z,\quad G_{t}^{\psi^{*}}(x_{t-1},x_{t})=1,\quad t\in[1:T],
\end{align*}
hence Property 3 follows from the form of the estimator (\ref{eq:twisted_Z_estimator})
and $Z_{t}^{\psi^{*}}=Z$ for all $t\in[0:T]$. Using the latter,
(\ref{eq:relate_FK_Models}), and (\ref{eq:P_marginal}) establishes
Property 2.
\end{proof}

To build some intuition, we provide a characterization of the optimal
policy in a specific setting which guides the choice of the function classes considered in Section \ref{sec:coxprocess}. 
\begin{prop}
\label{prop:logconcavity}For any policy $\psi\in\Psi$ such that
the corresponding twisted potentials $(G_{t}^{\psi})_{t\in[0:T]}$
and transition densities of $(M_{t}^{\psi})_{t\in[1:T]}$ are log-concave
on their domain of definition, the optimal policy $\phi^{*}=(\phi_{t}^{*})_{t\in[0:T]}$
w.r.t. $\mathbb{Q}^{\psi}$ is a sequence of log-concave functions.
\end{prop}

\begin{proof}[Proof of Proposition \ref{prop:logconcavity}]
For $t=T$, log-concavity of $\phi_{T}^{*}=G_{T}^{\psi}$ follows
by assumption. For $t\in[0:T-1]$, we proceed with an inductive argument
on the backward recursion (\ref{eq:optimal_phi}). Assuming that $\phi_{t+1}^{*}$
is log-concave, note that $x_{t}\mapsto M_{t+1}^{\psi}(\phi_{t+1}^{*})(x_{t})$
is log-concave since the product $(x_{t},x_{t+1})\mapsto\phi_{t+1}^{*}(x_{t},x_{t+1})M_{t+1}^{\psi}(x_{t},x_{t+1})$
is and log-concavity is preserved by marginalization. Hence $\phi_{t}^{*}$
is log-concave as the product of log-concave functions is also log-concave.
\end{proof}

\section{Proofs of Section \ref{sec:PolicyLearning}}
\begin{proof}[Proof of Proposition \ref{prop:moment_bound}]
We begin by noting the semigroup property
\[
Q_{s,u}^{\psi}(\varphi)=Q_{s,t}^{\psi}\circ Q_{t,u}^{\psi}(\varphi),\quad0\leq s<t<u\leq T,
\]
where we recall that we have defined $Q_{T}^{\psi}(\varphi)=G_{T}^{\psi}$ for any $\varphi$ for notational convenience.

Define the approximate Bellman operators as $\hat{Q}_{t}^{\psi}\varphi=P_{t}^{\psi,N}Q_{t}^{\psi}\varphi$
for $\varphi\in L^{2}(\nu_{t+1}^{\psi}),t\in[0:T]$.
The measures
$\nu_{t}^{\psi}$ for $t\in[0:T]$ have been introduced in Definition \ref{def:bellman_operators}.
By defining $\hat{\phi}_{T+1}=1$ for notational convenience and using (\ref{eq:backward_phi}), we obtain the following telescoping decomposition
\begin{align*}
\hat{\phi}_{t}-\phi_{t}^{*} & = \sum_{u=t}^{T} Q_{t-1,u-1}^{\psi}\circ\hat{Q}_{u}^{\psi}(\hat{\phi}_{u+1})-Q_{t-1,u-1}^{\psi}\circ Q_{u}^{\psi}(\hat{\phi}_{u+1}).
\end{align*}
Hence by the triangle inequality, we have
\begin{align*}
\|\hat{\phi}_{t}-\phi_{t}^{*}\|_{L^{2}(\nu_{t}^{\psi})} & \leq\sum_{u=t}^{T}\|Q_{t-1,u-1}^{\psi}\circ \hat{Q}_{u}^{\psi}(\hat{\phi}_{u+1})-Q_{t-1,u-1}^{\psi}\circ Q_{u}^{\psi} (\hat{\phi}_{u+1})\|_{L^{2}(\nu_{t}^{\psi})}
\end{align*}
for any $t\in[0:T]$. Under Assumption A1, (\ref{eq:feynmanKac_semigroup})
are linear bounded operators, hence
\begin{align*}
\|\hat{\phi}_{t}-\phi_{t}^{*}\|_{L^{2}(\nu_{t}^{\psi})} & \leq\sum_{u=t}^{T}C_{t-1,u-1}^{\psi}\|P_{u}^{\psi,N}Q_{u}^{\psi}(\hat{\phi}_{u+1})-Q_{u}^{\psi}(\hat{\phi}_{u+1})\|_{L^{2}(\nu_{t}^{\psi})}.
\end{align*}
Taking expectations and applying Assumption A2 yields (\ref{eq:ADP_bound}).
\end{proof}

\begin{proof}[Proof of Proposition \ref{prop:stability}]
It follows from (\ref{eq:twisted_FK_models})
that for any $r\in[1:T]$ and $\varphi\in L^{1}(\eta_{r}^{\psi})$ we have
\begin{equation}
\eta_{0}^{\psi}(\varphi)=\frac{\mu^{\psi}(G_{0}^{\psi}\varphi)}{\mu^{\psi}(G_{0}^{\psi})},
\quad\eta_{r}^{\psi}(\varphi)=\frac{\eta_{r-1}^{\psi}(M_{r}^{\psi}(G_{r}^{\psi}\varphi))}{\eta_{r-1}^{\psi}(M_{r}^{\psi}(G_{r}^{\psi}))}
, \quad \eta_{r-1}^{\psi}(M_{r}^{\psi}(G_{r}^{\psi})) = \frac{Z_r^\psi}{Z_{r-1}^\psi}.\label{eq:time_evolution_FKmodels}
\end{equation}
Now for $r\in[1:T-1]$ and
$\varphi\in L^{2}(\nu_{r+1}^{\psi})$, using Jensen's inequality and
the above identity
\begin{align*}
\|Q_{r}^{\psi}(\varphi)\|_{L^{2}(\nu_{r}^{\psi})}^{2} & =\int_{\mathsf{X}^{2}}G_{r}^{\psi}(x,y)^{2}M_{r+1}^{\psi}(\varphi)^{2}(y)\eta_{r-1}^{\psi}(\mathrm{d}x)M_{r}^{\psi}(x,\mathrm{d}y)\\
 & \leq\|G_{r}^{\psi}\|_{\infty}\int_{\mathsf{X}^{2}}G_{r}^{\psi}(x,y)M_{r+1}^{\psi}(\varphi^{2})(y)\eta_{r-1}^{\psi}(\mathrm{d}x)M_{r}^{\psi}(x,\mathrm{d}y)\\
 & =\|G_{r}^{\psi}\|_{\infty}\eta_{r-1}^{\psi}(M_{r}^{\psi}(G_{r}^{\psi}))\int_{\mathsf{X}}M_{r+1}^{\psi}(\varphi^{2})(y)\eta_{r}^{\psi}(\mathrm{d}y)\\
 & =\frac{Z_{r}^{\psi}}{Z_{r-1}^{\psi}}\|G_{r}^{\psi}\|_{\infty}\|\varphi\|_{L^{2}(\nu_{r+1}^{\psi})}^{2}.
\end{align*}
The result for $r = 0$ follows the same arguments.
Letting $\varphi\in L^2(\nu_{t+1}^\psi)$, whence $Q_{s+1,t}^{\psi}(\varphi)\in L^2(\nu_{s+2}^\psi)$, the above bound with $r=s+1$  implies that
\begin{align*}
\|Q_{s,t}^{\psi}(\varphi)\|_{L^{2}(\nu_{s+1}^{\psi})}^{2}
=\|Q_{s+1}^{\psi}Q_{s+1,t}^{\psi}(\varphi)\|_{L^{2}(\nu_{s+1}^{\psi})}^{2}
 & \leq \frac{Z_{s+1}^{\psi}}{Z_{s}^{\psi}}\|G_{s+1}^{\psi}\|_{\infty} \|Q_{s+1,t}^{\psi}(\varphi)\|_{L^{2}(\nu_{s+2}^{\psi})}^{2}.
\end{align*}
Iterating we establish (\ref{eq:stability_naive}).

When $G_r^\psi (x,y) = G_r^\psi(y)$ for all $x,y \in \mathsf{X}$ and $r\in[1:T]$,
\begin{align*}
\|Q_{s,t}^\psi(\varphi)\|_{L^{2}(\eta_{s}^{\psi}M_{s+1}^{\psi})}^{2}
&= \int \left[ Q_{s,t}^\psi(\varphi)(x) \right]^2 \eta_{s}^{\psi} M_{s+1}^{\psi} ( \mathrm{d} x)\\
&= \int \left[ \frac{ Q_{s,t}^\psi(\varphi)(x)}{ Q_{s,t}^\psi (1)(x)} \right]^2  (Q_{s,t}^\psi (1))^2(x) \eta_{s}^{\psi}M_{s+1}^{\psi} ( \mathrm{d} x)\\
&\leq \int \frac{ Q_{s,t}^\psi(\varphi^2)(x)}{ Q_{s,t}^\psi (1)(x)}   (Q_{s,t}^\psi (1))^2(x) \eta_{s}^{\psi}M_{s+1}^{\psi} ( \mathrm{d} x),
\end{align*}
by Jensen's inequality applied to the Markov operator $\varphi \mapsto Q_{s,t}^\psi(\varphi)/Q_{s,t}^\psi(1)$.
From Assumption A3 in \eqref{eq:mixing_assumption}, and the boundedness of $(G_t^{\psi})_{t\in[0:T]}$ it follows that
 \begin{align*}
 Q_{s,t}^\psi (1)(x)
  &= G^\psi_{s+1}(x) \int M_{s+2}^\psi (x, \mathrm{d}y ) Q_{s+1,t}^\psi (1)(y) \leq \kappa_{s+2}^\psi \|G_{s+1}^\psi\|_{\infty}\ \sigma_{s+2}^\psi ( Q_{s+1,t}^\psi (1) )<\infty.
 \end{align*}
Therefore we can write
 \begin{align*}
 &\|Q_{s,t}^\psi(\varphi)\|_{L^{2}(\eta_{s}^{\psi}M_{s+1}^{\psi})}^{2}
 \leq \int  Q_{s,t}^\psi(\varphi^2)(x)  Q_{s,t}^\psi (1) (x) \eta_{s}^{\psi}M_{s+1}^{\psi} ( \mathrm{d} x) \\
 &\leq   \kappa_{s+2}^\psi \,\|G_{s+1}^{\psi}\|_{\infty}\,\sigma_{s+2}^{\psi} \big( Q_{s+1,t}^\psi (1) \big) \,\int Q_{s,t}^\psi(\varphi^2)(x) \eta_{s}^{\psi}M_{s+1}^{\psi} ( \mathrm{d} x)\\
  &\leq   \kappa_{s+2}^\psi \,\|G_{s+1}^{\psi}\|_{\infty} \,\sigma_{s+2}^{\psi} \big( Q_{s+1,t}^\psi (1) \big) \,\eta_{s}^{\psi}M_{s+1}^{\psi}(Q_{s,t}^\psi (1))
    \int\frac{ Q_{s,t}^\psi(\varphi^2)(x) }
		{\eta_{s}^{\psi}M_{s+1}^{\psi}(Q_{s,t}^\psi (1))} \eta_{s}^{\psi}M_{s+1}^{\psi} ( \mathrm{d} x)\\
 &=\left[	\kappa_{s+2}^\psi \,\|G_{s+1}^{\psi}\|_{\infty}\,\sigma_{s+2}^{\psi} \big( Q_{s+1,t}^\psi (1) \big) \frac{Z_t^\psi}{Z_s^\psi}\right]
	\|\varphi\|_{L^{2}(\eta_{t}^{\psi}M_{t+1}^{\psi})}^{2},
  \end{align*}
 since one can check that for any function $f$
\[\frac{\eta_s^\psi M_{s+1}^\psi Q_{s,t}^\psi (f)}{\eta_s^\psi M_{s+1}^\psi Q_{s,t}^\psi (1)} = \eta_t^\psi M_{t+1}^\psi (f),  \qquad \eta_s^\psi M_{s+1}^\psi Q_{s,t}^\psi (1) =\frac{Z_t^\psi}{Z_s^\psi}.\qedhere
\]
 \end{proof}
\section{Proofs of Section \ref{sec:LimitTheorems}}\label{sec:supp_limittheorems}
Given $\gamma\in\mathcal{S}(\mathsf{E})$ and matrix-valued $\varphi:\mathsf{E}\rightarrow\mathbb{R}^{p\times d}$
with $\varphi_{i,j}\in\mathcal{B}(\mathsf{E})$ for all $i\in[1:p],j\in[1:d]$,
we extend the definition of $\gamma(\varphi)$ element-wise, i.e. $\gamma(\varphi)_{i,j}=\gamma(\varphi_{i,j})$.
Assuming that the Gram matrices
\begin{equation}
A_{t}^{\psi,N}:=\nu_{t}^{\psi,N}(\Phi_{t}\Phi_{t}^{T}),\quad t\in[0:T],\label{eq:LLS_AMatrix}
\end{equation}
are invertible, under (\ref{eq:linear_functionclass}) the estimated policy has the form $\hat{\phi}_{t}=\exp(-\Phi_{t}^{T}\beta_{t}^{\psi,N}),t\in[0:T],$
where the least squares estimators $\beta_{t}^{\psi,N}=(A_{t}^{\psi,N})^{-1}b_{t}^{\psi,N},t\in[0:T]$
are defined by the backward recursion
\begin{align}
b_{T}^{\psi,N} & =-\nu_{T}^{\psi,N}(\log G_{T}^{\psi}\cdot\Phi_{T}),\label{eq:LLS_bVector}\\
b_{t}^{\psi,N} & =-\nu_{t}^{\psi,N}(\{\log G_{t}^{\psi}+\log M_{t+1}^{\psi}(\exp(-\Phi_{t+1}^{T}(A_{t+1}^{\psi,N})^{-1}b_{t+1}^{\psi,N}))\}\:\Phi_{t}),\nonumber
\end{align}
for $t\in[0:T-1]$. To prove the claims in Theorem \ref{thm:LimitShorten}, we first establish
convergence of $\beta_{t}^{\psi,N}$ to $\beta_{t}^{\psi}:=(A_{t}^{\psi})^{-1}b_{t}^{\psi}$,
given by the Gram matrix $A_{t}^{\psi}:=\nu_{t}^{\psi}(\Phi_{t}\Phi_{t}^{T})$
and vector $b_{t}^{\psi}$ defined by the backward recursion
\begin{align}
b_{T}^{\psi} & =-\nu_{T}^{\psi}(\log G_{T}^{\psi}\cdot\Phi_{T}),\label{eq:limit_bvector}\\
b_{t}^{\psi} & =-\nu_{t}^{\psi}(\{\log G_{t}^{\psi}+\log M_{t+1}^{\psi}(\exp(-\Phi_{t+1}^{T}(A_{t+1}^{\psi})^{-1}b_{t+1}^{\psi}))\}\:\Phi_{t}),\nonumber
\end{align}
for $t\in[0:T-1]$.
\begin{prop}
\label{prop:LimitTheorems}Consider ADP algorithm (\ref{eq:ADP}),
with current policy $\psi\in\Psi$, under linear least squares approximations
(\ref{eq:linear_functionclass}) with basis functions $(\Phi_{t})_{t\in[0:T]}$
chosen so that:

\textbf{{[}A7{]}} the Gram matrices $(A_{t}^{\psi})_{t\in[0:T]}$
are invertible;

\textbf{{[}A8{]}} the function $x\mapsto M_{t}^{\psi}(\exp(-\Phi_{t}^{T}\beta))(x)$
is $\mathcal{X}$-measurable for all $\beta\in\mathbb{R}^{M},t\in[1:T]$
and the integrals in (\ref{eq:limit_bvector}) are finite;

\textbf{{[}A9{]}} for each $t\in[0:T-1]$, there exist a $\mathcal{X}$-measurable
function $C_{t}:\mathsf{X}\rightarrow\mathbb{R}_{+}$ and a continuous
function $\delta_{t}:\mathbb{R}_{+}\rightarrow\mathbb{R}_{+}$ satisfying
$\nu_{t}^{\psi}(C_{t}|\Phi_{t}|)<\infty$ and $\lim_{x\rightarrow0}\delta_{t}(x)=0$
respectively such that
\[
\left|\log M_{t+1}^{\psi}(\exp(-\Phi_{t+1}^{T}\beta))(x)-\log M_{t+1}^{\psi}(\exp(-\Phi_{t+1}^{T}\beta'))(x)\right|\leq C_{t}(x)\delta_{t}(|\beta-\beta'|)
\]
holds for all $x\in\mathsf{X}$ and $\beta,\beta'\in\mathbb{R}^{M}$.
As $N\rightarrow\infty,$ the least squares estimators $\beta^{\psi,N}:=(\beta_{t}^{\psi,N})_{t\in[0:T]}$
converge in probability to $\beta^{\psi}:=(\beta_{t}^{\psi})_{t\in[0:T]}$;

\textbf{{[}A10{]}} (i) for each $t\in[0:T-1]$, the function $\beta\mapsto\log M_{t+1}^{\psi}(\exp(-\Phi_{t+1}^{T}\beta))(x)$
is continuously differentiable for all $x\in\mathsf{X}$;

(ii) its gradient $x\mapsto g_{t+1}^{\psi}(\beta,x):=\nabla_{\beta}\log M_{t+1}^{\psi}(\exp(-\Phi_{t+1}^{T}\beta))(x)$
is $\mathcal{X}$-measurable for all $\beta\in\mathbb{R}^{M}$,
satisfies $\nu_{t}^{\psi}(|\Phi_{t}g_{t+1}^{\psi}(\beta_{t+1}^{\psi},\cdot)^{T}|)<\infty$
and for each $t\in[0:T-1]$, there exists a positive, $\mathcal{X}$-measurable
function $C'_{t}:\mathsf{X}\rightarrow\mathbb{R}_{+}$ satisfying
$\nu_{t}^{\psi}(C'_{t}|\Phi_{t}|)<\infty$ such that
\[
\big|g_{t+1}^{\psi}(\beta,x)-g_{t+1}^{\psi}(\beta',x)\big|\leq C'_{t}(x)|\beta-\beta'|
\]
holds for all $x\in\mathsf{X}$ and $\beta,\beta'\in\mathbb{R}^{M}$;


\textbf{{[}A11{]}} the vector-valued function $\xi^{\psi}=(\xi_{t}^{\psi})_{t\in[0:T]}:\mathsf{X}^{2T+1}\rightarrow\mathbb{R}^{(T+1)M}$
defined componentwise as
\begin{align}
\xi_{t}^{\psi} & =-(A_{t}^{\psi})^{-1}\{\log G_{t}^{\psi}+\log M_{t+1}^{\psi}(\exp(-\Phi_{t+1}^{T}\beta_{t+1}^{\psi}))\}\Phi_{t}-(A_{t}^{\psi})^{-1}\Phi_{t}\Phi_{t}^{T}\beta_{t}^{\psi},\quad t\in[0:T-1],\label{eq:CLT_vectorfunc}\\
\xi_{T}^{\psi} & =-(A_{T}^{\psi})^{-1}(\log G_{T}^{\psi}\cdot\Phi_{T}+\Phi_{T}\Phi_{T}^{T}\beta_{T}^{\psi}),\nonumber
\end{align}
satisfies $\xi^{\psi}\in L^{2}(\nu^{\psi})$ with $\nu^{\psi}:=\otimes_{t=0}^{T}\,\nu_{t}^{\psi}\in\mathcal{P}(\mathsf{X}^{2T+1})$
and the following central limit theorem
\begin{equation}
\sqrt{N}\left(\nu^{\psi,N}(\xi^{\psi})-\nu^{\psi}(\xi^{\psi})\right)\stackrel{\mathtt{d}}{\longrightarrow}\mathcal{N}\left(0_{(T+1)M},\Gamma^{\psi}\right)\label{eq:assume_CLT}
\end{equation}
with $\nu^{\psi,N}:=\otimes_{t=0}^{T}\,\nu_{t}^{\psi,N}$.
\newline
Then we have
\begin{equation}
\sqrt{N}\left(\beta^{\psi,N}-\beta^{\psi}\right)\stackrel{\mathtt{d}}{\longrightarrow}\mathcal{N}\left(0_{(T+1)M},\Sigma^{\psi}\right)\label{eq:LLS_CLT}
\end{equation}
where $\Sigma^{\psi}=U^{\psi}\Gamma^{\psi}(U^{\psi})^{T}$ is given
by a block upper triangular matrix $U^{\psi}\in\mathbb{R}^{(T+1)M\times(T+1)M}$
defined by blocks of size $M\times M$
\begin{equation}
U_{s,t}^{\psi}=\begin{cases}
\prod_{u=s-1}^{t-2}E_{u}^{\psi}, & s<t,\\
I_{M}, & s=t,\\
0_{M\times M}, & s>t,
\end{cases}\label{eq:blockmatrix_U}
\end{equation}
for $s,t\in[1:T+1]$, with $E_{t}^{\psi}:=-(A_{t}^{\psi})^{-1}\nu_{t}^{\psi}(\Phi_{t}g_{t+1}^{\psi}(\beta_{t+1}^{\psi},\cdot)^{T}),t\in[0:T-1]$ and
$0_{M\times M}$ as the $M\times M$ matrix of zeros.
\end{prop}

\begin{proof}[Proof of Proposition \ref{prop:LimitTheorems}]
Note that for each $t\in[0:T]$, by the strong law of large numbers (LLN)
for the particle approximation $\nu_{t}^{\psi,N}$ (see \cite{DelMoral_2004})
$A_{t}^{\psi,N}\rightarrow A_{t}^{\psi}$ almost surely as $N\rightarrow\infty$,
therefore using continuity of matrix inversion and the continuous
mapping theorem, we have $(A_{t}^{\psi,N})^{-1}\rightarrow(A_{t}^{\psi})^{-1}$
almost surely. Using continuity of the spectral matrix norm and another
application of the continuous mapping theorem, we see that the minimum
eigenvalue of $A_{t}^{\psi,N}$ converges to that of $A_{t}^{\psi}$,
which is strictly positive under Assumption A7. Hence for sufficiently
large values of $N$, we have invertibility of $A_{t}^{\psi,N}$ with
probability one.

Starting with time $t=T$, by LLN $b_{T}^{\psi,N}\rightarrow b_{T}^{\psi}$
in probability, so by Slutsky's lemma it follows that $\beta_{T}^{\psi,N}\rightarrow\beta_{T}^{\psi}$
in probability. Consider the difference
\[
\beta_{T}^{\psi,N}-\beta_{T}^{\psi}=(A_{T}^{\psi,N})^{-1}(b_{T}^{\psi,N}-A_{T}^{\psi,N}\beta_{T}^{\psi})=((A_{T}^{\psi})^{-1}+o_{p}(1))\,(b_{T}^{\psi,N}-A_{T}^{\psi,N}\beta_{T}^{\psi}).
\]
Since $(A_{T}^{\psi})^{-1}(b_{T}^{\psi,N}-A_{T}^{\psi,N}\beta_{T}^{\psi})=\nu_{T}^{\psi,N}(\xi_{T}^{\psi})$
and $\nu_{T}^{\psi}(\xi_{T}^{\psi})=0_{M}$ with $\xi_{T}^{\psi}$
defined in (\ref{eq:CLT_vectorfunc}), it follows from (\ref{eq:assume_CLT})
that $b_{T}^{\psi,N}-A_{T}^{\psi,N}\beta_{T}^{\psi}=O_{p}(N^{-1/2})$.
Therefore
\begin{equation}
\beta_{T}^{\psi,N}-\beta_{T}^{\psi}=\nu_{T}^{\psi,N}(\xi_{T}^{\psi})+o_{p}(N^{-1/2})\label{eq:beta_terminaltime}
\end{equation}
and applying the central limit theorem (CLT) in Assumption A11 gives
\[
\sqrt{N}\left(\beta_{T}^{\psi,N}-\beta_{T}^{\psi}\right)\stackrel{\mathtt{d}}{\longrightarrow}\mathcal{N}\left(0_{M},\Gamma_{T+1,T+1}^{\psi}\right)
\]
where $\Gamma_{T+1,T+1}^{\psi}\in\mathbb{R}^{M\times M}$ refers to
the lowest right block of $\Gamma^{\psi}$.

We now argue inductively: for time $t\in[0:T-1]$, we decompose $b_{t}^{\psi,N}=c_{t}^{\psi,N}+d_{t}^{\psi,N}$
where
\begin{align*}
c_{t}^{\psi,N} & :=-\nu_{t}^{\psi,N}(\{\log G_{t}^{\psi}+\log M_{t+1}^{\psi}(\exp(-\Phi_{t+1}^{T}\beta_{t+1}^{\psi}))\}\Phi_{t}),\\
d_{t}^{\psi,N} & :=\nu_{t}^{\psi,N}(\{\log M_{t+1}^{\psi}(\exp(-\Phi_{t+1}^{T}\beta_{t+1}^{\psi}))-\log M_{t+1}^{\psi}(\exp(-\Phi_{t+1}^{T}\beta_{t+1}^{\psi,N}))\}\Phi_{t}).
\end{align*}
Assumption A8 implies $c_{t}^{\psi,N}\rightarrow b_{t}^{\psi}$ in
probability. If $\beta_{t+1}^{\psi,N}\rightarrow\beta_{t+1}^{\psi}$
in probability, by Assumption A9 we have
\[
|d_{t}^{\psi,N}|\leq\nu_{t}^{\psi,N}(C_{t}|\Phi_{t}|)\delta_{t}(|\beta_{t+1}^{\psi,N}-\beta_{t+1}^{\psi}|)=o_{p}(1),
\]
hence $\beta_{t}^{\psi,N}\rightarrow\beta_{t}^{\psi}$ in probability.
We now examine the difference
\begin{equation}
\beta_{t}^{\psi,N}-\beta_{t}^{\psi}=((A_{t}^{\psi})^{-1}+o_{p}(1))\,(c_{t}^{\psi,N}+d_{t}^{\psi,N}-A_{t}^{\psi,N}\beta_{t}^{\psi}).\label{eq:decompose_beta}
\end{equation}
Since $(A_{t}^{\psi})^{-1}(c_{t}^{\psi,N}-A_{t}^{\psi,N}\beta_{t}^{\psi})=\nu_{t}^{\psi,N}(\xi_{t}^{\psi})$
and $\nu_{t}^{\psi}(\xi_{t}^{\psi})=0_{M}$ with $\xi_{t}^{\psi}$
defined in (\ref{eq:CLT_vectorfunc}), it follows from (\ref{eq:assume_CLT})
that $c_{t}^{\psi,N}-A_{t}^{\psi,N}\beta_{t}^{\psi}=O_{p}(N^{-1/2})$.
To study the term $d_{t}^{\psi,N}$, we use Assumption A10(i) and apply
Taylor's theorem to obtain
\[
d_{t}^{\psi,N}=-\nu_{t}^{\psi,N}((\beta_{t+1}^{\psi,N}-\beta_{t+1}^{\psi})^{T}g_{t+1}^{\psi}(\beta_{t+1}^{\psi},\cdot)\Phi_{t})+r_{t}^{\psi,N}
\]
with remainder
\[
r_{t}^{\psi,N}=-\nu_{t}^{\psi,N}\left((\beta_{t+1}^{\psi,N}-\beta_{t+1}^{\psi})^{T} \left[ g_{t+1}^{\psi}(\tilde{\beta}_{t+1}^{N},\cdot)
- g_{t+1}^{\psi}({\beta}_{t+1}^{\psi},\cdot) \right] \Phi_{t}\right)
\]
for some $\tilde{\beta}_{t+1}^{N}$ lying on the line segment between $\beta_{t+1}^{\psi,N}$
and $\beta_{t+1}^{\psi}$.
Applying Assumption A10(ii) we have that
 \begin{align*}
  \big| r_{t}^{\psi,N}\big|
  &\leq |\tilde{\beta}_{t+1}^{\psi,N}-\beta_{t+1}^{\psi}| |\beta_{t+1}^{\psi,N}-\beta_{t+1}^{\psi}| \nu_t^{\psi,N} \left( C_t'(\cdot) |\Phi_t |\right)\\
   &\leq |\beta_{t+1}^{\psi,N}-\beta_{t+1}^{\psi}|^2 \nu_t^{\psi,N} \left( C_t'(\cdot) |\Phi_t |\right)\\
   &= |\beta_{t+1}^{\psi,N}-\beta_{t+1}^{\psi}|^2 \left[ \nu_t^{\psi} \left(C_t'(\cdot) |\Phi_t |\right) + o_p(1)\right]
 \end{align*}
where the second inequality follows from the definition of $\tilde{\beta}_{t+1}^{N}$ and the final equality by the LLN.
By the inductive hypothesis we have that
 \[
 \sqrt{N}\left( \beta_{t+1}^{\psi,N}-\beta_{t+1}^{\psi}\right) \stackrel{\mathtt{d}}{\longrightarrow}\mathcal{N}\left(0_{M},\Sigma_{t+1,t+1}^{\psi}\right)
 \]
for some $\Sigma_{t+1,t+1}^{\psi}\in\mathbb{R}^{M\times M}$,
and since by assumption $\nu_t^{\psi} \left( C_t'(\cdot) |\Phi_t |\right)<\infty$ we conclude that
$r_{t}^{\psi,N}=O_{p}(N^{-1})$. From Assumption A10(ii) and the LLN we conclude that $d_{t}^{\psi, N}=O_{p}(N^{-1/2})$ and we can thus write
\[
(A_{t}^{\psi})^{-1}d_{t}^{\psi,N}=E_{t}^{\psi}(\beta_{t+1}^{\psi,N}-\beta_{t+1}^{\psi})+o_{p}(N^{-1/2})
\]
where $E_{t}^{\psi}:=-(A_{t}^{\psi})^{-1}\nu_{t}^{\psi}(\Phi_{t}g_{t+1}^{\psi}(\beta_{t+1}^{\psi},\cdot)^{T})$.
Combining these observations with (\ref{eq:decompose_beta}) gives
\begin{equation}
\beta_{t}^{\psi,N}-\beta_{t}^{\psi}-E_{t}^{\psi}(\beta_{t+1}^{\psi,N}-\beta_{t+1}^{\psi})=\nu_{t}^{\psi,N}(\xi_{t}^{\psi})+o_{p}(N^{-1/2}).\label{eq:beta_difference}
\end{equation}
Stacking (\ref{eq:beta_difference}) for $t\in[0:T-1]$ and (\ref{eq:beta_terminaltime})
as a $(T+1)M$-dimensional vector yields
\[
\zeta^{\psi,N}:=\left(\begin{array}{c}
(\beta_{0}^{\psi,N}-\beta_{0}^{\psi})-E_{0}^{\psi}(\beta_{1}^{\psi,N}-\beta_{1}^{\psi})\\
(\beta_{1}^{\psi,N}-\beta_{1}^{\psi})-E_{1}^{\psi}(\beta_{2}^{\psi,N}-\beta_{2}^{\psi})\\
\vdots\\
(\beta_{T-1}^{\psi,N}-\beta_{T-1}^{\psi})-E_{T-1}^{\psi}(\beta_{T}^{\psi,N}-\beta_{T}^{\psi})\\
\beta_{T}^{\psi,N}-\beta_{T}^{\psi}
\end{array}\right)=\nu^{\psi,N}(\xi^{\psi})+o_{p}(N^{-1/2}).
\]
Noting that the block matrix $U^{\psi}$ defined in (\ref{eq:blockmatrix_U})
is such that $U^{\psi}\zeta^{\psi,N}=\beta^{\psi,N}-\beta^{\psi}$
for any $N\in\mathbb{N}$, (\ref{eq:LLS_CLT}) follows from the CLT
in Assumption A11 and an application of the continuous mapping theorem.
\end{proof}

We first make some remarks about the assumptions required in
Proposition \ref{prop:LimitTheorems}. Assumptions A7 and A8
ensure that the least squares estimators converge to a well-defined
limit. Assumptions A9 and A10 are made to deal with the intractability
of the function $(\beta,x)\mapsto\log M_{t+1}^{\psi}(\exp(-\Phi_{t+1}^{T}\beta))(x)$,
which can be verified when its form is known. Lastly, Assumption A11,
which asserts existence of a path central limit theorem for the function
(\ref{eq:CLT_vectorfunc}), can be deduced in the case of multinomial
resampling  from \cite[Theorem 9.7.1]{DelMoral_2004}.
In the following, we will write $A_{s,t}\in\mathbb{R}^{M\times M}$ to denote the $s,t\in[1:T+1]$
submatrix of a block matrix $A\in\mathbb{R}^{(T+1)M\times(T+1)M}$.

\begin{thm}\label{thm:LimitTheorems}
Consider ADP algorithm (\ref{eq:ADP}), with current policy $\psi\in\Psi$,
under linear least squares approximations (\ref{eq:linear_functionclass}) with
basis functions $(\Phi_{t})_{t\in[0:T]}$ chosen so that Assumptions A7-A11
in Proposition \ref{prop:LimitTheorems} are satisfied. Then as $N\rightarrow\infty$, for all $x\in\mathsf{X}^{2T+1}$,
the estimated policy $\hat{\phi}(x)$ converges in probability to the policy
$\tilde{\phi}(x)$ generated by the idealized algorithm (\ref{eq:ideal_ADP}). Moreover, for all $x\in\mathsf{X}^{2T+1}$,
we have
\begin{equation}
\sqrt{N}\left(\hat{\phi}(x)-\tilde{\phi}(x)\right)\stackrel{\mathtt{d}}{\longrightarrow}\mathcal{N}\left(0_{(T+1)},\Omega^{\psi}(x)\right),
\label{eq:CLT_policy}
\end{equation}
where $\Omega^{\psi}:\mathsf{X}^{2T+1}\rightarrow\mathbb{R}^{(T+1)\times(T+1)}$
is given by
\begin{equation}\label{eqn:CLT_var}
\Omega_{s,t}^{\psi}=\tilde{\phi}_{s}\tilde{\phi}_{t}\Phi_{s}^{T}\sum_{k=s}^{T+1}\sum_{\ell=t}^{T+1}U_{s,k}^{\psi}\Gamma_{k,\ell}^{\psi}(U_{\ell,t}^{\psi})^{T}\Phi_{t}
\end{equation}
for $s,t\in[1:T+1]$.
\end{thm}

\begin{proof}[Proof of Theorem \ref{thm:LimitTheorems}]
Appealing to the continuous mapping theorem allows us to conclude
from Proposition \ref{prop:LimitTheorems} that $\hat{\phi}_{t}$ converges
(pointwise) in probability to $\tilde{\phi}_{t}:=\exp(-\Phi_{t}^{T}\beta_{t}^{\psi}),t\in[0:T]$.
Applying the delta method on (\ref{eq:LLS_CLT}) establishes that
the (pointwise) fluctuations satisfy (\ref{eq:CLT_policy}), where
$\Omega_{s,t}^{\psi}=\tilde{\phi}_{s}\tilde{\phi}_{t}\Phi_{s}^{T}\Sigma_{s,t}^{\psi}\Phi_{t}$
for $s,t\in[1:T+1]$.
The form of the asymptotic variance (\ref{eqn:CLT_var}) follows from
the block upper triangular structure of (\ref{eq:blockmatrix_U}).
\end{proof}

\section{Proofs of Section \ref{sec:iADP}}\label{sec:iterativethm}
\begin{proof}[Proof of Theorem \ref{thm:iADP}]
Under Assumptions A4 and A5, existence of a unique invariant distribution
$\pi\in\mathcal{P}(\Psi)$ and geometric convergence (\ref{eq:geometric_convergence})
follow from \cite[Theorem 1.1]{Diaconis_Freedman_1999}. Let $\varphi^{*}$
denote a fixed point of $F$ and define the backward process $\varphi^{(I)}=F_{U^{(1)}}^{N}\circ\cdots\circ F_{U^{(I)}}^{N}(\varphi^{*})$
for $I\in\mathbb{N}$. Noting from \cite[Proposition 1.1]{Diaconis_Freedman_1999}
that the limit $\varphi^{(\infty)}:=\lim_{I\rightarrow\infty}\varphi^{(I)}$
does not depend on $\varphi^{*}$ and is distributed according to
$\pi$, we shall construct the random policy $\psi\sim\pi$ by taking
$\psi=\varphi^{(\infty)}$.

By the triangle inequality,
\begin{equation}
\rho(\psi,\varphi^{*})\leq\rho(\varphi^{(\infty)},\varphi^{(I)})+\rho(\varphi^{(I)},\varphi^{*})\label{eq:iADP_twoterms}
\end{equation}
for any $I\in\mathbb{N}$. To examine the first term in (\ref{eq:iADP_twoterms}),
we consider the decomposition in the proof of \cite[Proposition 5.1]{Diaconis_Freedman_1999}:
\[
\rho(\varphi^{(I+J)},\varphi^{(I)})\leq\sum_{i=0}^{J-1}\prod_{j=1}^{I+i}L_{U^{(j)}}^{N}\rho(F_{U^{(I+i+1)}}^{N}(\varphi^{*}),\varphi^{*})
\]
for $I,J\in\mathbb{N}$. By the monotone convergence theorem, taking the
limit $J\rightarrow\infty$ gives
\[
\mathbb{E}\left[\rho(\varphi^{(\infty)},\varphi^{(I)})\right]\leq\sum_{i=0}^{\infty}\prod_{j=1}^{I+i}\mathbb{E}\left[L_{U^{(j)}}^{N}\right]\mathbb{E}\left[\rho(F_{U^{(I+i+1)}}^{N}(\varphi^{*}),\varphi^{*})\right].
\]
Under Assumptions A4 and A5, it follows that $\zeta:=\mathbb{E}\left[\rho(F_{U}^{N}(\varphi^{*}),\varphi^{*})\right]<\infty$
since by the triangle inequality
\begin{align*}
\rho(F_{U}^{N}(\varphi^{*}),\varphi^{*})
&\leq \rho(F_{U}^{N}(\varphi^{*}),F_{U}^{N}(\varphi_0)) + \rho(F_{U}^{N}(\varphi_0),\varphi_0)
+ \rho(\varphi_0,\varphi^{*}) \\
&\leq(1+L_{U}^{N})\rho(\varphi^{*},\varphi_{0})+\rho(F_{U}^{N}(\varphi_{0}),\varphi_{0}).
\end{align*}
Applying Assumption A5, the triangle inequality and the fact that
$\varphi^{(I)} \to \varphi^{(\infty)}$ as $I\to \infty$ establishes that
\begin{align*}
\mathbb{E}\left[\rho(\varphi^{(\infty)},\varphi^{(I)})\right]
&\leq \sum_{j=0}^\infty \mathbb{E}\left[\rho(\varphi^{(I+j)},\varphi^{(I+j+1)})\right]
\leq\zeta\alpha^{I}(1-\alpha)^{-1}
\end{align*}

and hence
\begin{equation}
\lim_{I\rightarrow\infty}\mathbb{E}\left[\rho(\varphi^{(\infty)},\varphi^{(I)})\right]=0.\label{eq:iADP_firstterm}
\end{equation}
For the second term in (\ref{eq:iADP_twoterms}), using the fact that
$\varphi^{*}$ is a fixed point of $F$, the triangle inequality and Assumptions
A5 and A6
\begin{align*}
\rho(\varphi^{(I)},\varphi^{*}) & =\rho(F_{U^{(1)}}^{N}\circ\cdots\circ F_{U^{(I)}}^{N}(\varphi^{*}),F(\varphi^{*}))\\
 & \leq\sum_{i=1}^{I}\rho(F_{U^{(1)}}^{N}\circ\cdots\circ F_{U^{(i)}}^{N}(\varphi^{*}),F_{U^{(1)}}^{N}\circ\cdots\circ F_{U^{(i-1)}}^{N}\circ F(\varphi^{*}))\\
 & \leq\sum_{i=1}^{I}\prod_{j=1}^{i-1}L_{U^{(j)}}^{N}\rho(F_{U^{(i)}}^{N}(\varphi^{*}),F(\varphi^{*}))\\
 & \leq N^{-1/2}\sum_{i=1}^{I}\prod_{j=1}^{i-1}L_{U^{(j)}}^{N}E_{U^{(i)}}^{\varphi^{*},N}
\end{align*}
with the convention that $(F_{U^{(1)}}^{N}\circ F_{U^{(0)}}^{N})(\varphi)=\varphi$.
Taking expectations and the limit $I\rightarrow\infty$ gives
\begin{equation}
\lim_{I\rightarrow\infty}\mathbb{E}\left[\rho(\varphi^{(I)},\varphi^{*})\right]\leq N^{-1/2}\mathbb{E}\left[E_{U}^{\varphi^{*},N}\right](1-\alpha)^{-1}.\label{eq:iADP_secondterm}
\end{equation}
Combining (\ref{eq:iADP_twoterms}), (\ref{eq:iADP_firstterm}) and
(\ref{eq:iADP_secondterm}) allows us to conclude (\ref{eq:characterize_invariant}).
\end{proof}

The following discussion offers some insights into when and why contraction (Assumption A5) happens.
Let $\rho$ denote a metric under which the set of all admissible policies $\Psi$ is
a complete separable metric space.
Let $\psi^{*}$ denote the optimal policy w.r.t. $\mathbb{Q}$ that we want to approximate.
Given two policies $\varphi,\xi\in\Psi$, by triangle inequality,
the ADP algorithm $F^{N}:\mathsf{U}\times\Psi\rightarrow\Psi$ satisfies
\begin{align}
\rho(F_{U}^{N}(\varphi),F_{U}^{N}(\xi)) & \leq\rho(F_{U}^{N}(\varphi),F(\varphi))+\rho(F_{U}^{N}(\xi),F(\xi))+\rho(F(\varphi),F(\xi))\nonumber \\
 & \leq\rho(F_{U}^{N}(\varphi),F(\varphi))+\rho(F_{U}^{N}(\xi),F(\xi))+\rho(F(\varphi),\psi^{*})+\rho(F(\xi),\psi^{*})\label{eq:decompose_errors}
\end{align}
where $F:\Psi\rightarrow\Psi$ denotes the idealized ADP algorithm
with exact projections. We consider the first and second terms of
(\ref{eq:decompose_errors}) that concern the Monte Carlo error of
the ADP algorithm. Under Assumption A6, we have
\begin{equation}
\rho(F_{U}^{N}(\varphi),F(\varphi))+\rho(F_{U}^{N}(\xi),F(\xi))\leq N^{-1/2}\left(E_{U}^{\varphi,N}+E_{U}^{\xi,N}\right)\label{eq:monte_carlo_error}
\end{equation}
where $(E_{U}^{\varphi,N})_{N\in\mathbb{N}}$ and $(E_{U}^{\xi,N})_{N\in\mathbb{N}}$
are uniformly integrable sequences of non-negative random variables
with finite mean that converge in distribution to a limit with support
on $\mathbb{R}_{+}$.
Assumption A6 is necessary to quantify the Monte Carlo error involved when employing approximate projections and can be deduced for example using the central limit theorem in Theorem \ref{thm:LimitTheorems}.
The third and fourth terms of (\ref{eq:decompose_errors})
concern the mis-specification error of the chosen function classes.
In particular, we have
\begin{equation}
\rho(F(\varphi),\psi^{*})+\rho(F(\xi),\psi^{*})=\rho(\tilde{\varphi},\varphi^{*})+\rho(\tilde{\xi},\xi^{*})\label{eq:misspecification_error}
\end{equation}
where $\tilde{\varphi}$ and $\tilde{\xi}$ denote idealized ADP approximations
of the optimal policies $\varphi^{*}$ and $\xi^{*}$ w.r.t. $\mathbb{Q}^{\varphi}$
and $\mathbb{Q}^{\xi}$ respectively. In the well-specified case,
by consistency of least squares, the errors $e(\varphi):=\rho(\tilde{\varphi},\varphi^{*})$
and $e(\xi):=\rho(\tilde{\xi},\xi^{*})$ would be equal to zero.

Combining (\ref{eq:decompose_errors}),
(\ref{eq:monte_carlo_error}) and (\ref{eq:misspecification_error})
gives
\[
\rho(F_{U}^{N}(\varphi),F_{U}^{N}(\xi))\leq N^{-1/2}\left(E_{U}^{\varphi,N}+E_{U}^{\xi,N}\right)+e(\varphi)+e(\xi).
\]
Therefore if $\rho(\varphi,\xi)\geq 1$, we have
\begin{equation}
\rho(F_{U}^{N}(\varphi),F_{U}^{N}(\xi))\leq L_{U}^{N}\rho(\varphi,\xi)\label{eq:result}
\end{equation}
with
\begin{equation}
L_{U}^{N}=N^{-1/2}\left(E_{U}^{\varphi,N}+E_{U}^{\xi,N}\right)+e(\varphi)+e(\xi).\label{eq:lipschitz_constant}
\end{equation}
If $\Psi$ is compact, which may be imposed by truncating our least squares estimators,
the expectation of (\ref{eq:lipschitz_constant}) is bounded by
\begin{align*}
\mathbb{E}\left[L_{U}^{N}\right] &= N^{-1/2}\left\{ \mathbb{E}\left[E_{U}^{\varphi,N}\right]+\mathbb{E}\left[E_{U}^{\xi,N}\right]\right\} +e(\varphi)+e(\xi)\\
&\leq 2N^{-1/2}\sup_{\varphi\in\Psi}\mathbb{E}\left[E_{U}^{\varphi,N}\right]+2\sup_{\varphi\in\Psi}e(\varphi).
\end{align*}
In the well-specified case, we have $\sup_{\varphi\in\Psi}e(\varphi)=0$ so
$\mathbb{E}[L_{U}^{N}]<1$ when the number of particles $N$ is sufficiently
large. In the mis-specified case, we also require that the mis-specification
error $\sup_{\varphi\in\Psi}e(\varphi)$ be sufficiently small.
If $\varphi=\xi$, (\ref{eq:result}) also holds since $\rho(F_{U}^{N}(\varphi),F_{U}^{N}(\xi))=0$.
Although the above arguments explain why one can expect contraction for policies $\varphi$ and $\xi$
that are distant, it does not capture the case $\rho(\varphi,\xi)\in(0,1)$, corresponding to when these policies are close.

The following example illustrates contraction in a simple setting with mis-specification.

\begin{eg}\label{eg:simple_example_contraction}
Let the state space be $\mathsf{X}=[0,1]$, equipped with its Borel $\sigma$-algebra $\mathcal{X}=\mathfrak{B}([0,1])$.
For simplicity, we consider a single time step and an initial distribution $\mu$ that is given by the uniform distribution on $\mathsf{X}$.
The potential function of interest is $G_0(x_0)=\exp(-x_0^2)$.

The function class we specify for the ADP algorithm is
$\mathsf{F}=\{\varphi(x)= a x: a\in\mathbb{R}\}$. Given a current policy $\psi_0(x_0)=\exp(-a_0x_0)$, the $\psi$-twisted SMC method
(Algorithm \ref{alg:twistedSMC}) will sample $N$ independent samples $X_0^n\sim\mu^{\psi}$ for $n\in[1:N]$. In terms of independent
uniform random variables $(U_0^n)_{n\in[1:N]}$, these samples can be generated using
\begin{align}\label{eqn:gen_samples_unif}
X_0^n=-{a_0}^{-1}\log\left(1-U_0^n\{1-\exp(-a_0)\}\right).
\end{align}
In this setting, ADP (Algorithm \ref{alg:ADP}) would consider the following least squares problem
\begin{align*}
\alpha_0&=\arg\min_{\varphi\in\mathsf{F}}\sum_{n=1}^N\left(\varphi(X_0^n)+\log G_0(X_0^n)-\log\psi_0(X_0^n)\right)^2\\
&=\arg\min_{\alpha\in\mathbb{R}}\sum_{n=1}^N\left(\alpha X_0^n-\{(X_0^n)^2-a_0X_0^n\}\right)^2\\
&=\frac{\sum_{n=1}^N(X_0^n)^3}{\sum_{n=1}^N(X_0^n)^2}-a_0.
\end{align*}
Therefore, the ADP algorithm can be represented as the iterated random function
$F_U^N(a_0) = a_0+\alpha_0$. By considering $a_0=0$, we see that Assumption A4 is satisfied since 
\begin{align*}
\mathbb{E}\left[F_U^N(0)\right] = \mathbb{E}\left[\frac{\sum_{n=1}^N(U_0^n)^3}{\sum_{n=1}^N(U_0^n)^2}\right] < \infty.
\end{align*}
As the number of particles $N\rightarrow\infty$,
\begin{align}\label{eqn:limiting_func_eg}
F_U^N(a_0)\rightarrow \frac{\int_0^1x_0^3\exp(-a_0x_0)\,\mathrm{d}x_0}{\int_0^1x_0^2\exp(-a_0x_0)\,\mathrm{d}x_0}=:F(a_0).
\end{align}
The limiting function $F$ corresponds to the idealized ADP algorithm with exact projections. 
We note that Assumption A6 holds since the above convergence rate is $O(N^{-1/2})$ by the central limit theorem.
In the left panel of Figure \ref{fig:simple_contraction}, we illustrate the distribution of the iterates $a_0^{(i)}=F_U^N(a_0^{(i-1)})$ with
initialization $a_0^{(0)}=0$ for different number of particles. This plot shows how the estimates produced by the ADP algorithm
concentrate around the fixed point iteration defined by $F$.

We now turn our attention to Assumption A5. The derivative of $F_U^N$ with respect to $a_0$ is
\begin{align*}
\frac{\mathrm{d}}{\mathrm{d}a_0}F_U^N(a_0) = \frac{\sum_{n=1}^N3(X_0^n)^2d_0^n}{\sum_{n=1}^N(X_0^n)^2}
- \frac{\sum_{n=1}^N(X_0^n)^3\cdot\sum_{n=1}^N2X_0^nd_0^n}{\left\lbrace\sum_{n=1}^N(X_0^n)^2\right\rbrace^2}
\end{align*}
where
\begin{align*}
d_0^n&=\frac{U_0^n\exp(-a_0)}{a_0(1-U_0^n\{1-\exp(-a_0)\})} + {a_0}^{-2}\log\left(1-U_0^n\{1-\exp(-a_0)\}\right)
\end{align*}
denotes the derivative of (\ref{eqn:gen_samples_unif}) with respect to $a_0$.
As $N\rightarrow\infty$, we have
\begin{align}\label{eqn:limiting_deriv_eg}
&\frac{\mathrm{d}}{\mathrm{d}a_0}F_U^N(a_0)\rightarrow \frac{\int_0^13x_0^2d(a_0,x_0)\exp(-a_0x_0)\,\mathrm{d}x_0}
{\int_0^1x_0^2\exp(-a_0x_0)\,\mathrm{d}x_0}\notag\\
&-\frac{\int_0^1x_0^3\exp(-a_0x_0)\,\mathrm{d}x_0
\cdot\int_0^12x_0d(a_0,x_0)\exp(-a_0x_0)\,\mathrm{d}x_0}
{\left\lbrace\int_0^1x_0^2\exp(-a_0x_0)\,\mathrm{d}x_0\right\rbrace^2}
\end{align}
where
\begin{align*}
d(a_0,X_0^n)=a_0^{-1}\left\lbrace\frac{\exp(-a_0)}{(1-\exp(-a_0))}(\exp(a_0X_0^n)-1)-X_0^n\right\rbrace=d_0^n.
\end{align*}

It is apparent from the right panel of Figure \ref{fig:simple_contraction} that the idealized ADP algorithm with exact projection is a contraction.
Moreover, for this particular example, the ADP algorithm is also a contraction (on average) even with a small number of particles.
\begin{figure}
\begin{centering}
\includegraphics[scale=0.4]{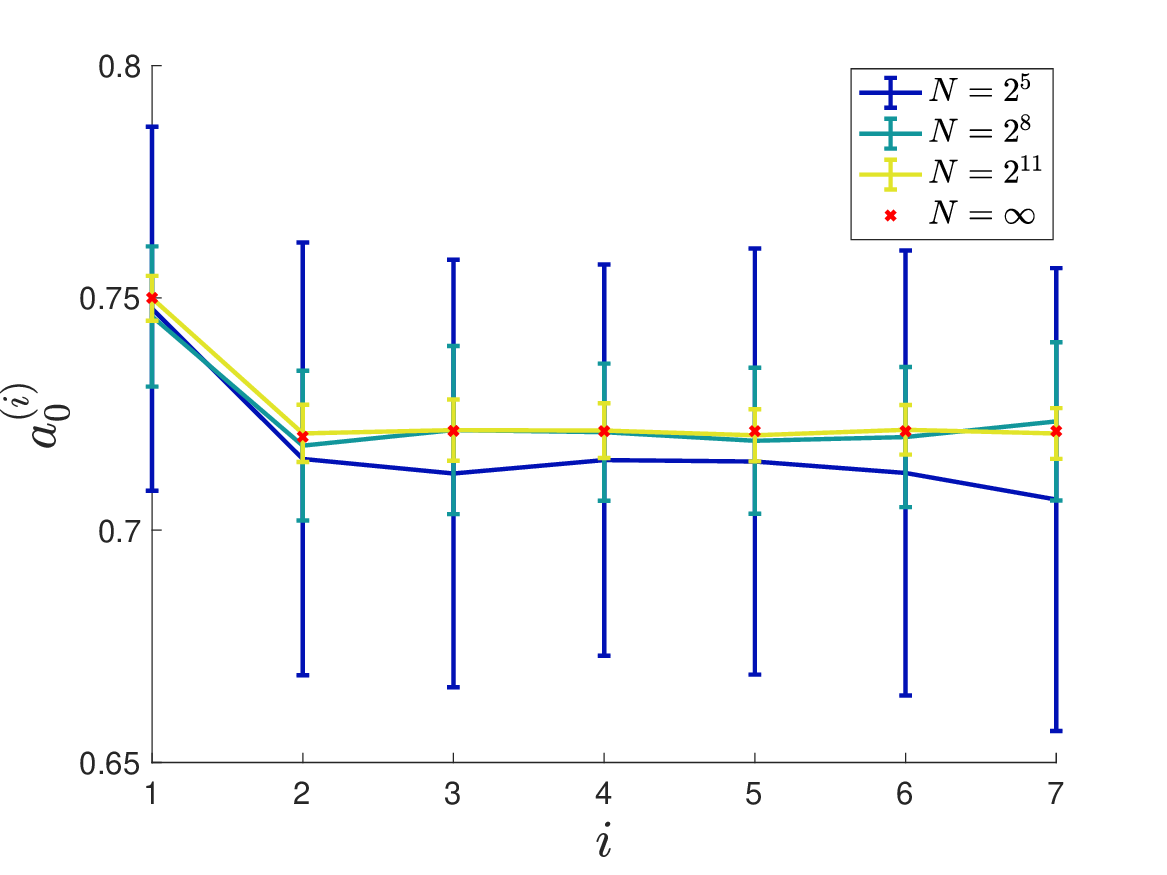}\includegraphics[scale=0.4]{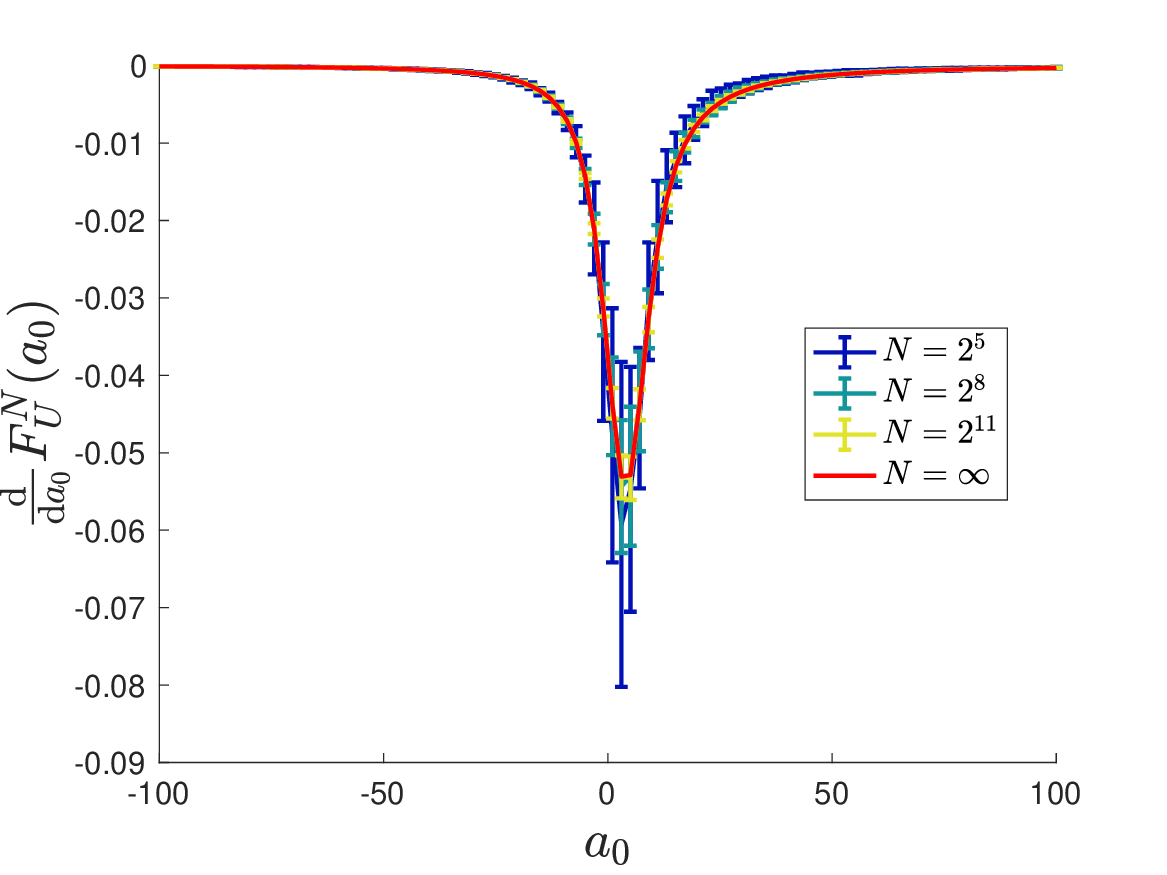}
\par\end{centering}
\caption{\label{fig:simple_contraction}Error bars illustrating the mean ($\pm$ one standard deviation) of the iterates (\emph{left}) and the derivative
of the iterated random function (\emph{right}) considered in Example \ref{eg:simple_example_contraction}. Each colour corresponds to
a specific number of particles $N$ in the ADP algorithm; the $N=\infty$ case corresponds to the expressions in (\ref{eqn:limiting_func_eg}) and (\ref{eqn:limiting_deriv_eg}).}
\end{figure}

\end{eg}

\section{Connection to Kullback-Leibler control\label{sec:control_connection}}
The Kullback-Leibler (KL) divergence from $\nu\in\mathcal{P}(\mathsf{E})$ to $\mu\in\mathcal{P}(\mathsf{E})$
is defined as $\mathrm{KL}(\mu|\nu)=\int_{\mathsf{E}}\log(\mathrm{d}\mu/\mathrm{d}\nu)(x)\mu(\mathrm{d}x)$
if the integral is finite and $\mu\ll\nu$, and $\mathrm{KL}(\mu|\nu)=\infty$ otherwise.
The intent of this section is to show that $\phi^{*}$ defined in
(\ref{eq:optimal_phi}) is the optimal policy of an associated KL optimal control problem \cite{Todorov_2009,Kappen_Gomez_Opper_2012}. Making
this connection allows us to leverage existing methodology and analysis
developed in the approximate dynamic programming literature \cite{Bertsekas_Tsitsiklis_1996,Tsitsiklis_VanRoy_2001}
in Sections \ref{sec:ADP} and \ref{sec:PolicyLearning} respectively.

Suppose that the current policy is $\psi\in\Psi$ and consider the
following optimal control problem
\begin{equation}
\inf_{\phi\in\Phi}\mathrm{KL}\left((\mathbb{Q}^{\psi})^{\phi}|\mathbb{P}\right)=\inf_{\phi\in\Phi}\mathbb{E}_{(\mathbb{Q}^{\psi})^{\phi}}\left[C(X_{0:T})\right]\label{eq:optimal_control_problem}
\end{equation}
where the set of admissible policies for the control problem is
\[
\Phi:=\left\{ \phi\in\Psi:\mathrm{KL}\left((\mathbb{Q}^{\psi})^{\phi}|\mathbb{Q}^{\psi}\right)<\infty\right\}
\]
and the cost functional $C:\mathsf{X}^{T+1}\rightarrow\mathbb{R}$
can be written as
\begin{align}
C(x_{0:T}) & :=\log\frac{\mathrm{d}(\mathbb{Q}^{\psi})^{\phi}}{\mathrm{d}\mathbb{Q}^{\psi}}(x_{0:T})-\log\frac{\mathrm{d}\mathbb{P}}{\mathrm{d}\mathbb{Q}^{\psi}}(x_{0:T}).\label{eq:cost_functional}
\end{align}
Using properties of KL divergence, it follows from Property 1 of Proposition
\ref{prop:Optimality} that $\phi^{*}$ defined in (\ref{eq:optimal_phi})
solves the optimal control problem (\ref{eq:optimal_control_problem}).
Rewriting (\ref{eq:cost_functional}) gives
\begin{align*}
\mathbb{E}_{(\mathbb{Q}^{\psi})^{\phi}}\left[C(X_{0:T})\right] & =\mathrm{KL}\left((\mu^{\psi})^{\phi}|\mu^{\psi}\right)+\sum_{t=1}^{T}\mathbb{E}_{(\mathbb{Q}^{\psi})^{\phi}}\left[\mathrm{KL}\left((M_{t}^{\psi})^{\phi}|M_{t}^{\psi})(X_{t-1})\right)\right]\\
 & -\mathbb{E}_{(\mu^{\psi})^{\phi}}\left[\log G_{0}^{\psi}(X_{0})\right]-\sum_{t=1}^{T}\mathbb{E}_{(\mathbb{Q}^{\psi})^{\phi}}\left[\log G_{t}^{\psi}(X_{t-1},X_{t})\right]+\log Z.
\end{align*}
We shall henceforth redefine the cost functional (\ref{eq:cost_functional})
to remove the intractable constant $\log Z$ that does not affect
the minimizer of (\ref{eq:optimal_control_problem}).

Given a policy $\phi\in\Phi$, the corresponding value functions $V^{\phi}=(V_{t}^{\phi})_{t\in[0:T]}$
of the control problem are given by the expected cost-to-go from a
fixed time and state (see for example \cite[Section 2.1]{Bertsekas_Tsitsiklis_1996})
\begin{align}
 & V_{0}^{\phi}(x_{0}):=\mathrm{KL}\left((M_{1}^{\psi})^{\phi}|M_{1}^{\psi})(x_{0})\right)+\sum_{s=1}^{T-1}\mathbb{E}_{(\mathbb{Q}^{\psi})^{\phi}}^{0,x_{0}}\left[\mathrm{KL}\left((M_{s+1}^{\psi})^{\phi}|M_{s+1}^{\psi})(X_{s})\right)\right]\nonumber \\
 & \qquad\qquad-\log G_{0}^{\psi}(x_{0})-\sum_{s=1}^{T}\mathbb{E}_{(\mathbb{Q}^{\psi})^{\phi}}^{0,x_{0}}\left[\log G_{s}^{\psi}(X_{s-1},X_{s})\right],\label{eq:value_functions}\\
 & V_{t}^{\phi}(x_{t-1},x_{t}):=\mathrm{KL}\left((M_{t+1}^{\psi})^{\phi}|M_{t+1}^{\psi})(x_{t})\right)+\sum_{s=t+1}^{T-1}\mathbb{E}_{(\mathbb{Q}^{\psi})^{\phi}}^{t,x_{t}}\left[\mathrm{KL}\left((M_{s+1}^{\psi})^{\phi}|M_{s+1}^{\psi})(X_{s})\right)\right]\nonumber \\
 & \qquad\qquad-\log G_{t}^{\psi}(x_{t-1},x_{t})-\sum_{s=t+1}^{T}\mathbb{E}_{(\mathbb{Q}^{\psi})^{\phi}}^{t,x_{t}}\left[\log G_{s}^{\psi}(X_{s-1},X_{s})\right],\quad t\in[1:T-1],\nonumber \\
 & V_{T}^{\phi}(x_{T-1},x_{T}):=-\log G_{T}^{\psi}(x_{T-1},x_{T}).\nonumber
\end{align}
In this notation, the total value of policy $\phi$ is given by
\[
v(\phi):=(\mu^{\psi})^{\phi}(V_{0}^{\phi})+\mathrm{KL}\left((\mu^{\psi})^{\phi}|\mu^{\psi}\right)=\mathrm{KL}\left((\mathbb{Q}^{\psi})^{\phi}|\mathbb{P}\right)-\log Z.
\]
We now define the optimal value $v^{*}$ and optimal value functions
$V^{*}=(V_{t}^{*})_{t\in[0:T]}$ w.r.t. $\mathbb{Q}^{\psi}$ by taking
the infimum over the set $\Phi$
\begin{align}
 & v^{*}:=\inf_{\phi}v(\phi),\label{eq:optimal_value}\\
 & V_{0}^{*}(x_{0}):=\inf_{\phi_{s},s\in[1:T]}V_{0}^{\phi}(x_{0}),\nonumber \\
 & V_{t}^{*}(x_{t-1},x_{t}):=\inf_{\phi_{s},s\in[t+1:T]}V_{t}^{\phi}(x_{t-1},x_{t}),\quad t\in[1:T-1],\nonumber \\
 & V_{T}^{*}(x_{T-1},x_{T}):=-\log G_{T}^{\psi}(x_{T-1},x_{T}),\nonumber
\end{align}
and denote the minimizer (if it exists) as $\phi^{*}=(\phi_{t})_{t\in[0:T]}$.
We stress the dependence of both $V^{*}$ and $\phi^{*}$ on the current
policy $\psi\in\Psi$ as it is omitted notationally. These minimization
problems can be solved using a backward dynamic programming approach.
From (\ref{eq:value_functions}) and (\ref{eq:optimal_value}), we
have the dynamic programming recursion
\begin{align}
 & V_{T}^{*}(x_{T-1},x_{T})=-\log G_{T}^{\psi}(x_{T-1},x_{T}),\label{eq:optimal_value_functions}\\
 & V_{t}^{*}(x_{t-1},x_{t})=-\log G_{t}^{\psi}(x_{t-1},x_{t})+\inf_{\phi_{t+1}}\left\{ (M_{t+1}^{\psi})^{\phi}(V_{t+1}^{*})(x_{t})\right.\nonumber \\
 & \left.\qquad\qquad+\,\mathrm{KL}\left((M_{t+1}^{\psi})^{\phi}|M_{t+1}^{\psi}\right)(x_{t})\right\} ,\quad t\in[1:T-1],\nonumber \\
 & V_{0}^{*}(x_{0})=-\log G_{0}^{\psi}(x_{0})+\inf_{\phi_{1}}\left\{ (M_{1}^{\psi})^{\phi}(V_{1}^{*})(x_{0})+\mathrm{KL}\left((M_{1}^{\psi})^{\phi}|M_{1}^{\psi}\right)(x_{0})\right\} ,\nonumber \\
 & v^{*}=\inf_{\phi_{0}}\left\{ (\mu^{\psi})^{\phi}(V_{0}^{*})+\mathrm{KL}\left((\mu^{\psi})^{\phi}|\mu^{\psi}\right)\right\} .\nonumber
\end{align}
The above is commonly referred to as the discrete time Bellman recursion.

Owing to the use of KL costs, the minimizations in (\ref{eq:optimal_value_functions})
are tractable: assuming that the current policy $\psi\in\Psi$ satisfies
$\mathrm{KL}(\mathbb{P}|\mathbb{Q}^{\psi})<\infty$, applying \cite[Proposition 2.3]{DaiPra_Meneghini_Runggaldier_1996}
gives
\begin{align}
 & V_{T}^{*}(x_{T-1},x_{T})=-\log G_{T}^{\psi}(x_{T-1},x_{T}),\label{eq:backward_recursion_value}\\
 & V_{t}^{*}(x_{t-1},x_{t})=-\log G_{t}^{\psi}(x_{t-1},x_{t})-\log M_{t+1}^{\psi}(e^{-V_{t+1}^{*}})(x_{t}),\quad t\in[1:T-1],\nonumber \\
 & V_{0}^{*}(x_{0})=-\log G_{0}^{\psi}(x_{0})-\log M_{1}^{\psi}(e^{-V_{1}^{*}})(x_{0}),\nonumber \\
 & v^{*}=-\log\mu^{\psi}(e^{-V_{0}^{*}})=-\log Z,\nonumber
\end{align}
with infimum attained at $\phi_{t}^{*}=e^{-V_{t}^{*}}$ for $t\in[0:T]$.
Observe that the optimal value functions are simply logarithmic transformations
of the optimal policy, and the dynamic programming recursion (\ref{eq:backward_recursion_value})
corresponds to (\ref{eq:optimal_phi}) in logarithmic scale. The optimal
value is $v^{*}=-\log Z$ as we have adjusted the cost functional
(\ref{eq:cost_functional}). Lastly, the finite KL condition guarantees
existence of a unique minimizer $\phi^{*}$ that lies in $\Phi$.
It should be clear from Proposition \ref{prop:Optimality} that working
with the subset $\Phi\subset\Psi$ is not necessary, i.e. such a condition
is only required when we formulate $\phi^{*}$ as the optimal policy
of a Kullback-Leibler control problem.

\section{A non-linear multimodal state space model\label{sec:kitagawa}}

We consider a popular toy non-linear state space model \cite{Gordon_Salmond_Smith_1993,Kitagawa_1996}
which corresponds to working on $(\mathsf{X},\mathcal{X})=(\mathbb{R},\mathfrak{B}(\mathbb{R})),\mathsf{Y=\mathbb{R}}$
and having
\begin{align}
\nu(\mathrm{d}x_{0}) & =\mathcal{N}(x_{0};0,5)\mathrm{d}x_{0},\label{eq:kitagawa_model}\\
f_{t}(x_{t-1},\mathrm{d}x_{t}) & =\mathcal{N}\left(x_{t};\frac{1}{2}x_{t-1}+\frac{25x_{t-1}}{1+x_{t-1}^{2}}+8\cos(1.2t),\sigma_{f}^{2}\right)\mathrm{d}x_{t},\nonumber \\
g_{t}(x_{t},y_{t}) & =\mathcal{N}\left(y_{t};\frac{1}{20}x_{t}^{2},\sigma_{g}^{2}\right),\nonumber
\end{align}
for $t\in[1:T]$, where $\theta=(\sigma_{f}^{2},\sigma_{g}^{2})\in\mathbb{R}_{+}\times\mathbb{R}_{+}$.
We will employ the BPF as uncontrolled SMC, i.e. set $\mu=\nu$
and $M_{t}=f_{t}$ for $t\in[1:T]$. As the smoothing distribution
(\ref{eq:smoothing_distribution}) is highly multimodal, owing to
the uncertainty of the sign of the latent process, this example is
commonly used as a benchmark to assess the performance of SMC methods.
Moreover, we observe from Figure \ref{fig:kitagawa_optimalpolicy}
that this problem also induces complex multimodal optimal policies.

\subsection{Approximate dynamic programming}

To approximate these policies, we rely on the following flexible function
classes
\begin{align}
	\mathsf{F}_t &= \biggl\{ \varphi(x_t) = -\log\biggl(\,\sum_{m=1}^M\alpha_{t,m}\exp\left(-\beta_t(x_t-\xi_{t,m})^2\right)\biggr)\notag\\
				 &\quad\quad : (\alpha_{t,m},\beta_t,\xi_{t,m})\in \mathbb{R}_+\times\mathbb{R}_+\times\mathbb{R}, \, m\in[1:M] \biggr\}\notag
\end{align}for all $t\in[0:T]$, which corresponds to a radial basis function
(RBF) approximation of the optimal policy in the natural scale. With
this choice of function classes, the approximate projections (\ref{eq:ADP})
can be implemented using non-linear least squares.

Given the output of a twisted SMC method based on the current policy,
we adopt the following approach which is computationally more efficient.
Firstly, we fix $\beta_{t}$ as a pre-specified bandwidth factor $\tau\in\mathbb{R}_{+}$
multiplied by the sample standard deviation of particles $(X_{t}^{n})_{n\in[1:N]}$
at time $t\in[0:T]$. Instead of performing the above logarithmic
projections to learn the associated value functions, we fit the RBF
approximation directly at the natural scale with $\xi_{t,n}=X_{t}^{n}$
for $n\in[1:N]$, as this can be efficiently implemented \cite[p. 161]{Lawson_Hanson_1974}
as a linear least squares problem with non-negativity constraints
in $(\alpha_{t,n})_{n\in[1:N]}$. We note that care has to be taken
to ensure that these computations are numerically stable. We then
sort the estimated weights $(\alpha_{t,n})_{n\in[1:N]}$ and keep
as knots $(\xi_{t,m})_{m\in[1:M]}$ particles with the $M$ largest
weights, as this avoids having to retain components with low weights.
This selection procedure allows us to adaptively focus our computational
effort on approximating the optimal policy at appropriate regions
of the state space.

Writing $(\alpha_{t,m}^{i+1})_{m\in[1:M]}$ as the weights, $\beta_{t}^{i+1}$
as the bandwidth and $(\xi_{t,m}^{i+1})_{m\in[1:M]}$ as the knots
estimated by cSMC at iteration $i\in[0:I-1]$ for $t\in[0:T]$, the
policy $\psi^{(i)}=(\psi_{t}^{(i)})_{t\in[0:T]}$ at iteration
$i\in[1:I]$ has the form
\begin{equation}
\psi_{t}^{(i)}(x_{t})=\sum_{m\in[1:M]^{i}}\alpha_{t,m}^{(i)}\exp\left(-\beta_{t}^{(i)}(x_{t}-\xi_{t,m}^{(i)})^{2}\right),\quad t\in[0:T],\label{eq:kitagawa_currentpolicy}
\end{equation}
where $m=(m_{j})_{j\in[1:i]}\in[1:M]^{i}$ is a multi-index, $\beta_{t}^{(i)}:=\sum_{j=1}^{i}\beta_{t}^{j}$,
$\xi_{t,m}^{(i)}:=\sum_{j=1}^{i}\beta_{t}^{j}\xi_{t,m_{j}}^{j}/\beta_{t}^{(i)}$
and
\[
\alpha_{t,m}^{(i)}:=\prod_{j=1}^{i}\alpha_{t,m_{j}}^{j}\exp\left(-\sum_{j=1}^{i}\beta_{t}^{j}(\xi_{t,m_{j}}^{j})^{2}+\beta_{t}^{(i)}(\xi_{t,m}^{(i)})^{2}\right).
\]
It follows that under policy (\ref{eq:kitagawa_currentpolicy}), the
initial distribution $\mu^{\psi^{(i)}}$ is a mixture of Gaussian
distributions, Markov transition kernels $(M_{t}^{\psi^{(i)}})_{t\in[1:T]}$
are given by mixtures of Gaussian transition kernels and evaluation
of the twisted potentials $(G_{t}^{\psi^{(i)}})_{t\in[0:T]}$ defined
in (\ref{eq:twisted_potentials}) is tractable; exact expressions
are given in Section \ref{appendix:kitagawa} of the Supplementary Material.
Figure \ref{fig:kitagawa_optimalpolicy}
shows that such a parameterization is flexible enough to provide an
adequate approximation of the optimal policy.

\begin{figure}
\begin{centering}
\includegraphics[scale=0.4]{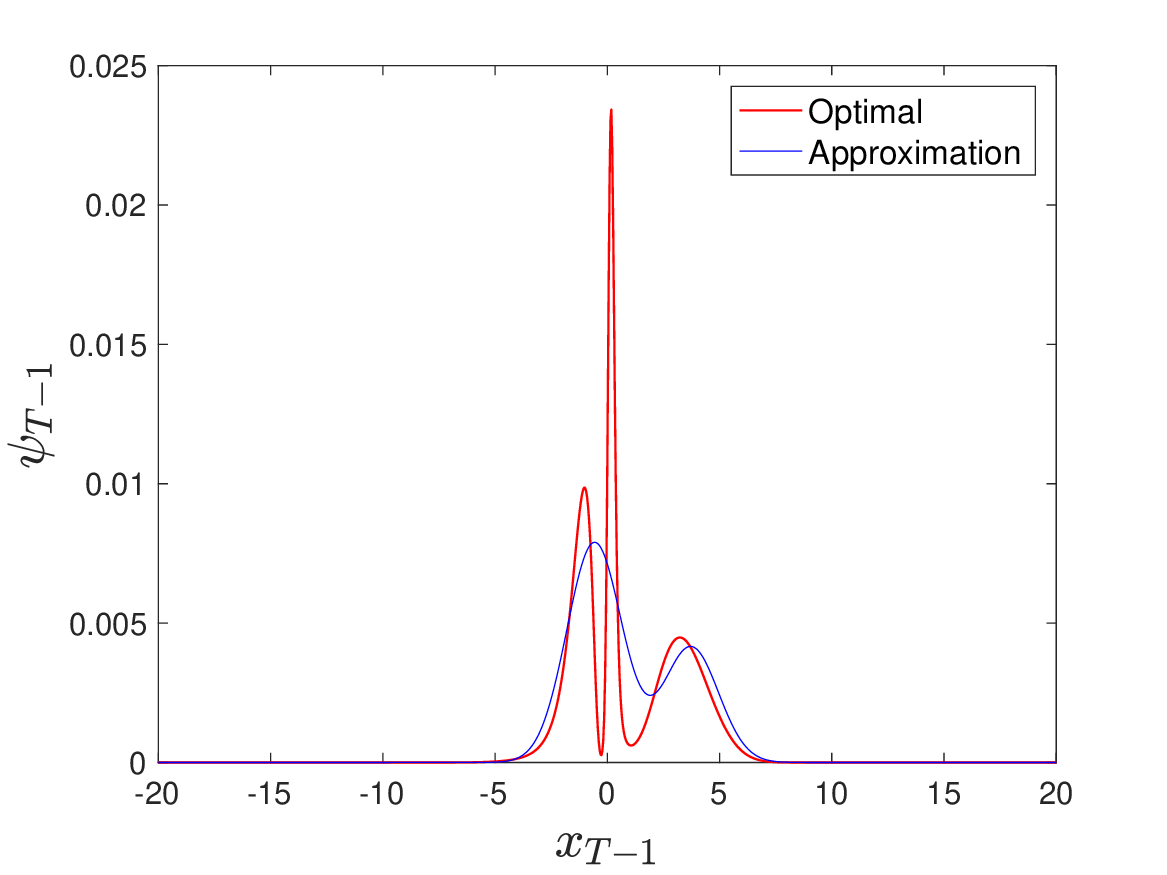}\includegraphics[scale=0.4]{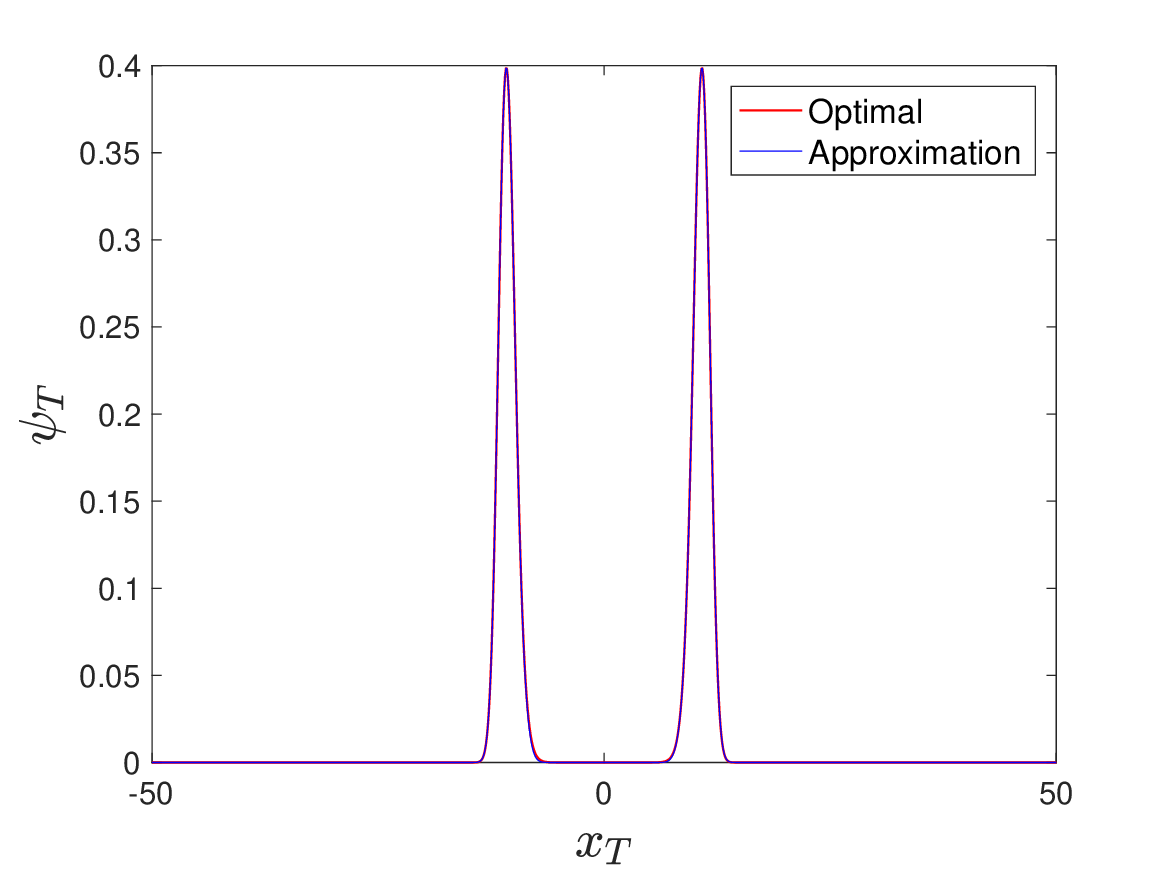}
\par\end{centering}
\caption{\label{fig:kitagawa_optimalpolicy}Optimal policy of non-linear multimodal
state space model (\ref{eq:kitagawa_model}) at terminal times. The problem setting
corresponds to $T=100,\sigma_{f}^{2}=10,\sigma_{g}^{2}=1$
and the algorithmic settings of cSMC is $I=1,N=512,M=16.$}

\end{figure}

\subsection{Comparison of algorithmic performance}

We investigate the use of cSMC when the observation noise is small,
i.e. high signal-to-noise ratio, since this is the regime where BPF
exhibits poor performance. To do so, we fix $\sigma_{f}^{2}=10$ and
simulate three sets of observations $y_{0:T}\in\mathsf{Y}^{T+1}$
of length $T+1=100$ according to (\ref{eq:kitagawa_model}) as $\sigma_{g}^{2}$
takes values in $\{0.1,0.5,1\}$. We use $N=512$ particles in cSMC
and $I=1$ iteration as preliminary runs indicate that policy refinement
under the parameterization (\ref{eq:kitagawa_currentpolicy}) provides little improvement,
especially when additional computing time is taken into account. The
number of particles in BPF is then chosen to match computational
time.
The number of components $M$ and bandwidth factor $\tau$ were tuned
using preliminary runs. These algorithmic settings and the results obtained
in $100$ independent repetitions of each method are summarized in
Table \ref{tab:kitagawa}. As expected, although the performance gains
over BPF diminish as the observation noise increases, it can be substantial
when $\sigma_{g}^{2}$ is small.

\begin{table}
\begin{centering}
\hspace{-0cm}%
\begin{tabular}{|c|c|c|c|c|c|}
\cline{4-6}
\multicolumn{1}{c}{} & \multicolumn{1}{c}{} &  & \multicolumn{3}{c|}{\textbf{\footnotesize{}Observation noise}}\tabularnewline
\cline{4-6}
\multicolumn{1}{c}{} & \multicolumn{1}{c}{} &  & \textit{\footnotesize{}$\sigma_{g}^{2}=0.1$} & \textit{\footnotesize{}$\sigma_{g}^{2}=0.5$} & \textit{\footnotesize{}$\sigma_{g}^{2}=1$}\tabularnewline
\hline
\multirow{11}{*}{\begin{turn}{90}
\textbf{\footnotesize{}Algorithm}
\end{turn}} & \multirow{4}{*}{{\footnotesize{}BPF}} & {\footnotesize{}$N$} & {\footnotesize{}$2252$} & {\footnotesize{}$4710$} & {\footnotesize{}$6553$}\tabularnewline
\cline{3-6}
 &  & {\footnotesize{}$\mathrm{ESS\%}$} & {\footnotesize{}$15.07\%$} & {\footnotesize{}$30.24\%$} & {\footnotesize{}$39.12\%$}\tabularnewline
 &  & {\footnotesize{}$\log Z$} & {\footnotesize{}$-281.6185\pm1.0054$} & {\footnotesize{}$-262.9861\pm0.6037$} & {\footnotesize{}$-250.6369\pm0.2845$}\tabularnewline
 &  & {\footnotesize{}$\mathrm{RVAR}$} & {\footnotesize{}$1.27\times10^{-5}$} & {\footnotesize{}$5.27\times10^{-6}$} & {\footnotesize{}$1.29\times10^{-6}$}\tabularnewline
\cline{2-6}
 & \multirow{7}{*}{{\footnotesize{}cSMC}} & {\footnotesize{}$I$} & {\footnotesize{}$1$} & {\footnotesize{}$1$} & {\footnotesize{}$1$}\tabularnewline
 &  & {\footnotesize{}$N$} & {\footnotesize{}$512$} & {\footnotesize{}$512$} & {\footnotesize{}$512$}\tabularnewline
 &  & {\footnotesize{}$M$} & {\footnotesize{}$16$} & {\footnotesize{}$16$} & {\footnotesize{}$16$}\tabularnewline
 &  & {\footnotesize{}$\tau$} & {\footnotesize{}$0.5$} & {\footnotesize{}$0.4$} & {\footnotesize{}$0.3$}\tabularnewline
\cline{3-6}
 &  & {\footnotesize{}$\mathrm{ESS\%}$} & {\footnotesize{}$82.35\%$} & {\footnotesize{}$92.51\%$} & {\footnotesize{}$94.66\%$}\tabularnewline
 &  & {\footnotesize{}$\log Z$} & {\footnotesize{}$-281.1483\pm0.2295$} & {\footnotesize{}$-262.7223\pm0.2425$} & {\footnotesize{}$-250.6949\pm0.1439$}\tabularnewline
 &  & {\footnotesize{}$\mathrm{RVAR}$} & {\footnotesize{}$6.67\times10^{-7}$ $(\mathbf{19.1})$} & {\footnotesize{}$8.52\times10^{-7}$ $(\mathbf{6.18})$} & {\footnotesize{}$3.29\times10^{-7}$ $(\mathbf{3.91})$}\tabularnewline
\hline
\end{tabular}
\par\end{centering}
\caption{\label{tab:kitagawa}
Non-linear multimodal state space model (\ref{eq:kitagawa_model}):
algorithmic settings and performance of BPF and cSMC for each observation
noise considered. Notationally, $N$ refers to the number of particles,
$I$ is the number of iterations taken by cSMC, $M$ denotes the number
of components and $\tau$ the bandwidth factor used in the ADP approximation.
Results were obtained using $100$ independent repetitions each of
method. The shorthand $\mathrm{ESS\%}$ denotes the percentage of
effective sample size averaged over time and repetitions, $\log Z$
refers to the estimation of the normalizing constant in logarithmic
scale ($\pm$ a standard deviation), $\mathrm{RVAR}$ is the sample
relative variance of these estimates over the repetitions. Shown in
bold is the gain that cSMC offers relative to BPF. }
\end{table}

\section{Linear quadratic Gaussian control\label{sec:LQG_control}}
 This section considers a Gaussian static model (Section \ref{sec:static_models}) which will allow
us to draw connections to concepts from the linear quadratic Gaussian
(LQG) control literature \cite{Anderson_Moore_2007}.
Consider $\mu(\mathrm{d}x_{0})=\mathcal{N}(x_{0};\mu_{0},\Sigma_{0})\mathrm{d}x_{0}$ on $(\mathsf{X},\mathcal{X})=(\mathbb{R}^{d},\mathfrak{B}(\mathbb{R}^{d}))$
and $\ell(x,y)=\exp(-(y-x)^{T}R^{-1}(y-x)/2)$ for some $y\in\mathsf{Y}=\mathbb{R}^{d}$
and symmetric positive definite $R\in\mathbb{R}^{d\times d}$. By
conjugacy, the models (\ref{eq:geometric_path}) are Gaussian and
for $t\in[0:T]$ we have $\eta_{t}(\mathrm{d}x_{t})=\mathcal{N}(x_{t};\mu_{t},\Sigma_{t})\mathrm{d}x_{t}$
with
\[
\mu_{t}:=\Sigma_{t}(\Sigma_{0}^{-1}\mu_{0}+\lambda_{t}R^{-1}y),\quad\Sigma_{t}:=(\Sigma_{0}^{-1}+\lambda_{t}R^{-1})^{-1}
\]
and
\[
Z_{t}=\det(\Sigma_{0})^{-1/2}\det(\Sigma_{t})^{1/2}\exp\left(-\frac{1}{2}\left\{ \mu_{0}^{T}\Sigma_{0}^{-1}\mu_{0}+\lambda_{t}y^{T}R^{-1}y-\mu_{t}^{T}\Sigma_{t}^{-1}\mu_{t}\right\} \right).
\]

\subsection{Riccati equation}

\begin{figure}
\begin{centering}
\includegraphics[scale=0.4]{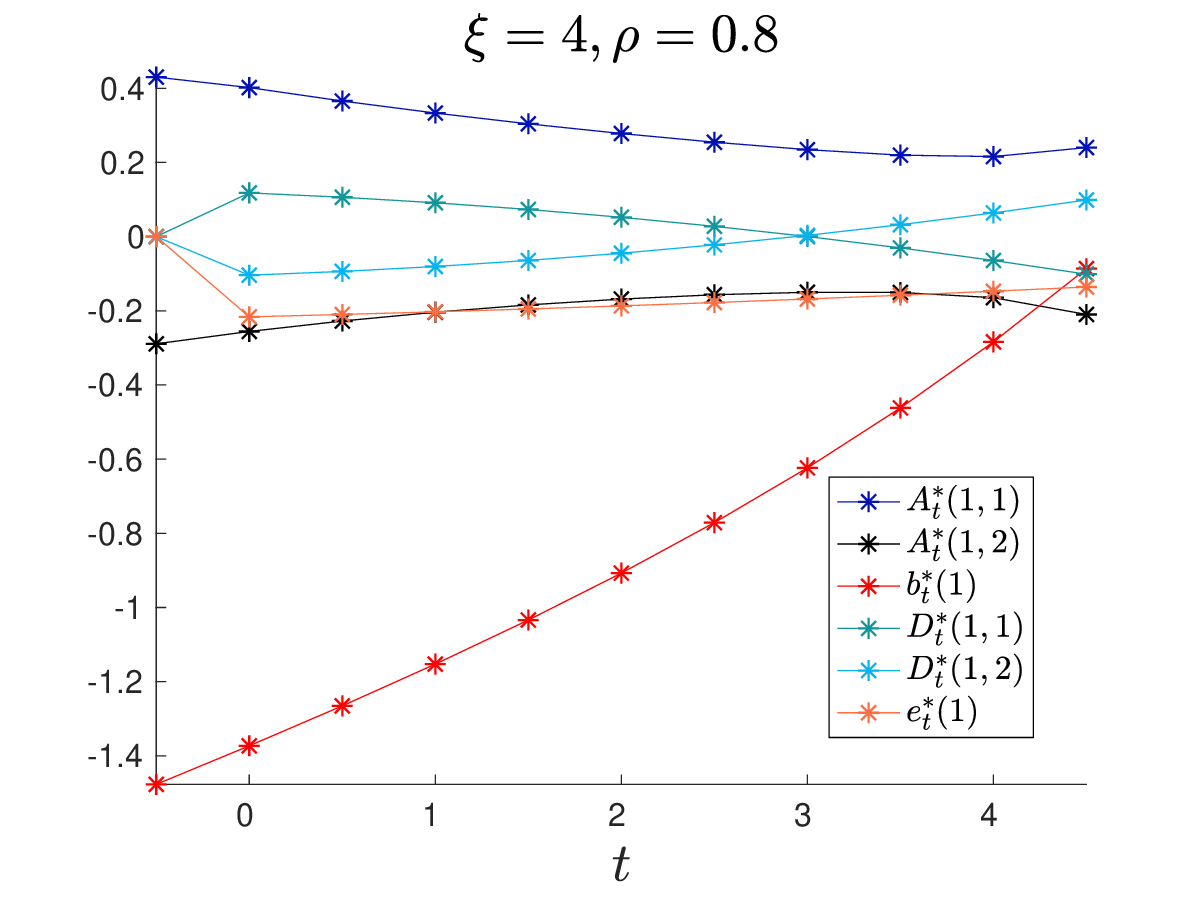}\includegraphics[scale=0.4]{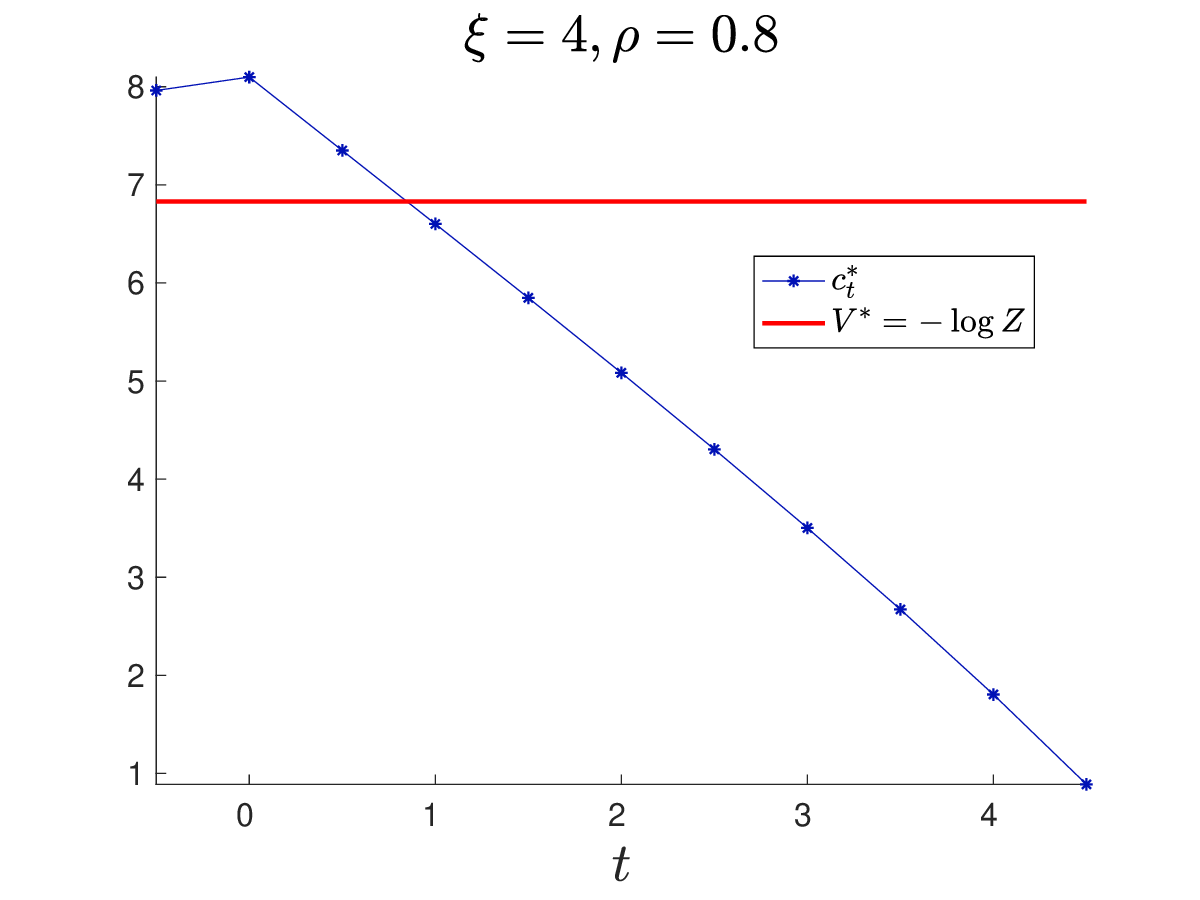}
\par\end{centering}
\begin{centering}
\includegraphics[scale=0.4]{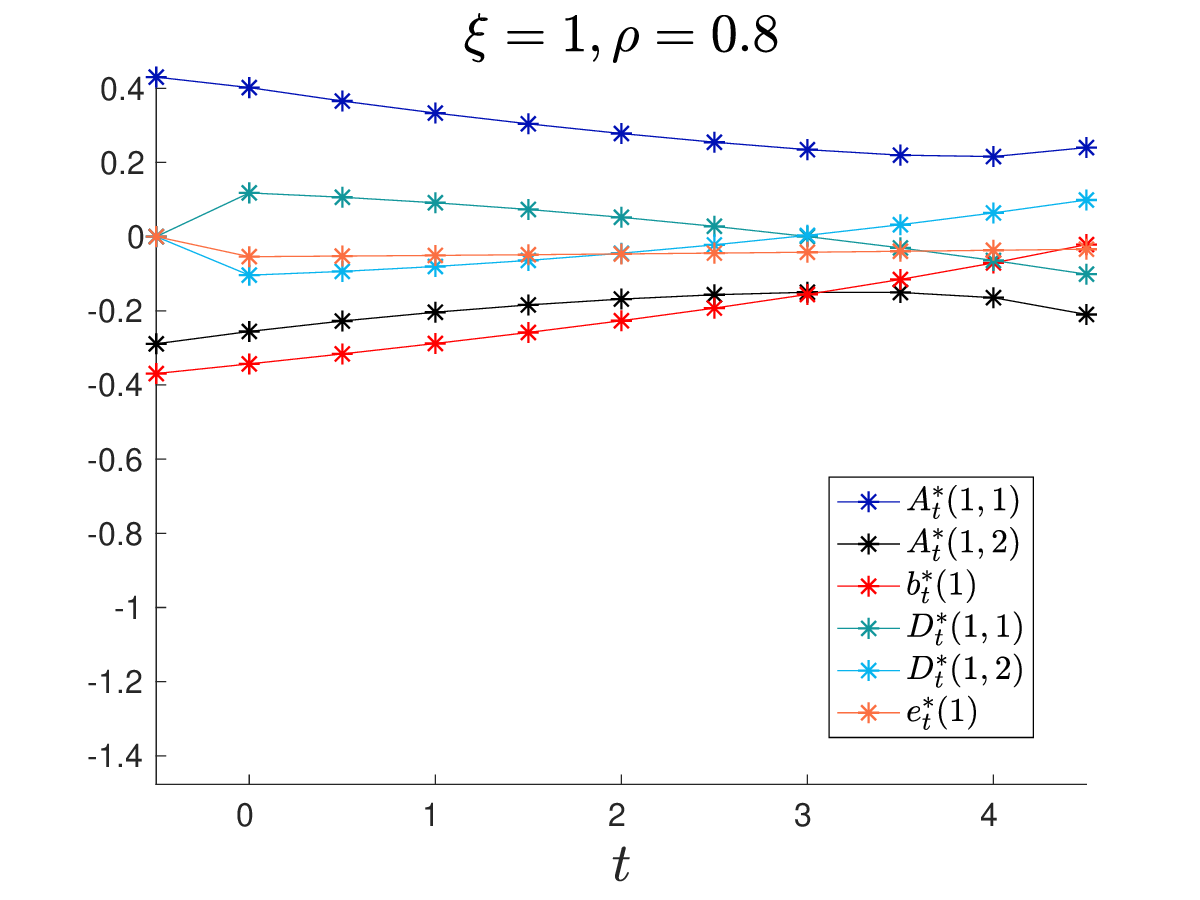}\includegraphics[scale=0.4]{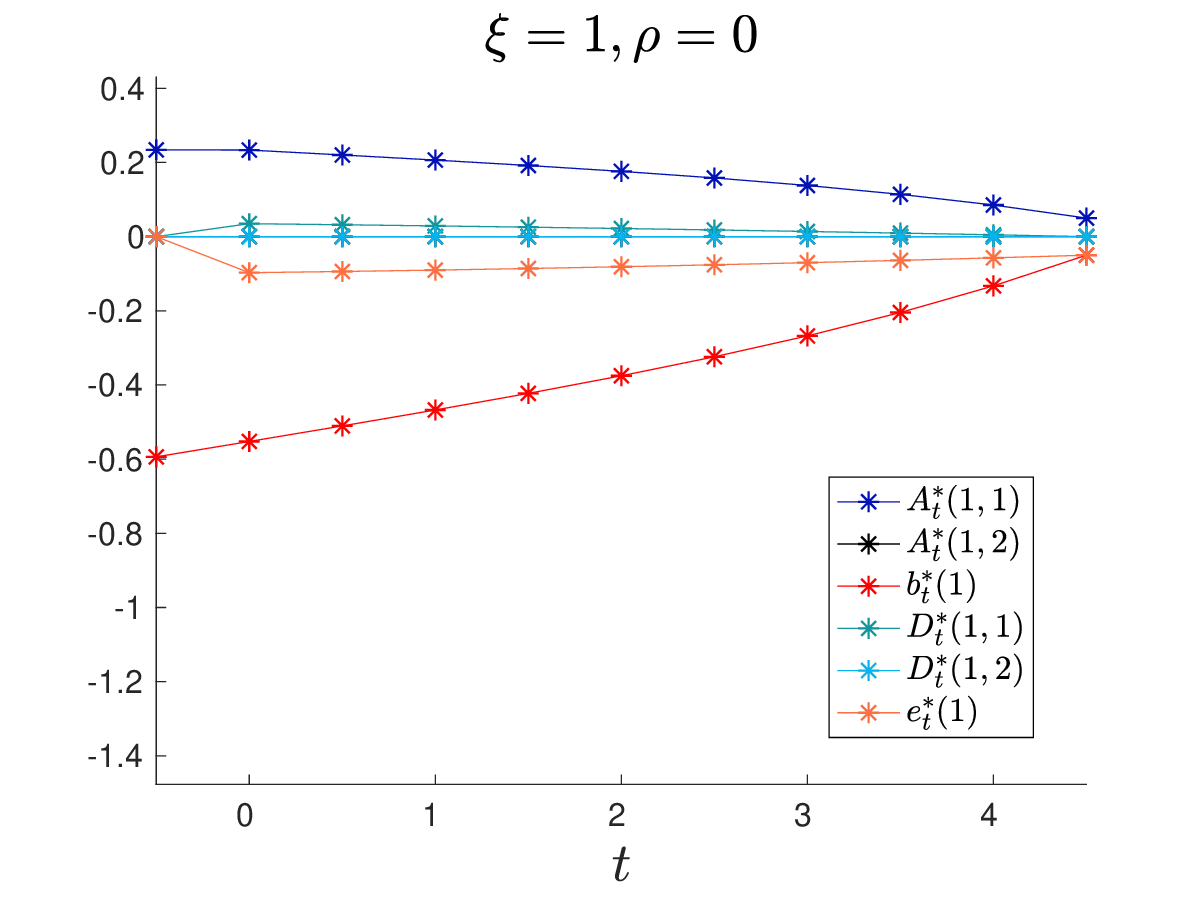}
\par\end{centering}
\caption{\label{fig:LQG_coefficients}Coefficients of the optimal policy w.r.t.
$\mathbb{Q}$ in LQG control under various problem settings. The algorithmic
settings of cSMC are $T=10,h=0.1,\lambda_{t}=t/T$.
Note that all except the top right plot share the same axes.}
\end{figure}

\begin{figure}
\begin{centering}
\includegraphics[scale=0.4]{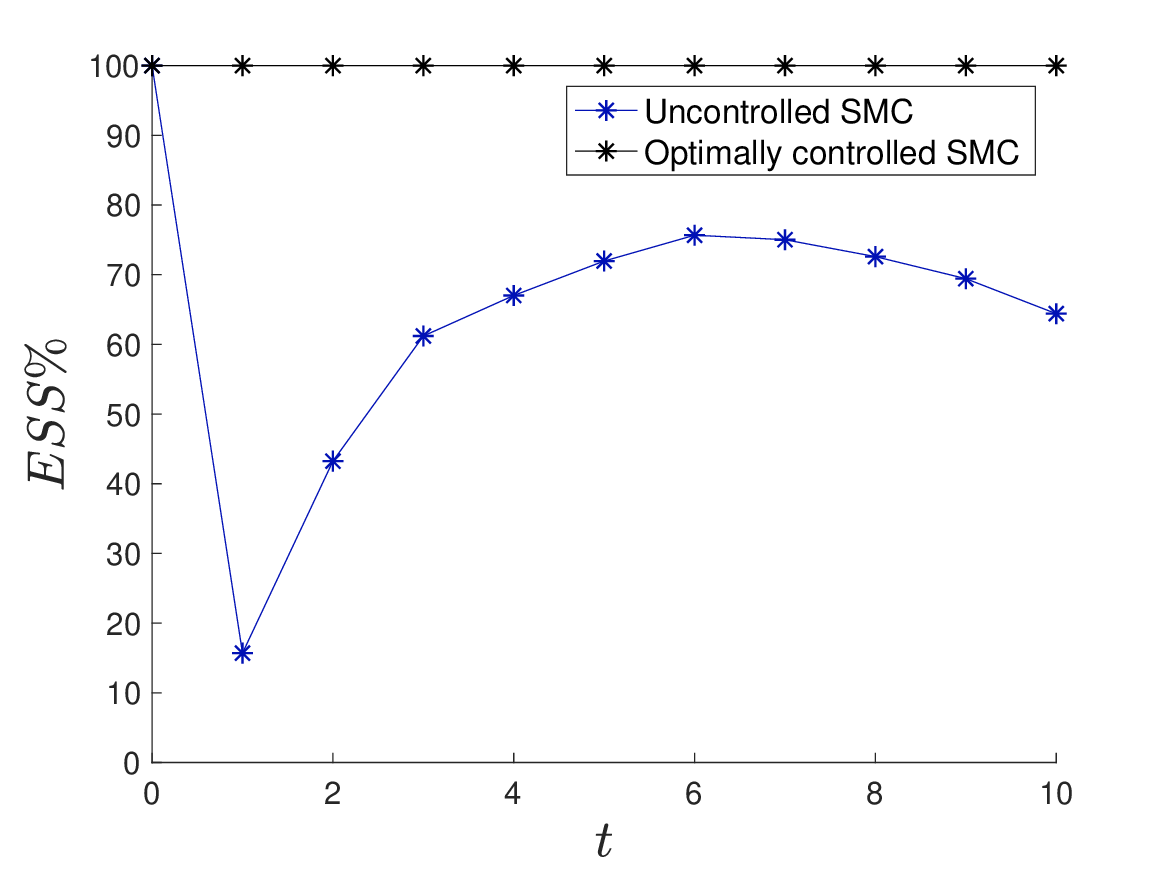}\includegraphics[scale=0.4]{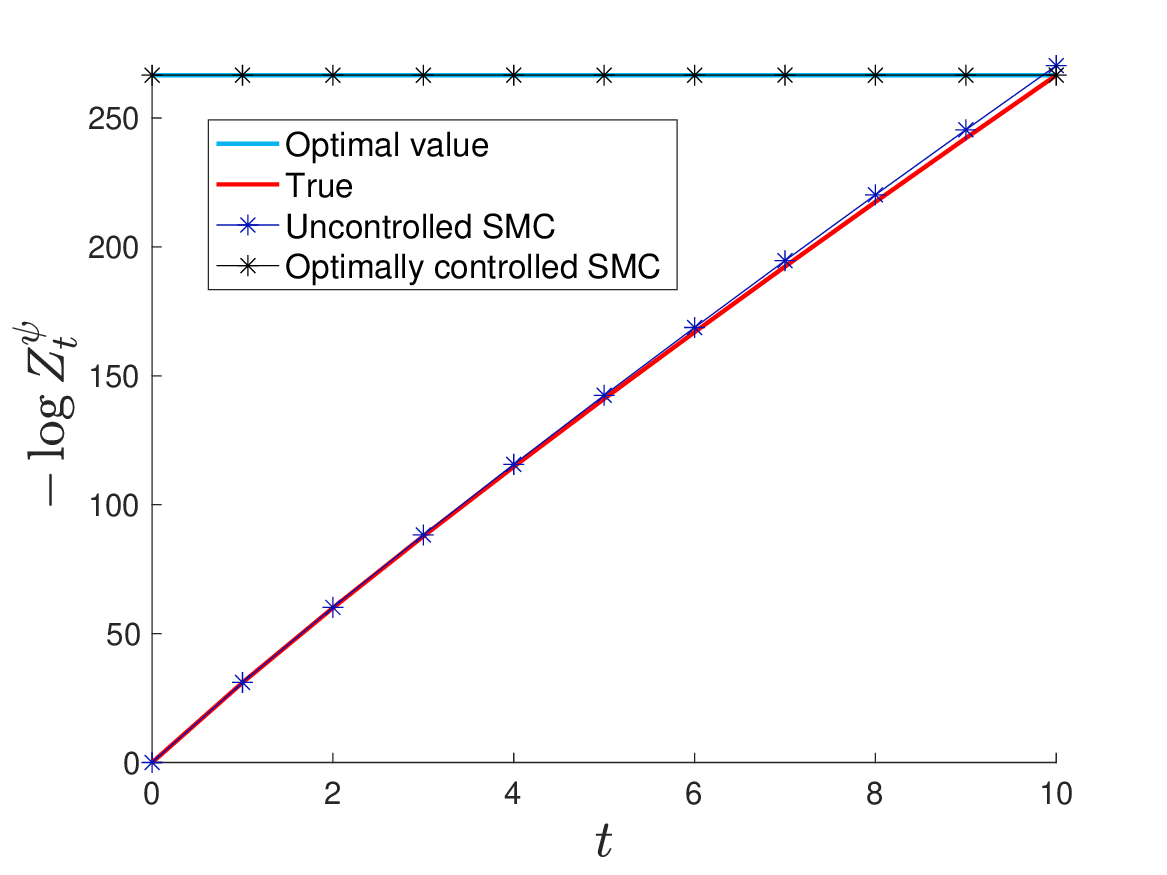}
\par\end{centering}
\caption{\label{fig:LQG_compare}Comparison of uncontrolled SMC and optimal LQG controlled
SMC in terms of effective sample size (\emph{left}) and normalizing
constant estimation (\emph{right}). The problem setting considered
here is $d=32,\xi=20,\rho=0.8$ and the algorithmic settings
of uncontrolled SMC are $N=2048,T=10,h=0.1,\lambda_{t}=t/T$. }
\end{figure}

We now show that the backward recursion (\ref{eq:optimal_phi}) with
$\psi$ initialized as a policy of constant one functions can be performed
exactly to obtain analytic expressions of the optimal policy w.r.t.
$\mathbb{Q}$. First note that under the choice of forward and
backward Markov transition
kernels specified in Section \ref{sec:application_staticmodels} with pre-conditioner $\Gamma=I_d$,
the potentials (\ref{eq:SMCsampler_potentials}) have the form
\begin{align}
-\log G_{t}(x_{t-1},x_{t})=x_{t}^{T}\tilde{A}_{t}x_{t}+x_{t}^{T}\tilde{b}_{t}+\tilde{c}_{t}+x_{t-1}^{T}\tilde{D}_{t}x_{t-1}+x_{t-1}^{T}\tilde{e}_{t},\label{eq:LQG_logpotential}
\end{align}
where
\begin{align}
 & \quad\quad\tilde{A}_{t}:=\frac{1}{8}h\Sigma_{t}^{-2},\quad\tilde{b}_{t}:=-\frac{1}{4}h\Sigma_{t}^{-2}\mu_{t},\quad\tilde{c}_{t}:=\frac{1}{2}(\lambda_{t}-\lambda_{t-1})y^{T}R^{-1}y,\label{eq:LQG_potential_coeff}\\
 & \tilde{D}_{t}:=-\frac{1}{8}h\Sigma_{t}^{-2}+\frac{1}{2}(\lambda_{t}-\lambda_{t-1})R^{-1},\quad\tilde{e}_{t}:=-(\lambda_{t}-\lambda_{t-1})R^{-1}y+\frac{1}{4}h\Sigma_{t}^{-2}\mu_{t},\nonumber
\end{align}
for $t\in[1:T]$. For sufficiently small step size, observe that dropping
$O(h)$ terms in (\ref{eq:LQG_potential_coeff}) gives $\log G_{t}(x_{t-1},x_{t})\approx(\lambda_{t}-\lambda_{t-1})\log\ell(x_{t-1},y)$
which, as expected, recovers the AIS potentials (\ref{eq:AIS_potentials}).
For notational convenience, we set $(\tilde{A}_{0},\tilde{b}_{0},\tilde{c}_{0},\tilde{D}_{0},\tilde{e}_{0})$
as the zero matrix or vector of the appropriate size and write the
mean of the Euler-Maruyama move as $x_{t-1}+h\nabla\log\eta_{t}(x_{t-1})/2=P_{t}x_{t-1}+q_{t}$
with $P_{t}:=I_{d}-h\Sigma_{t}^{-1}/2$ and $q_{t}:=h\Sigma_{t}^{-1}\mu_{t}/2$.
\begin{prop}
The optimal policy $\psi^{*}=(\psi_{t}^{*})_{t\in[0:T]}$ w.r.t. $\mathbb{Q}$
is given by
\begin{align}
-\log\psi_{0}^{*}(x_{0}) & =x_{0}^{T}A_{0}^{*}x_{0}+x_{0}^{T}b_{0}^{*}+c_{0}^{*},\label{eq:LQG_optimalpolicy}\\
-\log\psi_{t}^{*}(x_{t-1},x_{t}) & =x_{t}^{T}A_{t}^{*}x_{t}+x_{t}^{T}b_{t}^{*}+c_{t}^{*}+x_{t-1}^{T}D_{t}^{*}x_{t-1}+x_{t-1}^{T}e_{t}^{*},\quad t\in[1:T],\nonumber
\end{align}
where the coefficients $(A_{t}^{*},b_{t}^{*},c_{t}^{*},D_{t}^{*},e_{t}^{*})_{t\in[0:T]}$
are determined by the backward recursion
\begin{align}
A_{t}^{*} & =\tilde{A}_{t}+\frac{1}{2}h^{-1}P_{t+1}(I_{d}-K_{t+1}^{*})P_{t+1}+D_{t+1}^{*},\label{eq:LQG_riccati}\\
b_{t}^{*} & =\tilde{b}_{t}+P_{t+1}K_{t+1}^{*}b_{t+1}^{*}+e_{t+1}^{*}+\frac{1}{2}P_{t+1}(I_{d}-K_{t+1}^{*})\Sigma_{t+1}^{-1}\mu_{t+1},\nonumber \\
c_{t}^{*} & =\tilde{c}_{t}+c_{t+1}^{*}-\frac{1}{2}\log\det(K_{t+1}^{*})+\frac{1}{2}h^{-1}q_{t+1}^{T}q_{t+1}\nonumber \\
 & -\frac{1}{2}h^{-1}(q_{t+1}-h b_{t+1}^{*})^{T}K_{t+1}^{*}(q_{t+1}-hb_{t+1}^{*}),\nonumber \\
D_{t}^{*} & =\tilde{D}_{t},\nonumber \\
e_{t}^{*} & =\tilde{e}_{t,}\nonumber
\end{align}
for $t\in[T-1:0]$, with $K_{t}^{*}:=(I_{d}+2hA_{t}^{*})^{-1},t\in[1:T]$
and initialization at $(A_{T}^{*},b_{T}^{*},c_{T}^{*},D_{T}^{*},e_{T}^{*})=(\tilde{A}_{T},\tilde{b}_{T},\tilde{c}_{T},\tilde{D}_{T},\tilde{e}_{T})$.
\end{prop}
\begin{proof}
We proceed by induction. Clearly, (\ref{eq:LQG_optimalpolicy}) holds
for $t=T$ since $\psi_{T}^{*}=G_{T}$. Assume that (\ref{eq:LQG_optimalpolicy})
holds for time $t+1$. The recursion (\ref{eq:optimal_phi}) can be
written as
\[
-\log\psi_{t}^{*}(x_{t-1},x_{t})=-\log G_{t}(x_{t-1},x_{t})-\log M_{t+1}(\psi_{t+1}^{*})(x_{t}).
\]
Some manipulations yield
\begin{align*}
 & -\log M_{t+1}(\psi_{t+1}^{*})(x_{t})=x_{t}^{T}\left(\frac{1}{2}h^{-1}P_{t+1}(I_{d}-K_{t+1}^{*})P_{t+1}+D_{t+1}^{*}\right)x_{t}\\
 & +x_{t}^{T}\left(P_{t+1}K_{t+1}^{*}b_{t+1}^{*}+e_{t+1}^{*}+\frac{1}{2}P_{t+1}(I_{d}-K_{t+1}^{*})\Sigma_{t+1}^{-1}\mu_{t+1}\right)-\frac{1}{2}\log\det(K_{t+1}^{*})\\
 & +c_{t+1}^{*}+\frac{1}{2}h^{-1}\left\{ q_{t+1}^{T}q_{t+1}-(q_{t+1}-hb_{t+1}^{*})^{T}K_{t+1}^{*}(q_{t+1}-hb_{t+1}^{*})\right\} .
\end{align*}
Adding this to (\ref{eq:LQG_logpotential}) establishes that $-\log\psi_{t}^{*}$
has the desired form (\ref{eq:LQG_optimalpolicy}) and equating coefficients
of the polynomial gives (\ref{eq:LQG_riccati}).
\end{proof}
The backward recursion (\ref{eq:LQG_riccati}) for the coefficients
is analogous to the Riccati equation in the context of LQG control.
To illustrate the behaviour of these coefficients, we set the prior
as $\mu_{0}=0_{d}$, $\Sigma_{0}=I_{d}$ and the likelihood as $y=(\xi,\ldots,\xi)^{T}$
for some $\xi\in\mathbb{R}$ and $R_{i,j}=\delta_{i,j}+(1-\delta_{i,j})\rho$
for $i,j\in[1:d]$ and some $\rho\in[-1,1]$ (here $\delta_{i,j}$
denotes the Kronecker delta).
The time evolution of these coefficients is plotted in the top row
of Figure \ref{fig:LQG_coefficients} for the problem setting $d=2$,
$\xi=4$, $\rho=0.8$. Noting that the optimal value of the Kullback-Leibler
control problem (\ref{eq:optimal_value}) is
\begin{align*}
v^{*} & =-\log Z=c_{0}^{*}+\frac{1}{2}\log\det(\Sigma_{0})-\frac{1}{2}\log\det(K_{0}^{*})\\
&-\frac{1}{2}(\Sigma_{0}^{-1}\mu_{0}-b_{0}^{*})^{T}K_{0}^{*}(\Sigma_{0}^{-1}\mu_{0}-b_{0}^{*})+\frac{1}{2}\mu_{0}^{T}\Sigma_{0}^{-1}\mu_{0}
\end{align*}
with $K_{0}^{*}:=(\Sigma_{0}^{-1}+2A_{0}^{*})^{-1}$, the dominant
contribution that the constant $c_{0}^{*}$ has to $v^{*}$ suggests
that it is important to estimate the constants in (\ref{eq:LQG_optimalpolicy})
to learn good policies. Moving from the bottom left to top left plot,
observe that increasing the location parameter $\xi$ from $1$ to
$4$ increases the magnitude of $(b_{t}^{*},e_{t}^{*})_{t\in[0:T]}$
but leaves $(A_{t}^{*},D_{t}^{*})_{t\in[0:T]}$ unchanged. This behaviour
is evident from the expressions of $(D_{t}^{*},e_{t}^{*})_{t\in[0:T]}$
and is unsuprising for $(A_{t}^{*})_{t\in[0:T]}$ as the parameter
$\xi$ does not alter the `structure' of the problem. The increase
in the magnitude of $(b_{t}^{*})_{t\in[0:T]}$ shows that the optimally
controlled SMC method achieves the desired terminal distribution
by initializing
\begin{equation}
\mu^{\psi^{*}}(\mathrm{d}x_{0})=\mathcal{N}(x_{0};K_{0}^{*}(\Sigma_{0}^{-1}\mu_{0}-b_{0}^{*}),K_{0}^{*})\mathrm{d}x_{0}\label{eq:LQG_initial}
\end{equation}
closer to the posterior distribution and taking larger drifts in
\begin{equation}
M_{t}^{\psi^{*}}(x_{t-1},\mathrm{d}x_{t})=\mathcal{N}\left(x_{t};K_{t}^{*}(P_{t}x_{t-1}+q_{t}-hb_{t}^{*}),hK_{t}^{*}\right)\mathrm{d}x_{t},\quad t\in[1:T].\label{eq:LQG_transition}
\end{equation}
Comparing the plots in the bottom row reveals that the off-diagonal
elements of $(A_{t}^{*},D_{t}^{*})_{t\in[0:T]}$ vanish under independence.
Therefore these terms should be taken into account for posterior distributions
that are very correlated.
Having obtained the optimal policy w.r.t. $\mathbb{Q}$ in a backward
sweep, we may then simulate the optimally controlled SMC method in
a forward pass. In Figure \ref{fig:LQG_compare}, we contrast the
output of the uncontrolled SMC method with that of the optimally
controlled.

\subsection{Approximate dynamic programming}

The ability to compute the optimal policy in this setting allows us
to evaluate the effectiveness of ADP algorithm (\ref{eq:ADP}) under
correct parameterization, i.e. select the function classes
\begin{align*}
\mathsf{F}_{0} & =\left\{ \varphi(x_{0})=x_{0}^{T}A_{0}x_{0}+x_{0}^{T}b_{0}+c_{0}:(A_{0},b_{0},c_{0})\in\mathbb{S}_{d}\times\mathbb{R}^{d}\times\mathbb{R}\right\} ,\\
\mathsf{F}_{t} & =\left\{ \varphi(x_{t-1},x_{t})=x_{t}^{T}A_{t}x_{t}+x_{t}^{T}b_{t}+c_{t}+x_{t-1}^{T}D_{t}x_{t-1}+x_{t-1}^{T}e_{t}\right.\\
 & \left.\quad\quad:(A_{t},b_{t},c_{t},D_{t},e_{t})\in\mathbb{S}_{d}\times\mathbb{R}^{d}\times\mathbb{R}\times\mathbb{S}_{d}\times\mathbb{R}^{d}\right\} ,\quad t\in[1:T].
\end{align*}
This choice corresponds to function classes
of the form (\ref{eq:linear_functionclass}), hence we can use linear
least squares to estimate the coefficients at each iteration of cSMC \textendash{} see (\ref{eq:LLS_AMatrix}) and
(\ref{eq:LLS_bVector}). If $(A_{t}^{i+1},b_{t}^{i+1},c_{t}^{i+1},D_{t}^{i+1},e_{t}^{i+1})$
denote the coefficients estimated at iteration $i\in[0:I-1]$ of Algorithm
\ref{alg:cSMC} in step 2(b), it follows that the policy at
iteration $i\in[1:I]$ is given by
\begin{align*}
-\log\psi_{0}^{(i)}(x_{0}) & =x_{0}^{T}A_{0}^{(i)}x_{0}+x_{0}^{T}b_{0}^{(i)}+c_{0}^{(i)},\\
-\log\psi_{t}^{(i)}(x_{t-1},x_{t}) & =x_{t}^{T}A_{t}^{(i)}x_{t}+x_{t}^{T}b_{t}^{(i)}+c_{t}^{(i)}+x_{t-1}^{T}D_{t}^{(i)}x_{t-1}+x_{t-1}^{T}e_{t}^{(i)},
\end{align*}
for $t\in[1:T],$ where $A_{t}^{(i)}:=\sum_{j=1}^{i}A_{t}^{j},b_{t}^{(i)}:=\sum_{j=1}^{i}b_{t}^{j},c_{t}^{(i)}:=\sum_{j=1}^{i}c_{t}^{j},D_{t}^{(i)}:=\sum_{j=1}^{i}D_{t}^{j},e_{t}^{(i)}:=\sum_{j=1}^{i}e_{t}^{j}$.
Observe from (\ref{eq:LQG_initial}) and (\ref{eq:LQG_transition})
that we need to impose the following positive definite constraints
\[
\Sigma_{0}^{-1}+2A_{0}^{(i)}\succ0,\quad I_{d}+2hA_{t}^{(i)}\succ0,\quad t\in[1:T],
\]
which can be done by projecting onto the set of real symmetric positive
definite matrices \cite{Higham_1988}. In our numerical implementation,
we find that these constraints are already satisfied when the step
size $h$ is sufficiently small. Although
the computational complexity of this ADP procedure is $O(N)$, it
scales quite costly in dimension $d$ as computation of least squares
estimators require inversion of $p\times p$ matrices where $p=d^{2}+3d+1$.
For problems with large $d$, it might be worth considering the use
of iterative linear solvers which offer reduced complexity. We note
that it is possible to avoid learning the $x_{t-1}$ dependency in
the policy $\psi_{t}^{*}(x_{t-1},x_{t}),t\in[1:T]$ and hence reduce
computational complexity drastically; we do not exploit this observation
here for simplicity of presentation but will do so for other applications.

\begin{figure}
\begin{centering}
\includegraphics[scale=0.4]{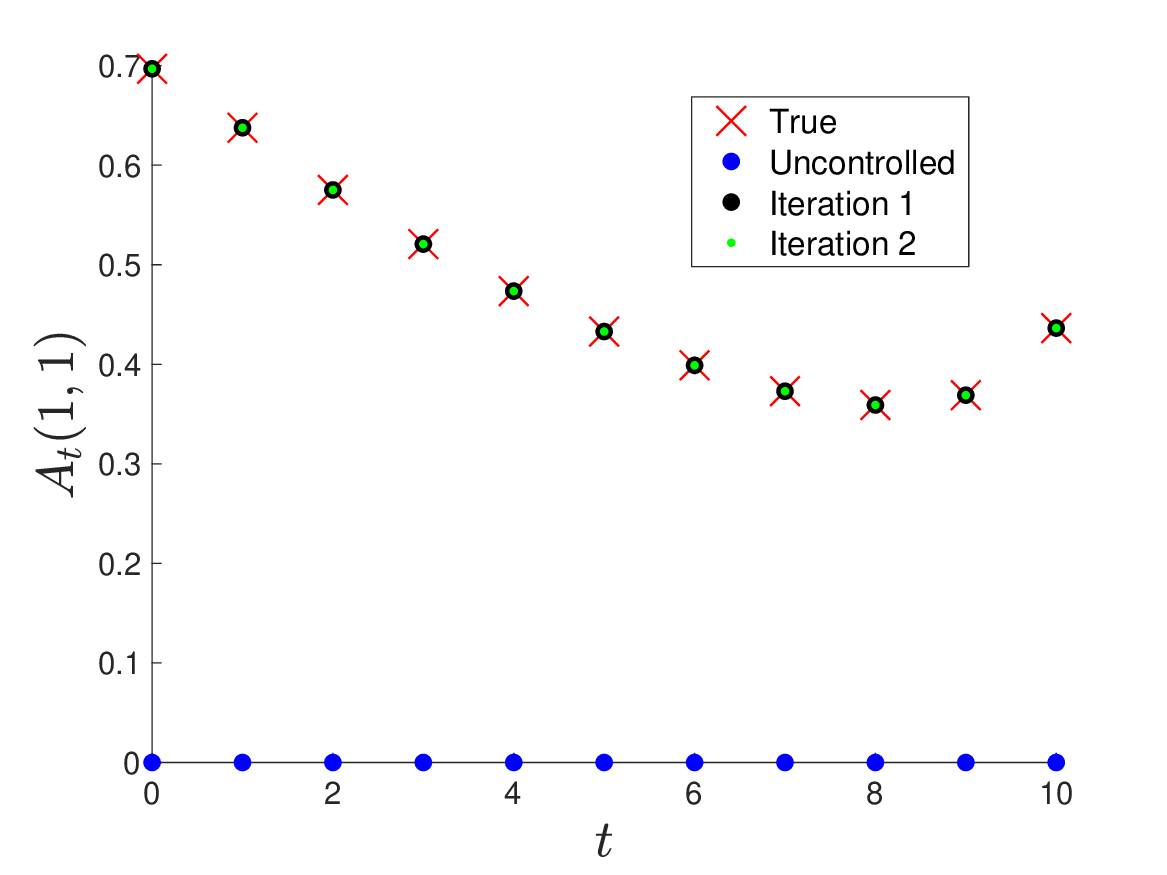}\includegraphics[scale=0.4]{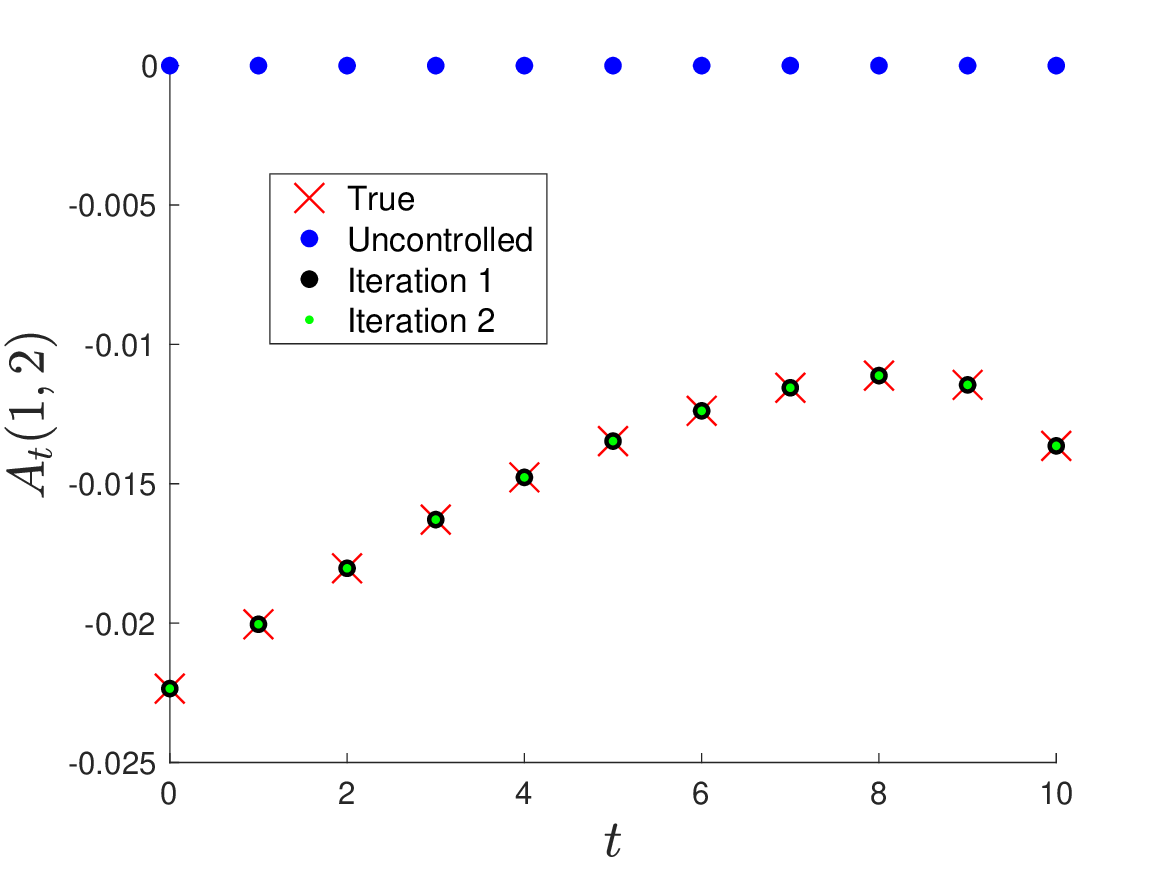}
\par\end{centering}
\begin{centering}
\includegraphics[scale=0.4]{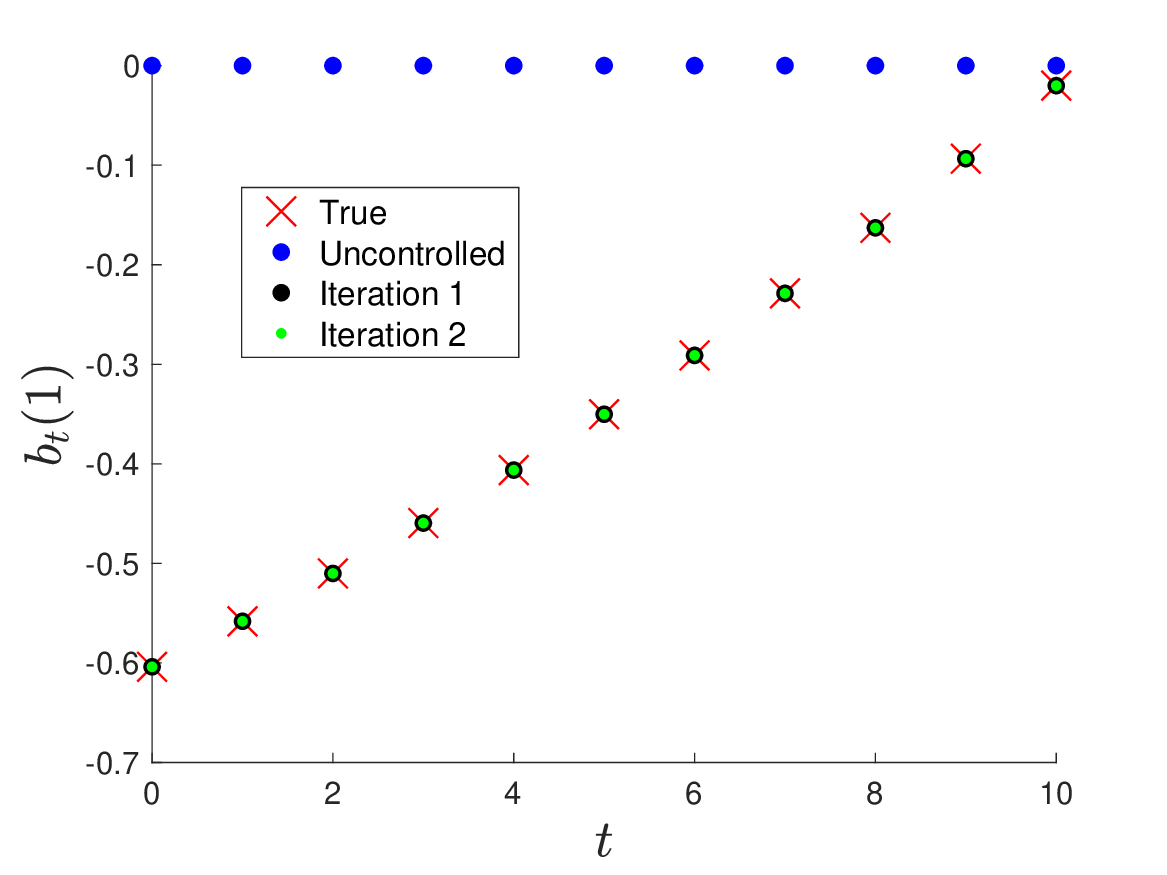}\includegraphics[scale=0.4]{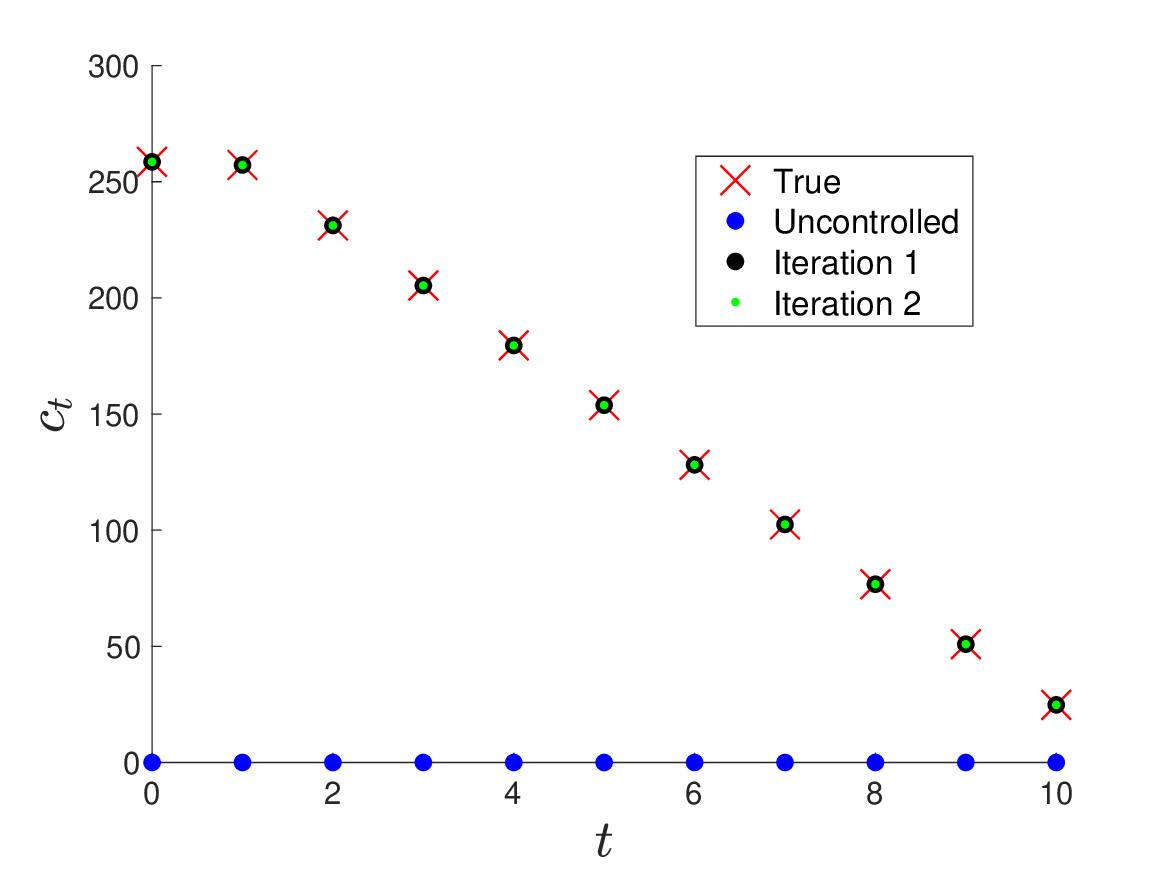}
\par\end{centering}
\caption{\label{fig:LQG_ADP}Coefficients of the optimal policy w.r.t. $\mathbb{Q}$
in LQG control against estimates obtained using ADP algorithm. The
problem setting is $d=32,\xi=20,\rho=0.8$
and the algorithmic settings of uncontrolled SMC are $N=2048,T=10,h=0.1,\lambda_{t}=t/T$. }
\end{figure}

Figure \ref{fig:LQG_ADP} displays the coefficients estimated by cSMC
with $I=2$ iterations. It is striking that with $N=2048$ particles,
we are able to accurately estimate, in a single ADP iteration, the
true coefficients in dimension $d=32$ (here $p=1121$). That said,
we typically need to increase $N$ with $d$ to prevent the Gram matrices
(\ref{eq:LLS_AMatrix}) from being ill-conditioned. Moreover, we find
that it is unnecessary to perform policy refinement in this example,
as the estimated policies converge immediately to an invariant distribution
that is very concentrated around the optimal policy (\ref{eq:LQG_optimalpolicy}),
which is the fixed point of the idealized algorithm in Theorem \ref{thm:iADP}
under correct parameterization. The performance of the resulting controlled
SMC method is indistinguishable from that in Figure \ref{fig:LQG_compare}.

\section{Bayesian logistic regression}

Consider a binary regression problem: each observation $y_{m}\in\{0,1\},m\in[1:M]$
is modelled as an independent Bernoulli random variable with probability
of success $\kappa(x^{T}X_{m})$, where $\kappa(u):=(1+\exp(-u))^{-1}$
for $u\in\mathbb{R}$ is the logistic link function, $x\in\mathsf{X}=\mathbb{R}^{d}$
denotes the unknown regression coefficients and $X_{m}\in\mathbb{R}^{d}$
the $m\in[1:M]$ row of a model matrix $X\in\mathbb{R}^{M\times d}$.
Hence the likelihood function and its gradient is given by
\[
\ell(x,y)=\exp\left(y^{T}Xx-\sum_{m=1}^{M}\log(1+\exp(x^{T}X_{m}))\right)
\]
and
\[
\nabla\log\ell(x,y)=X^{T}y-\sum_{m=1}^{M}(1+\exp(-x^{T}X_{m}))^{-1}X_{m}
\]
where $y=(y_{m})_{m\in[1:M]}\in\mathsf{Y}=\{0,1\}^{M}$ is a given
dataset of interest. Following \cite{Hanson_Branscum_Johnson_2014},
we specify a Gaussian prior distribution $\mu(\mathrm{d}x_{0})=\mathcal{N}(x_{0};\mu_{0},\Sigma_{0})\mathrm{d}x_{0}$
on $(\mathsf{X},\mathcal{X})=(\mathbb{R}^{d},\mathfrak{B}(\mathbb{R}^{d}))$
of the form $\mu_{0}=0_{d}$ and $\Sigma_{0}=\pi^{2}M/(3d)(X^{T}X)^{-1}$.

\subsection{Approximate dynamic programming\label{sec:logistic_ADP}}

\begin{figure}
\begin{centering}
\includegraphics[scale=0.4]{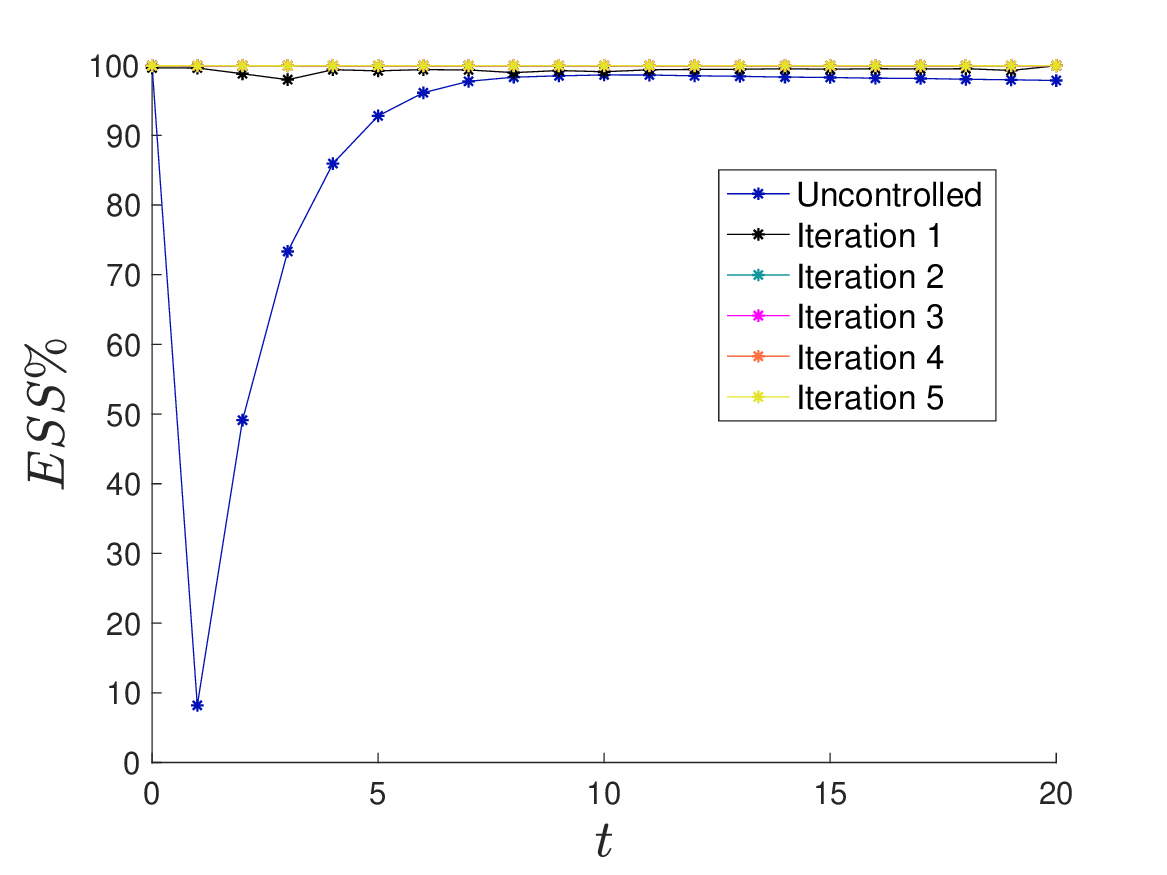}\includegraphics[scale=0.4]{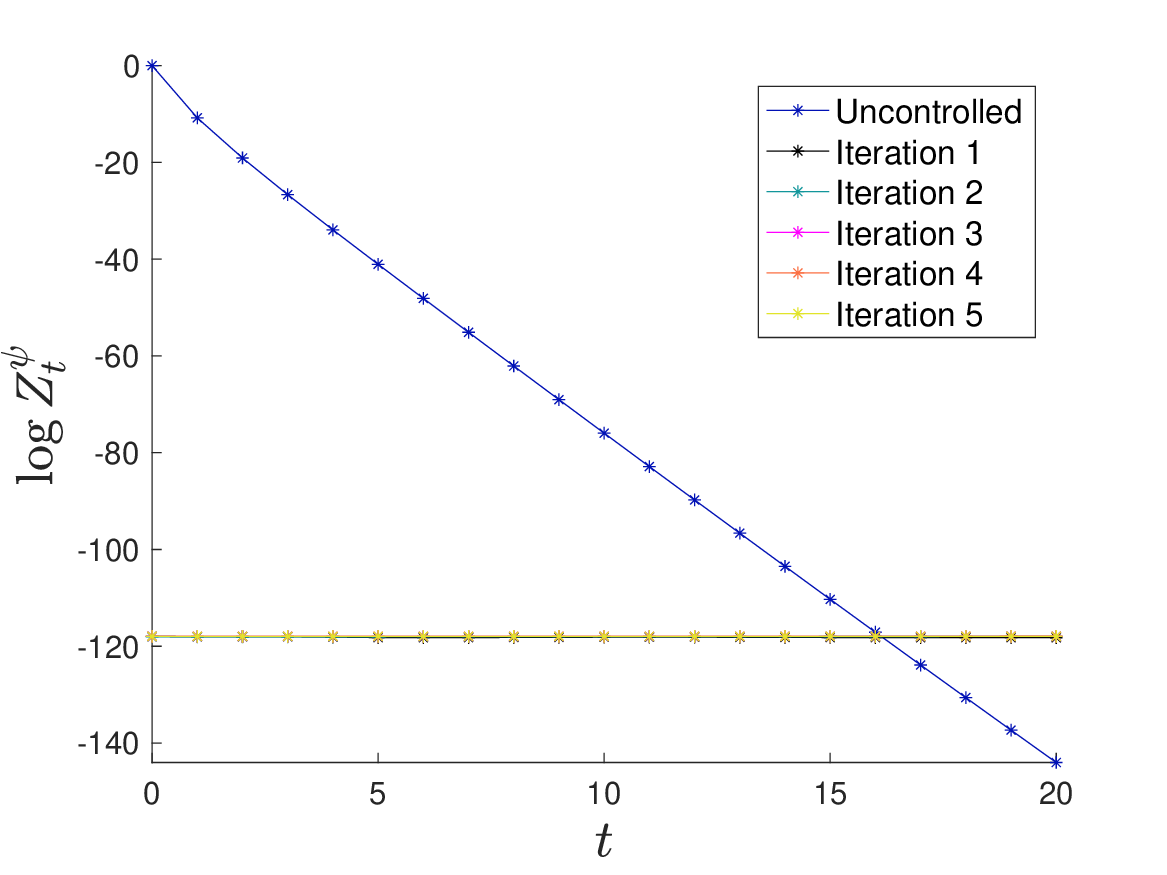}
\par\end{centering}
\begin{centering}
\includegraphics[scale=0.4]{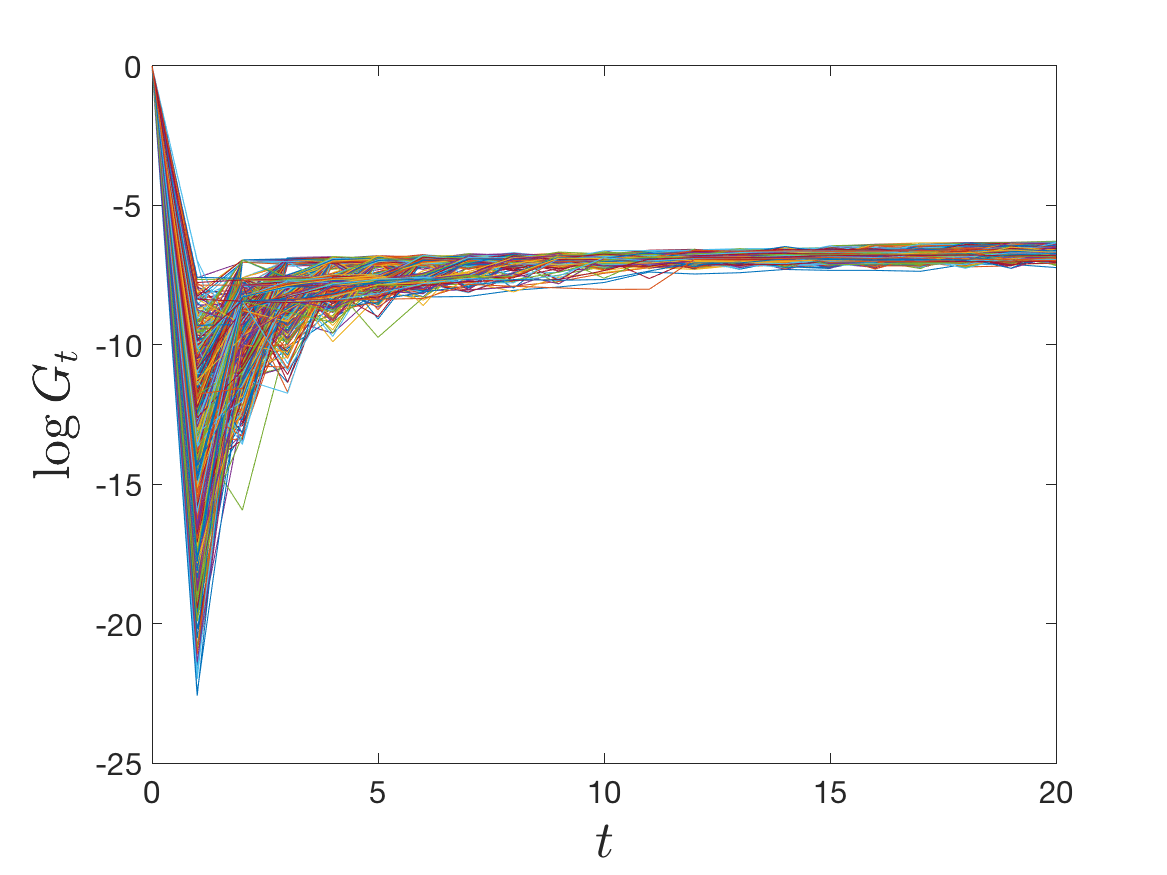}\includegraphics[scale=0.4]{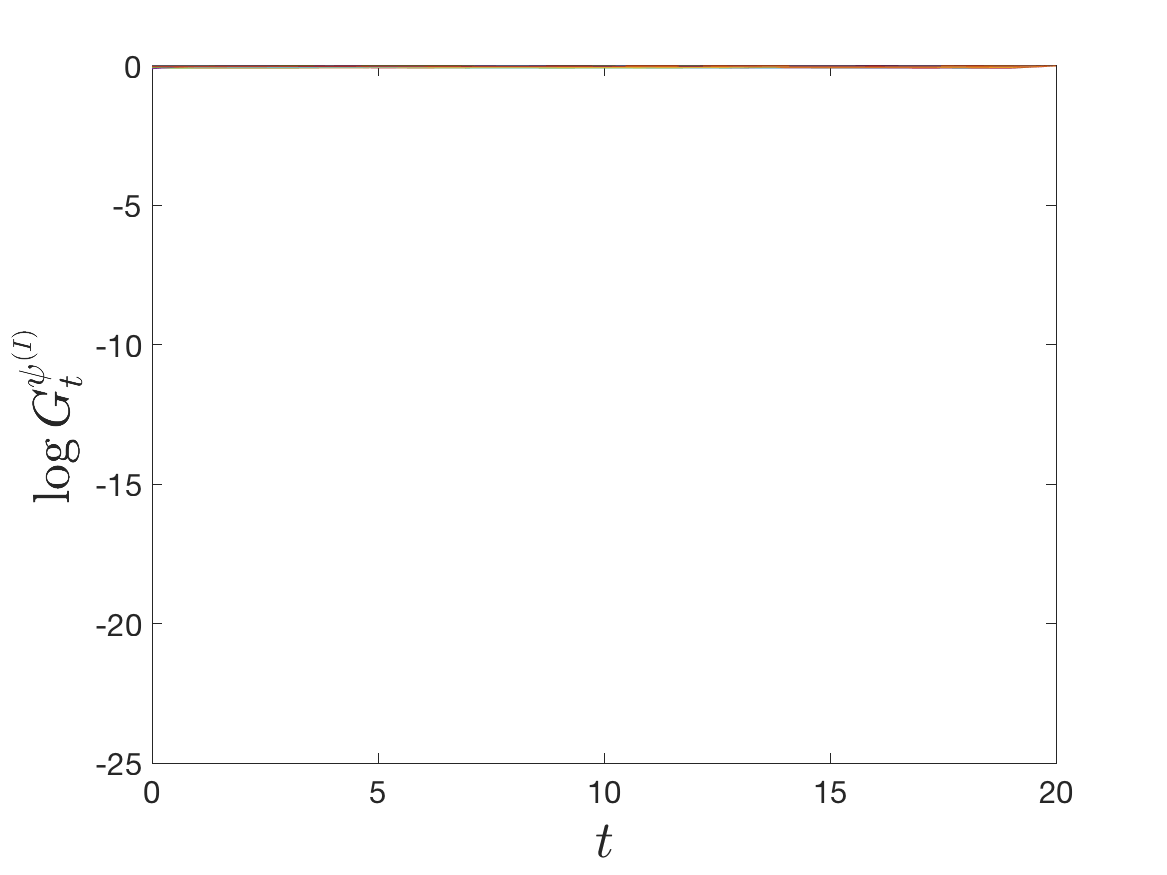}
\par\end{centering}
\caption{\label{fig:logistic_performance}Comparison of uncontrolled and controlled
SMC methods in terms of effective sample size (\emph{top left}),
normalizing constant estimation (\emph{top right}) and variance of
particle weights (\emph{bottom }row) when performing Bayesian logistic regression on
the Heart disease dataset. The algorithmic settings of cSMC are $I=5,N=1024,T=20,$
$h=1\times10^{-4},\lambda_{t}=t/T$. }
\end{figure}

In view of Proposition \ref{prop:logconcavity} and the previous section on LQG control,
we consider the function classes in (\ref{eqn:functionclass_staticmodels}).
As before, coefficients $(A_{t}^{i+1},b_{t}^{i+1},c_{t}^{i+1})_{t\in[0:T]}$
at each iteration $i\in[0:I-1]$ can be estimated by linear least
squares and the policy $\psi^{(i)}=(\psi_{t}^{(i)})_{t\in[0:T]}$
at iteration $i\in[1:I]$ has the form
\begin{align*}
-\log\psi_{0}^{(i)}(x_{0}) & =x_{0}^{T}A_{0}^{(i)}x_{0}+x_{0}^{T}b_{0}^{(i)}+c_{0}^{(i)},\\
-\log\psi_{t}^{(i)}(x_{t-1},x_{t}) & =x_{t}^{T}A_{t}^{(i)}x_{t}+x_{t}^{T}b_{t}^{(i)}+c_{t}^{(i)}-(\lambda_{t}-\lambda_{t-1})\log\ell(x_{t-1},y),
\end{align*}
for $t\in[1:T]$, where $A_{t}^{(i)}:=\sum_{j=1}^{i}A_{t}^{j},b_{t}^{(i)}:=\sum_{j=1}^{i}b_{t}^{j},c_{t}^{(i)}:=\sum_{j=1}^{i}c_{t}^{j}$
for $t\in[0:T]$. Assuming that the constraints $K_{0}^{(i)}:=(\Sigma_{0}^{-1}+2A_{0}^{(i)})^{-1}\succ0$,
$K_{t}^{(i)}:=(\Gamma^{-1}+2hA_{t}^{(i)})^{-1}\succ0$, $t\in[1:T]$
are satisfied or imposed, then sampling from
\[
\mu^{\psi^{(i)}}(\mathrm{d}x_{0})=\mathcal{N}\left(x_{0};K_{0}^{(i)}(\Sigma_{0}^{-1}\mu_{0}-b_{0}^{(i)}),K_{0}^{(i)}\right)\mathrm{d}x_{0}
\]
and
\begin{equation}
M_{t}^{\psi^{(i)}}(x_{t-1},\mathrm{d}x_{t})=\mathcal{N}\left(x_{t};K_{t}^{(i)}\{\Gamma^{-1}q_{t}(x_{t-1})-hb_{t}^{(i)}\},hK_{t}^{(i)}\right)\mathrm{d}x_{t},\label{eq:logistic_controlled_langevin}
\end{equation}
with $q_{t}(x_{t-1}):=x_{t-1}+h\Gamma\nabla\log\eta_{t}(x_{t-1})/2$
for $t\in[1:T]$ is feasible and evaluation of the twisted potentials
$(G_{t}^{\psi^{(i)}})_{t\in[0:T]}$ defined in (\ref{eq:twisted_potentials})
is tractable since
\begin{align*}
\mu(\psi_{0}^{(i)}) =&\det(\Sigma_{0})^{-1/2}\det(K_{0}^{(i)})^{1/2}\exp\left(\frac{1}{2}(\Sigma_{0}^{-1}\mu_{0}-b_{0}^{(i)})^{T}K_{0}^{(i)}(\Sigma_{0}^{-1}\mu_{0}-b_{0}^{(i)})\right)\\
&\times\exp\left(-\frac{1}{2}\mu_{0}^{T}\Sigma_{0}^{-1}\mu_{0}-c_{0}^{(i)}\right)
\end{align*}
and
\begin{align*}
M_{t}(\psi_{t}^{(i)})(x_{t-1}) & =\det(\Gamma)^{-1/2}\det(K_{t}^{(i)})^{1/2}\\
&\quad\times\exp\left(\frac{1}{2}h^{-1}(\Gamma^{-1}q_{t}-hb_{t}^{(i)})^TK_{t}^{(i)}(\Gamma^{-1}q_{t}-hb_{t}^{(i)})(x_{t-1})\right)\\
 & \quad\times\exp\left(-\frac{1}{2}h^{-1}(q_{t}^{T}\Gamma^{-1}q_{t})(x_{t-1})-c_{t}^{(i)}+(\lambda_{t}-\lambda_{t-1})\log\ell(x_{t-1},y)\right)
\end{align*}
for $t\in[1:T]$. We note that imposing $A_{t}=0$ and letting $b_{t}$
depend on the argument $x_{t-1}$ in (\ref{eqn:functionclass_staticmodels})
is related to the approach in \cite{Kappen_Ruiz_2016,Ruiz_Kappen_2017},
as (\ref{eq:logistic_controlled_langevin}) then corresponds to an
Euler-Maruyama discretization of a controlled diffusion with an additive
control $x_{t-1}\mapsto b_{t}^{(i)}(x_{t-1})$. For this application, we set
the pre-conditioner as $\Gamma=I_d$ and we illustrate in Figure
\ref{fig:logistic_performance} that
the parameterization (\ref{eqn:functionclass_staticmodels}) provides a
good approximation of the optimal policy on a particular dataset concerning
modeling of heart diseases.

%
%

%

%

\subsection{Comparison of algorithmic performance\label{sec:logistic_comparison}}

We now perform a comparison of algorithms on the analysis of three
real datasets\footnote{Datasets were downloaded from the \href{http://archive.ics.uci.edu/ml/}{UCI machine learning repository}
and standardized before analysis.} with different characteristics, in the same manner as Section \ref{sec:coxprocess}.
We use $N=1024$ number of particles in cSMC and select
the number of iterations using preliminary
runs \textendash{} see Figure \ref{fig:logistic_performance}. The
number of particles used in AIS is then chosen to match computational
cost, measured in terms of run time. These algorithmic settings
and the results obtained using $100$ independent repetitions each
of method are summarized in Table \ref{tab:logistic}. Although AIS
provides state-of-the-art results in complex scenarios for these models
\cite{Chopin_2017}, the comparison shows that for all datasets considered,
cSMC outperforms it and particularly so for the task of marginal likelihood
estimation by several orders of magnitude.

\hspace{-2cm}
\begin{table}
\begin{centering}
\begin{tabular}{|c|c|c|c|c|c|}
\cline{4-6}
\multicolumn{1}{c}{} & \multicolumn{1}{c}{} &  & \multicolumn{3}{c|}{\textbf{\footnotesize{}Dataset}}\tabularnewline
\cline{4-6}
\multicolumn{1}{c}{} & \multicolumn{1}{c}{} &  & \textit{\footnotesize{}Heart disease} & \textit{\footnotesize{}Australian credit} & \textit{\footnotesize{}German credit}\tabularnewline
\multicolumn{1}{c}{} & \multicolumn{1}{c}{} &  & {\footnotesize{}$\left(M=270,d=14\right)$} & {\footnotesize{}$\left(M=690,d=15\right)$} & {\footnotesize{}$\left(M=1000,d=25\right)$}\tabularnewline
\hline
\multirow{13}{*}{\begin{turn}{90}
\textbf{\footnotesize{}Algorithm}
\end{turn}} & \multirow{6}{*}{{\footnotesize{}AIS}} & {\footnotesize{}$N$} & {\footnotesize{}$1843$} & {\footnotesize{}$1843$} & {\footnotesize{}$2048$}\tabularnewline
 &  & {\footnotesize{}$h$} & {\footnotesize{}$5\times10^{-2}$} & {\footnotesize{}$3\times10^{-2}$} & {\footnotesize{}$1\times10^{-2}$}\tabularnewline
\cline{3-6}
 &  & {\footnotesize{}$\mathrm{ESS\%}$} & {\footnotesize{}$82.95\%$} & {\footnotesize{}$79.75\%$} & {\footnotesize{}$74.95\%$}\tabularnewline
 &  & {\footnotesize{}$\log Z$} & {\footnotesize{}$-118.0198\pm0.4383$} & {\footnotesize{}$-252.8699\pm1.5128$} & {\footnotesize{}$-527.4392\pm3.3088$}\tabularnewline
 &  & {\footnotesize{}$\mathrm{VAR}$} & {\footnotesize{}$1.92\times10^{-1}$} & {\footnotesize{}$2.29$} & {\footnotesize{}$10.95$}\tabularnewline
 &  & {\footnotesize{}$\mathrm{RMSE}$} & {\footnotesize{}$4.40\times10^{-1}$} & {\footnotesize{}$2.60$} & {\footnotesize{}$10.06$}\tabularnewline
\cline{2-6}
 & \multirow{7}{*}{{\footnotesize{}cSMC}} & {\footnotesize{}$I$} & {\footnotesize{}$3$} & {\footnotesize{}$4$} & {\footnotesize{}$3$}\tabularnewline
 &  & {\footnotesize{}$N$} & {\footnotesize{}$1024$} & {\footnotesize{}$1024$} & {\footnotesize{}$1024$}\tabularnewline
 &  & {\footnotesize{}$h$} & {\footnotesize{}$1\times10^{-4}$} & {\footnotesize{}$1\times10^{-3}$} & {\footnotesize{}$5\times10^{-4}$}\tabularnewline
\cline{3-6}
 &  & {\footnotesize{}$\mathrm{ESS\%}$} & {\footnotesize{}$99.99\%$} & {\footnotesize{}$99.95\%$} & {\footnotesize{}$99.91\%$}\tabularnewline
 &  & {\footnotesize{}$\log Z$} & {\footnotesize{}$-117.9638\pm0.0117$} & {\footnotesize{}$-250.7504\pm0.0101$} & {\footnotesize{}$-517.9299\pm0.0092$}\tabularnewline
 &  & {\footnotesize{}$\mathrm{VAR}$} & {\footnotesize{}$1.36\times10^{-4}$ $(\mathbf{1.41\times10^{3}})$} & {\footnotesize{}$1.03\times10^{-4}$ $(\mathbf{2.23\times10^{4}})$} & {\footnotesize{}$8.39\times10^{-5}$ $(\mathbf{1.31\times10^{5}})$}\tabularnewline
 &  & {\footnotesize{}$\mathrm{RMSE}$} & {\footnotesize{}$1.16\times10^{-2}$ $(\mathbf{37.86})$} & {\footnotesize{}$1.02\times10^{-2}$ $(\mathbf{2.55\times10^{2}})$} & {\footnotesize{}$9.11\times10^{-3}$ $(\mathbf{1.10\times10^{3}})$}\tabularnewline
\hline
\end{tabular}
\par\end{centering}
\caption{\label{tab:logistic}Algorithmic settings and performance of AIS and cSMC
when performing Bayesian logistic regression for each dataset.
Notationally, $N$ refers to the number of particles, $h$
the step size used in MALA for AIS and ULA for cSMC, and $I$ is the
number of iterations taken by cSMC. Both algorithms take $T=20$ time
steps for all datasets. Results were obtained using $100$ independent
repetitions each of method. The shorthand $\mathrm{ESS\%}$ denotes
the percentage of effective sample size averaged over time and repetitions,
$\log Z$ refers to the estimation of the normalizing constant in
logarithmic scale ($\pm$ a standard deviation), $\mathrm{VAR}$ is
the sample variance of these estimates over the repetitions and $\mathrm{RMSE}$
the corresponding root mean squared error, which we computed by taking
reference to an estimate obtained using many repetitions of a SMC
method with a large number of particles. Shown in bold are the gains
that cSMC offers relative to AIS.}
\end{table}

\section{Model specific expressions}
\subsection{Expressions for non-linear multimodal state space model\label{appendix:kitagawa}}
For notational simplicity, we write $\mu_{0}=0,\sigma_{0}^{2}=5$
and $\mu_{t}(x_{t-1}):=x_{t-1}/2+25x_{t-1}/(1+x_{t-1}^{2})+8\cos(1.2t)$
for $t\in[1:T].$ Assume that the policy $\psi^{(i)}=(\psi_{t}^{(i)})_{t\in[0:T]}$
at iteration $i\in[1:I]$ has the form (\ref{eq:kitagawa_currentpolicy}).
The initial distribution is given by
\[
\mu^{\psi^{(i)}}(\mathrm{d}x_{0})=\sum_{m\in[1:M]^{i}}A_{0,m}^{(i)}\mathcal{N}\left(x_{0};\mu_{0,m}^{(i)},(\sigma_{0}^{(i)})^{2}\right)\mathrm{d}x_{0}
\]
with
\[
\mu_{0,m}^{(i)}:=(\sigma_{0}^{(i)})^{2}\left(2\beta_{0}^{(i)}\xi_{0,m}^{(i)}+\mu_{0}\sigma_{0}^{-2}\right),\quad(\sigma_{0}^{(i)})^{2}:=\left(2\beta_{0}^{(i)}+\sigma_{0}^{-2}\right)^{-1},
\]
and
\[
A_{0,m}^{(i)}:=\frac{\alpha_{0,m}^{(i)}\exp\left(-\beta_{0}^{(i)}(\xi_{0,m}^{(i)})^{2}+(\mu_{0,m}^{(i)})^{2}(\sigma_{0}^{(i)})^{-2}/2\right)}{\sum_{n\in[1:M]^{i}}\alpha_{0,n}^{(i)}\exp\left(-\beta_{0}^{(i)}(\xi_{0,n}^{(i)})^{2}+(\mu_{0,n}^{(i)})^{2}(\sigma_{0}^{(i)})^{-2}/2\right)}.
\]
For each $t\in[1:T]$, the Markov transition kernel
\[
M_{t}^{\psi^{(i)}}(x_{t-1},\mathrm{d}x_{t})=\sum_{m\in[1:M]^{i}}A_{t,m}^{(i)}(x_{t-1})\mathcal{N}\left(x_{t};\mu_{t,m}^{(i)}(x_{t-1}),(\sigma_{t}^{(i)})^{2}\right)\mathrm{d}x_{t}
\]
with
\[
\mu_{t,m}^{(i)}(x_{t-1}):=(\sigma_{t}^{(i)})^{2}\left(2\beta_{t}^{(i)}\xi_{t,m}^{(i)}+\mu_{t}(x_{t-1})\sigma_{f}^{-2}\right),\quad(\sigma_{t}^{(i)})^{2}:=\left(2\beta_{t}^{(i)}+\sigma_{f}^{-2}\right)^{-1},
\]
and
\[
A_{t,m}^{(i)}(x_{t-1}):=\frac{\alpha_{t,m}^{(i)}\exp\left(-\beta_{t}^{(i)}(\xi_{t,m}^{(i)})^{2}+\mu_{t,m}^{(i)}(x_{t-1}){}^{2}(\sigma_{t}^{(i)})^{-2}/2\right)}{\sum_{n\in[1:M]^{i}}\alpha_{t,n}^{(i)}\exp\left(-\beta_{t}^{(i)}(\xi_{t,n}^{(i)})^{2}+\mu_{t,n}^{(i)}(x_{t-1}){}^{2}(\sigma_{t}^{(i)})^{-2}/2\right)}.
\]
Evaluation of the twisted potentials $(G_{t}^{\psi^{(i)}})_{t\in[0:T]}$
defined in (\ref{eq:twisted_potentials}) is tractable since
\begin{align*}
\mu(\psi_{0}^{(i)}) & =\frac{\sigma_{0}^{(i)}}{\sigma_{0}}\exp\left(-\frac{1}{2}\mu_{0}^{2}\sigma_{0}^{-2}\right)\sum_{m\in[1:M]^{i}}\alpha_{0,m}^{(i)}\exp\left(-\beta_{0}^{(i)}(\xi_{0,m}^{(i)})^{2}+\frac{1}{2}(\mu_{0,m}^{(i)})^{2}(\sigma_{0}^{(i)})^{-2}\right)
\end{align*}
and
\begin{align*}
M_{t}(\psi_{t}^{(i)})(x_{t-1}) =&\frac{\sigma_{t}^{(i)}}{\sigma_{f}}\exp\left(-\frac{1}{2}\mu_{t}(x_{t-1})^{2}\sigma_{f}^{-2}\right)\\
&\times\sum_{m\in[1:M]^{i}}\alpha_{t,m}^{(i)}\exp\left(-\beta_{t}^{(i)}(\xi_{t,m}^{(i)})^{2}+\,\frac{1}{2}\mu_{t,m}^{(i)}(x_{t-1})^{2}(\sigma_{t}^{(i)})^{-2}\right)
\end{align*}
for $t\in[1:T]$.

\subsection{Expressions for Lorenz-96 model\label{appendix:lorenz96}}

Suppose that the current policy is given by (\ref{eq:lorenz_currentpolicy})
and write
\[
\tilde{A}_{t}^{(i)}:=A_{t}^{(0)}+A_{t}^{(i)},\quad\tilde{b}_{t}^{(i)}:=b_{t}^{(0)}+b_{t}^{(i)},\quad\tilde{c}_{t}^{(i)}:=c_{t}^{(0)}+c_{t}^{(i)},
\]
for $t\in[0:T]$, where $(A_{t}^{(0)},b_{t}^{(0)},c_{t}^{(0)})_{t\in[0:T]}$ are the coefficients corresponding to APF.
If the constraints $K_{0}^{(i)}:=(\sigma_{f}^{-2}I_{d}+2\tilde{A}_{0}^{(i)})^{-1}\succ0$,
$K_{t}^{(i)}:=(\sigma_{f}^{-2}h^{-1}I_{d}+2\tilde{A}_{t}^{(i)})^{-1}\succ0,t\in[1:T]$
are satisfied or imposed, then sampling from
\[
\mu^{\psi^{(i)}}(\mathrm{d}x_{0})=\mathcal{N}\left(x_{0};-K_{0}^{(i)}\tilde{b}_{0}^{(i)},K_{0}^{(i)}\right)\mathrm{d}x_{0}
\]
and
\[
M_{t}^{\psi^{(i)}}(x_{t-1},\mathrm{d}x_{t})=\mathcal{N}\left(K_{t}^{(i)}\{\sigma_{f}^{-2}h^{-1}q(x_{t-1})-\tilde{b}_{t}^{(i)}\},K_{t}^{(i)}\right)\mathrm{d}x_{t},\quad t\in[1:T],
\]
is feasible and evaluation of the twisted potentials $(G_{t}^{\psi^{(i)}})_{t\in[0:T]}$
defined in (\ref{eq:twisted_potentials}) is tractable since
\begin{align*}
\mu(\psi_{0}^{(i)}) & =\sigma_{f}^{-d}\det(K_{0}^{(i)})^{1/2}\exp\left(\frac{1}{2}(\tilde{b}_{0}^{(i)})^{T}K_{0}^{(i)}\tilde{b}_{0}^{(i)}-\tilde{c}_{0}^{(i)}\right)
\end{align*}
and
\begin{align*}
M_{t}(\psi_{t}^{(i)})(x_{t-1}) & =\sigma_{f}^{-d}h^{-d/2}\det(K_{t}^{(i)})^{1/2}\exp\left(-\frac{1}{2}\sigma_{f}^{-2}h^{-1}(q^{T}q)(x_{t-1})-\tilde{c}_{t}^{(i)}\right)\\
 & \quad\times\exp\left(\frac{1}{2}(\sigma_{f}^{-2}h^{-1}q-\tilde{b}_{t}^{(i)})^{T}K_{t}^{(i)}(\sigma_{f}^{-2}h^{-1}q-\tilde{b}_{t}^{(i)})(x_{t-1})\right)
\end{align*}
for $t\in[1:T]$.

\subsection{Expressions for neuroscience model\label{appendix:neuroscience}}
Assume that the constraints $k_{0}^{(i)}:=(1+2a_{0}^{(i)})^{-1}>0$,
$k_{t}^{(i)}:=(\sigma^{-2}+2a_{t}^{(i)})^{-1}>0$, $t\in[1:T]$ are
satisfied or imposed. Then the initial distribution
\[
\mu^{\psi^{(i)}}(\mathrm{d}x_{0})=\mathcal{N}\left(x_{0};-k_{0}^{(i)}b_{0}^{(i)},k_{0}^{(i)}\right)\mathrm{d}x_{0}
\]
and the Markov transition kernels
\[
M_{t}^{\psi^{(i)}}(x_{t-1},\mathrm{d}x_{t})=\mathcal{N}\left(x_{t};k_{t}^{(i)}(\alpha\sigma^{-2}x_{t-1}-b_{t}^{(i)}),k_{t}^{(i)}\right)\mathrm{d}x_{t}
\]
for $t\in[1:T].$ Moreover, the twisted potentials $(G_{t}^{\psi^{(i)}})_{t\in[0:T]}$
defined in (\ref{eq:twisted_potentials}) can be evaluated since
\[
\mu(\psi_{0}^{(i)})=(k_{0}^{(i)})^{1/2}\exp\left(\frac{1}{2}k_{0}^{(i)}(b_{0}^{(i)})^{2}-c_{0}^{(i)}\right)
\]
and
\[
M_{t}(\psi_{t}^{(i)})(x_{t-1})=(k_{t}^{(i)})^{1/2}\sigma^{-1}\exp\left(\frac{1}{2}k_{t}^{(i)}(\alpha\sigma^{-2}x_{t-1}-b_{t}^{(i)})^{2}-\frac{1}{2}\sigma^{-2}\alpha^{2}x_{t-1}^{2}-c_{t}^{(i)}\right)
\]
for $t\in[1:T]$.

\end{document}